\documentclass[sigconf]{acmart}

\usepackage{amsmath}
\usepackage{amsfonts}
\usepackage{tabularx}
\usepackage{algorithm,algpseudocode}
\usepackage{array,multirow}
\usepackage{xspace}
\usepackage{pifont}
\usepackage{adjustbox}
\usepackage{xcolor}
\usepackage{colortbl}
\usepackage{pbox}
\usepackage{graphicx}
\usepackage{subfigure}
\usepackage{tikz}

\def\halfcheckmark{\tikz\draw[scale=0.4,fill=black](0,.35) -- (.25,0) -- (1,.7) -- (.25,.15) -- cycle (0.75,0.2) -- (0.77,0.2)  -- (0.6,0.7) -- cycle;}

\newtheorem{lemma}{Lemma}
\def\ojoin{\setbox0=\hbox{$\bowtie$}%
	\rule[0.15ex]{.22em}{.6pt}\llap{\rule[0.9ex]{.22em}{.6pt}}}

\def\fullouterjoin{\mathbin{\ojoin\mkern-5.5mu\bowtie\mkern-5.5mu\ojoin}}

\def\halfcheckmark{\tikz\draw[scale=0.3,fill=black](0,.3) -- (.2,0) -- (1,.7) -- (.25,.15) -- cycle (0.75,0.2) -- (0.77,0.2)  -- (0.6,0.7) -- cycle;}

\newcommand{\Ours}{\emph{FactorJoin}\xspace}
\newcommand{\CE}{\textsf{CardEst}\xspace}

\mathchardef\mhyphen="2D
\newcommand\revise[1]{\textcolor{black}{#1}}

\definecolor{mygrey}{RGB}{230,230,240}

\newcommand{\ziniu}[1]{{{\color{olive}(Ziniu: #1)}}}

\begin{document}
		\title{FactorJoin: A New Cardinality Estimation Framework \break for Join Queries}
\author{
 %\alignauthor
 Ziniu Wu,
 Parimarjan Negi,
 Mohammad Alizadeh,
 Tim Kraska,
 Samuel Madden
}
\affiliation{%
  \institution{Massachusetts Institute of Technology, CSAIL}
  \country{Cambridge, MA, USA}
}
\email{{ziniuw, pnegi, alizadeh, kraska, madden}@mit.edu}

\begin{abstract}
Cardinality estimation is one of the most fundamental and challenging problems in query optimization. Neither classical nor learning-based methods yield satisfactory performance when estimating the cardinality of the join queries.
They either rely on simplified assumptions leading to ineffective cardinality estimates or build large models to understand the data distributions, leading to long planning times and a lack of generalizability across queries. 
		
In this paper, we propose a new framework \Ours for estimating join queries. \Ours combines the idea behind the classical \emph{join-histogram} method to efficiently handle joins with the learning-based methods to accurately capture attribute correlation.
Specifically, \Ours scans every table in a DB and builds single-table conditional distributions during an offline preparation phase. When a join query comes, \Ours translates it into a factor graph model over the learned distributions to effectively and efficiently estimate its cardinality.
%\Ours only needs to scan every table in a DB instance, build their statistics using practically any single-table estimator approach, and automatically produce the effective and efficient estimates of the join queries during query runtime.
%
Unlike existing learning-based methods, \Ours does not need to de-normalize joins upfront or require executed query workloads to train the model.
Since it only relies on single-table statistics, \Ours has small space overhead and is extremely easy to train and maintain.
In our evaluation, \Ours can produce more effective estimates than the previous state-of-the-art learning-based methods, with 40x less estimation latency, 100x smaller model size, and 100x faster training speed at comparable or better accuracy. In addition, \Ours can estimate 10,000 sub-plan queries within one second to optimize the query plan, which is very close to the traditional cardinality estimators in commercial DBMS.

\end{abstract}

\maketitle

\newcommand{\system}{Cardinal\xspace}
\newcommand{\cmark}{\ding{51}}%
\newcommand{\cmarkcur}{\color{blue}\ding{51}}%
\newcommand{\cmarktodo}{\color{red}\ding{51}}%
\newcommand{\xmark}{$-$}%
\newcolumntype{R}[2]{%
	>{\adjustbox{angle=#1,lap=\width-(#2)}\bgroup}%
	l%
	<{\egroup}%
}
\newcommand*\rot{\multicolumn{1}{R{90}{0em}|}}
\newcommand{\rott}{\multicolumn{1}{|R{60}{-1em}|}}

\makeatletter
\newcommand{\thickhline}{%
	\noalign {\ifnum 0=`}\fi \hrule height 1pt
	\futurelet \reserved@a \@xhline
}
\newcolumntype{"}{@{\hskip\tabcolsep\vrule width 1pt\hskip\tabcolsep}}
\makeatother

%\pagestyle{\vldbpagestyle}

%\begingroup
%\renewcommand\thefootnote{}\footnote{\noindent
%	$\#$ The first two authors contribute equally to this paper. \\
%	$*$ Corresponding author.
%}
%\addtocounter{footnote}{-1}
%\endgroup

\section{Introduction}
	Cardinality estimation (\CE) is a critical component of modern database query optimizers.
	The goal of \CE is to estimate the result size of each query operator (i.e., filters and joins), allowing the optimizer to select the most efficient join ordering and physical operator implementations. 
	\revise{An ideal \CE method satisfies several properties:  it is \emph{effective} at generating high-quality query plans, \emph{efficient} so that it minimizes estimation latency, and easy to {\it deploy} in that it has a small model size, fast training times, and the ability to scale with the number of tables and generalize to new queries.}
	Unfortunately, existing \CE techniques do not satisfy at least one of these properties. 
	The reason behind these failures can largely be reduced to the handling of two related problems: 
	(1) how to characterize attribute correlations and (2) how to model the distribution of join-keys, which directly determine the size of joins.
	%(1) correlations between attributes of the same table, which we call intra-table correlations and,
	%(2) correlation between different join keys, which we call inter-table correlations. 
	
	\smallskip
	\noindent \underline{\textbf{Background:}}
	Many deployed \CE approaches are still based on the Selinger model~\cite{selinger1979access} and assume \emph{attribute independence}, where the model ignores correlations between attributes, and \emph{join-key uniformity}, \revise{where the model further assumes that join keys have uniformly distributed values.}
	%, e.g., every tuple from one table is equally likely to join with any tuple from another table (seeFigure~\ref{fig: demo}(a)).
	%By assuming {attribute independence} the Selinger model ignores intra-table correlation between attributes.
	%(e.g., every user has the same number of followers). and thus there is no inter-table correlation. 
    These assumptions provide a simple way of estimating the cardinality of join queries using only single-table statistics (e.g., histograms) over single attributes, which are easy to create and maintain. 
	\revise{As a result, the Selinger-based \CE techniques are extremely efficient and easy to deploy but at the cost of effectiveness because most real-world datasets contain complex attribute correlations and skewed join-key distributions.}
	%(low latency) and easy to deploy (low storage and training overhead, etc.) but at the cost of effectiveness. 
	%Because most real-world datasets contain complex correlations between attributes and skewed join-key distributions, approaches based on these assumptions often produce severe \CE errors.

\begin{figure*}
\vspace{-.2in}
	\centering
	\includegraphics[width=16cm]{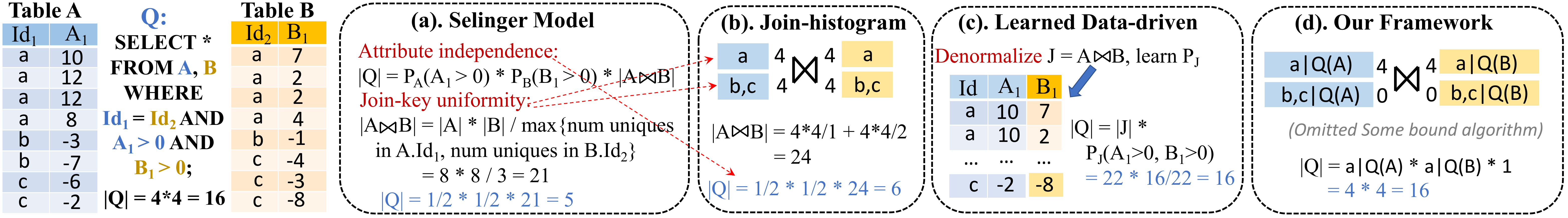}
	%\vspace{-1em}
	\caption{Existing \CE approaches for handling join queries. (a) The Selinger model writes the cardinality of a two-table join query as the product of filter selectivity $P_A, P_B$ on both tables and the estimated join size $|A \Join B|$. Using join-key uniformity, this is estimated as the product of table sizes $|A| * |B|$ divided by the maximal number of unique values of the join keys. (b) The join-histogram bin the domain of join keys $Id_1, Id_2$ as histograms, assumes join-key uniformity within each bin. (c) The learned data-driven methods de-normalize $J = A \Join B$ and learn the distribution $P_J$ to estimate the cardinality of $Q$. (d) Our framework captures the correlation between join keys and filter predicates to accurately estimate $|Q|$ without de-normalization.}
	\label{fig: demo}
	%\vspace{-1em}
\end{figure*}

	To overcome the limitations of these assumptions, many commercial database systems try to \revise{relax} them to increase their effectiveness.
    \revise{For example, past works ~\cite{ioannidis2003history, ioannidis1993optimal, dell2007join} propose and some systems (e.g. Oracle~\cite{oracle}) employ {\it join-histograms} that relax the \emph{join-key uniformity assumption}.}
    %by partially accounting for inter-table correlations.
	Specifically, these methods build frequency histograms on the join keys and then construct the join distribution by ``multiplying'' the histograms 
	%\tim{add ``to build the joined''? distributed. I am afraid that multiply will not be clear for most reviewers) }
	%\ziniu{but we have the figure to show the details}
	to estimate the join size (see Figure~\ref{fig: demo}(b)). In this way, these methods can more precisely capture the join-key distributions to generate more effective estimates of the join cardinality. However, they still assume that the join keys are uniformly distributed within each bin of the histogram.
	%and do not account for attribute correlations.
%instead of over the entire domain. \srm{One sentence on pros and cons of this method? I.e., avoids inter-table correlation but still has intra table?}
     
    Similarly, there have been works on relaxing the \emph{attribute independence assumption}. %assumptions made to handle intra-table correlations between attributes within a table.
    For example, multi-dimensional histograms~\cite{deshpande2001independence, gunopulos2000approximating, gunopulos2005selectivity, muralikrishna1988equi, wang2003multi, liu2021lhist}, self-tuning histograms~\cite{bruno2001stholes, srivastava2006isomer, khachatryan2015improving, fuchs2007compressed}, and singular value decomposition~\cite{poosala1997selectivity}, have been proposed to capture the correlation between attributes. 
    However, their effectiveness and/or efficiency are still not satisfactory~\cite{yang2019deep, han2021cardinality}.
    %\srm{What do we mean by performance here - accuracy, or runtime?}.
    More recently, learning-based methods which learn the underlying data distributions \cite{kipf2018learned, hilp2019deepdb, zhu2020flat, yang2019deep}  have been proposed to more accurately and compactly represent the attribute correlations within a {\emph single table}, but these still have efficiency concerns that we detail below.
    
    %However, they can not efficiently handle joins.
    
    %actually transform the inter-table correlation problem into an intra-table correlation problem.
    %\ziniu{I don't think it is true for query-driven methods.} \srm{Right - rather than saying they don't do inter-table, we should ding these methods based on their optimization time and lack of generalizability.  We say this below but the way this is written now makes it sound like all learning-based methods do this transformation.  The distinction between "query-driven learning" and "data-driven learning" is not clear - I don't even know what it means!}
    
    Specifically, the learned data-driven methods~\cite{getoor2001selectivity, tzoumas2011lightweight, yang2020neurocard, wu2020bayescard, zhu2020flat, hilprecht2019deepdb} analyze data and build distributions for all join patterns in a database.
    They need to de-normalize the joined tables and add a potentially exponential number of extra columns.
    %assume that the data is first de-normalized \footnote{Some data-driven techniques propose to use heuristics to only de-normalize a subset of tables.} into a single large table with a potentially exponential number of extra columns. 
    Then, they build distributions over the de-normalized tables to characterize all attribute correlations and handle joins (see also Figure~\ref{fig: demo}(c).
    \revise{This allows the learned data-driven methods to be highly accurate for join estimates at the cost of slow training time and large model size (i.e., worse deployability).} 
    %Moreover, these methods are often orders-of-magnitude slower than their traditional counterparts (i.e., worse efficiency) as they require complicated models to capture all the peculiarities of the large de-normalized tables. 
    Alternatively, learned query-driven methods~\cite{kipf2018learned, sun2019end, liu2021fauce} circumvent the \emph{join-key uniformity} assumption by building supervised models to map the join queries to their cardinalities. However, they require an impractical number of executed queries to train their models, which is unavailable to new DB instances and do not generalize well to unseen queries.
    %Alternatively, query-driven learning-based methods \cite{kipf2018learned, sun2019end, liu2021fauce} improve this assumption by taking the workload into consideration but still implicitly de-normalize the data for training.
    %\srm{Need to state why these methods do this - not clear why it is necessary? Give 1-2 sentences?}

%	The learned query-driven methods~\cite{kipf2018learned, sun2019end, liu2021fauce} require an impractical amount of executed queries to train their model, which is unavailable to new DB instances and not generalizable to unseen queries.
    \smallskip	
    \noindent \underline{\textbf{Our Approach:}}
    Surprisingly, we are not aware of any work which tries to combine the advances of learning-based methods for accurately characterizing correlations with the idea of \emph{join-histograms}. % to efficiently estimate cardinalities of joins. 
%    In fact, the original join-histogram approach  uses the \emph{attribute independence assumption} for efficiency~\cite{dell2007join}, whereas the modern learning-based methods can not efficiently handle joins.
%    \noindent \underline{\textbf{Contribution:}}	
    In this work, we develop a novel \CE framework called \Ours that combines these two approaches, i.e., using 
    \emph{join-histograms} to efficiently handle joins coupled with learning-based methods to accurately capture attribute correlation.
    %\emph{join-histogram} technique to capture inter-table correlation with learning-based methods to improve the intra-table correlation. 
    \revise{This is not a straightforward extension.}
    %While this might sound like a straightforward extension, it is not. 
    \revise{As we will show, adapting the idea of \emph{join-histograms} while preserving the benefits of using learning-based methods for attribute correlations is non-trivial}.
    \Ours only builds sophisticated models to capture the correlations within a single table.
    %Like other learning-based methods, \Ours builds sophisticated models to capture the correlations but only within a single table. 
    The choice of the model to capture correlation is orthogonal to the techniques of \Ours, however, we do require the model to be able to provide conditional distributions of one or more keys. 
    In our current implementation of \Ours, we use \revise{sampling~\cite{lipton1990practical} and Bayesian Networks~\cite{koller2009probabilistic}} as they are fast to train and execute. 
    
    These single-table conditional distributions are then combined using a {\it factor graph model}~\cite{loeliger2004introduction}.
    Specifically, \Ours translates a join query $Q$ into a factor graph over single-table data distributions. 
This allows us to formulate the problem of estimating the cardinality of $Q$ as a well-studied inference problem~\cite{koller2009probabilistic, loeliger2004introduction, lauritzen1996graphical} on this factor graph.
    Conceptually, there exist some similarities to the \emph{join-histogram} technique. 
    However, our formulation allows \Ours to generalize to cyclic and self-joins. 
    It also improves the computation efficiency by relying on existing inference algorithms~\cite{mackay2003information, kschischang2001factor}, and facilitates scaling to arbitrary join sizes.
    Finally, in contrast to the idea of \emph{join-histogram} and many other techniques in the space, we propose to actually not estimate the expected join cardinality, but rather use a probabilistic upper bound. 
    As shown by others~\cite{cai2019pessimistic, atserias2008size, abo2017shannon, hertzschuch2021simplicity}, under-estimation is often worse than over-estimation, and tailoring the estimate to a probabilistic upper bound can significantly improve the end-to-end query performance. 

	%In a nutshell, when compared with the traditional methods, \Ours can generate much more effective estimates that are comparable to the state-of-the-art learned methods. When compared with the learned methods, \Ours relies only on single-table statistics for efficiency, compactness, and generalizability.

    %Specifically, using only single table statistics, \Ours estimates the cardinality of join queries by joining histograms but eliminates their simplifying assumptions using machine learning techniques.

	As a result,  \Ours is able to estimate the cardinality of join queries using only single table statistics with similar or better estimation effectiveness than the state-of-the-art (SOTA) learning-based techniques. 
	When compared with the learned data-driven methods, \Ours has a much smaller model size, %(i.e., smaller tables are easier to learn),
	faster training/updating speed, and easier for system deployment because we do not need to denormalize the join tables or add exponentially many extra columns. 
	In addition, \Ours supports all forms of equi-joins, including cyclic joins and self joins, as well as complex base table filter predicates, including disjunctive filter clauses and string pattern matching predicates \revise{because it can flexibly plug-in various types of base-table estimators to support them.}
	These queries are not supported by the existing learned data-driven methods~\cite{yang2020neurocard, hilprecht2019deepdb, zhu2020flat}, which require the join template to be a tree.
	Unlikely the learned query-driven methods~\cite{kipf2018learned, sun2019end, liu2021fauce}, \Ours does not depend on the executed query workload, so it is robust against data updates and workload shifts, can be quickly adapted to a new DB instance, and is generalizable to new queries.
	However, in presence of query workload, \Ours can incorporate this information to further optimize the model construction process. 
	%More importantly, by relying on simple Bayesian models for single tables with a straightforward way to combine them using factor graphs, \Ours inference time is closer to the execution time of traditional methods, like the Selinger model, than deep-learning-based techniques. This allows \Ours to be used not only for long-running queries but for more realistic workloads, which consist of short and long-running queries. 
	%\ziniu{This claim might be too strong, the experiment does not show we can outperformance Postgre on short-running queries. So I'll just remove it.}

    We integrate \Ours into Postgres' query optimizer and evaluate its end-to-end query performance on two well-established and challenging real-world \CE benchmarks: STATS-CEB~\cite{han2021cardinality} and IMDB-JOB~\cite{leis2015good}. On the STATS-CEB, \Ours achieves near-optimal performance in end-to-end query time. It has comparable performance to the previous SOTA learned data-driven model (FLAT~\cite{zhu2020flat}) in terms of estimation effectiveness but with 40x lower estimation latency, 100x smaller model sizes, and 100x faster training times. On the IMDB-JOB, our framework achieves the SOTA performance in end-to-end query time (query execution plus planning time). Specifically, the pure execution time of our framework is comparable to the previous SOTA method (pessimistic estimator~\cite{cai2019pessimistic}), but our estimation latency is 100x lower, making our overall end-to-end query time significantly faster. 
	Specifically, our framework can estimate 10,000 sub-plan queries in one second to optimize the query plan, which is close to the planning time of the traditional \CE method used in Postgres. 
	We also carry out a series of controlled ablation studies to demonstrate the robustness and advantages of different technical novelties of \Ours.

	%\tim{why did you remove that it can incorporate it? I think it is important}

	In summary, our main contributions are:
	
	$\bullet$ We formulate the problem of estimating the cardinality of join queries in its general form as a factor graph inference problem involving only single-table data distributions (Section~\ref{sec: framework}).
	
	$\bullet$ We propose a new binning and upper bound-based algorithm to approximate the factor graph inference (Section~\ref{sec: bound}).
	
	$\bullet$ We design \revise{attribute causal relation exploration} and progressive estimation of sub-plan queries techniques to improve the efficiency of our framework (Section~\ref{sec: efficiency}).
	
	$\bullet$ We conduct extensive experiments to show the advantages of our framework (Section~\ref{sec: experiments}).
	
	Before describing the details of \Ours, we begin with a detailed description of existing \CE approaches.

\section{Background and analysis}
\begin{table*}
	\centering
	\includegraphics[width=17.5cm]{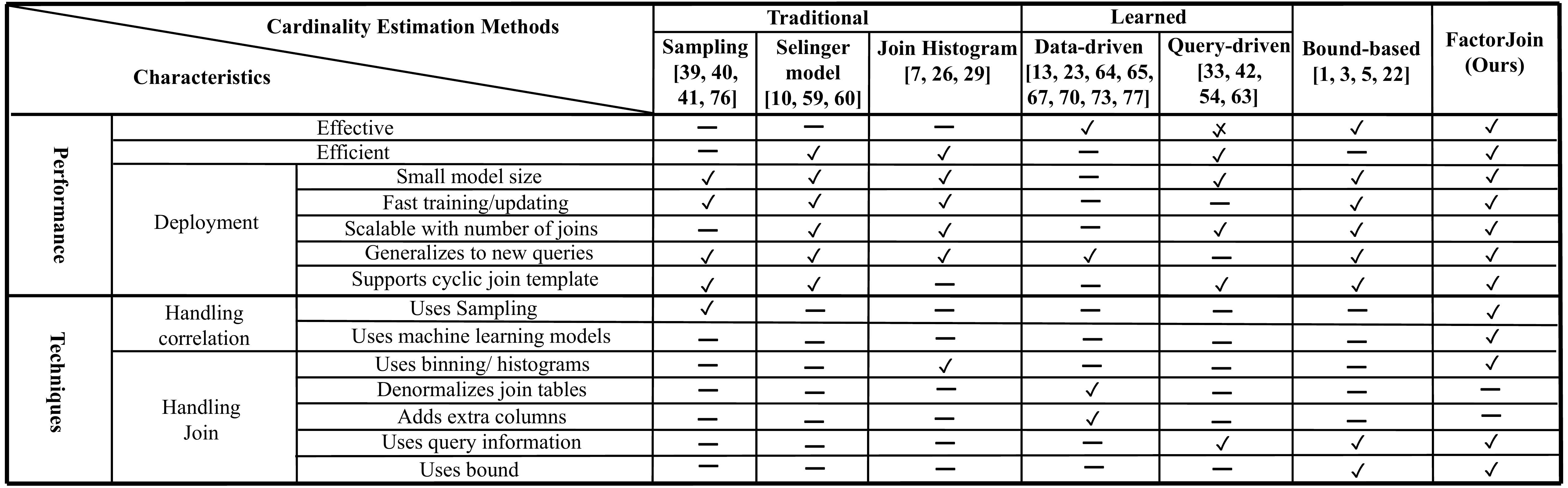}
	\caption{Summary of existing \CE methods.}
	\label{tab: summary}
	%\vspace{-2em}
\end{table*}

\label{sec: prelim}
In this section, we first define the  \CE problem and then analyze existing \CE methods and how our approach differs from them.

\subsection{\CE problem definition}
\noindent\underline{\textbf{Single table \CE:}} Let $A$ be a table with $n$ attributes $A_1, \ldots, A_n$. A single table selection query $Q(A)$ on $A$ can be viewed a conjunction or disjunction of filter predicates over each attribute, e.g.  $Q(A) = (A_1 \in [0, 5]) \ \lor $ $(A_2$ LIKE `\%An\%'$) \land (A_5 \leq -10)$.

Let $|Q(A)|$ denote the cardinality of $Q(A)$, i.e. the number of tuples in $A$ satisfying $Q(A)$. Let $P_A(Q(A))$ denotes the selectivity (probability) of tuples satisfying $Q(A)$. Then, we have $|Q(A)| = P_A(Q(A)) * |A|$, where $|A|$ denotes total number of tuples in $A$. 

\smallskip
\noindent\underline{\textbf{\CE for multi-table join queries:}} \revise{Consider a database instance $\mathcal{D}$ with m tables $A, B, \ldots, M$, a query $Q$ consists of a join graph representing the join conditions among the selected tables and a set of base-table filter predicates.}
\revise{For simplicity, we abuse the notation and use $Q(I)$ to denote the filter predicates of $Q$ on table $I$.}
%(denotes the filter on table $I$ as $Q(I)$). 

Let $\Omega$ denote the denormalized table resulting from the join conditions of $Q$. Then, the cardinality $|Q| = P_{\Omega}(Q(A), Q(B), \ldots, Q(M)) * |\Omega|$, where $Q(I)$ can be empty set if table $I$ is not touched by $Q$. Estimating $Q$ is particularly challenging because there can exponential number of join patterns in the database, each associated with a unique $\Omega$ and distinct probability distribution $P_{\Omega}$. Therefore, to accurately estimate different join queries, the \CE methods need to capture exponential many complicated data distributions.

\subsection{Analysis of existing \CE methods}
\label{subsect: related}
Since the focus of this paper is multi-table join queries, we defer the discussion of single-table estimators to Section~\ref{sec: related} and
focus on analyzing different approaches for handling join queries. We can categorize these approaches into four classes: \emph{traditional}, \emph{learned data-driven}, \emph{learned query-driven}, and \emph{bound-based}. We summarized their characteristics in Table~\ref{tab: summary} and provide their details as follows.

\smallskip
\noindent \underline{\textbf{Traditional methods:}}
The sampling-based methods~\cite{lipton1990practical, leis2017cardinality,zhao2018random, li2016wander} construct a small sample on each table and join these samples to estimate the cardinality of join queries. 
The traditional histogram-based methods generally adopt the attribute independence and join uniformity assumptions to decompose the join queries as a combination of single table estimates.
Taking a query $Q$ joining two tables $A$ and $B$ as an example, these assumptions will simplify the cardinality $|Q| = P_{\Omega}(Q(A), Q(B)) * |\Omega|$ as $P_A(Q(A)) * P_B(Q(B)) * \overline{|\Omega|}$, where $P_A(Q(A))$ is the estimated single table selectivity. There exist two widely-used approaches to estimate the size for the denormalized join table $\overline{|\Omega|}$. First, the Selinger models ~\cite{selinger1979access, psql2020, mysql2020} assume that join keys have uniformly distributed values, collect the number of distinct values (NDV) on the join keys from both tables, and estimate $\overline{|\Omega|}$ as $|A| * |B| / max\{NDV(A), NDV(B)\}$.
Second, the \emph{join-histogram} methods~\cite{ioannidis2003history, ioannidis1993optimal, dell2007join} first bin the domain of the join keys as histograms and assume join uniformity assumption for the values in each histogram bin. Then they apply the \emph{distinct values} methods for each bin and sum over all bins.

Our work follows the convention of the \emph{join-histogram} methods but avoids the simplifying assumptions, which can not be trivially achieved. Naive approaches such as building the full multi-dimensional histograms can resolve the attribute independence assumption but would introduce unaffordable storage overhead. 
Many follow-up works~\cite{ioannidis1993universality, ioannidis1995balancing, ioannidis1991propagation} explore different histogram strategies to reduce the error caused by the join uniformity assumption. They either take an impractical amount of time to construct the histograms or cannot provide high-quality estimates.

In short, the traditional methods combine the estimates from single tables to estimate the join queries. This approach is very efficient, easy to train, update and maintain, thus perfect for system deployment. However, their simplifying assumptions can generate erroneous estimates and poor query plans.

%Multi-dimensional histograms~\cite{poosala1997selectivity, deshpande2001independence, gunopulos2000approximating, gunopulos2005selectivity, muralikrishna1988equi, wang2003multi, liu2021lhist}, self-tuning histograms~\cite{bruno2001stholes, srivastava2006isomer, khachatryan2015improving, fuchs2007compressed}, and kernel density estimations~\cite{heimel2015self,kiefer2017estimating} are popular variants of the traditional methods. They can improve the estimation accuracy or efficiency of the traditional histogram and sampling algorithms on a single table. However, they still use simplified assumptions to estimate the cardinality of join queries.

\smallskip
\noindent \underline{\textbf{Learned data-driven methods:}} 
These methods circumvent the simplifying assumptions used in traditional methods by understanding the data distributions for the exponential number of join templates in a DB instance.
%The learned data-driven methods need to denormalize join tables up front and add a possibly exponential number of extra columns.
Specifically, some methods~\cite{getoor2001selectivity, tzoumas2011lightweight, tzoumas2013vldb} denormalize the join of all tables in a database and use Bayesian networks to model the distribution.
The current SOTA approaches for handling join queries try to model the data distributions of the denormalized join tables by using fanout-based techniques~\cite{hilp2019deepdb, zhu2020flat, wu2020bayescard, yang2020neurocard, wang2021face}.
These methods denormalize some tables and add a possibly exponential number of fanout columns, which are used to formulate the distributions for each join template. 
%The fanout method is first proposed to perform unbiased join sampling~\cite{zhao2018random} and later adopted by DeepDB~\cite{hilp2019deepdb} to perform \CE for PK-FK joins only. Later, FLAT~\cite{zhu2020flat} extends it to support general FK-FK joins. 
Using this method, the learned data-driven methods can produce effective estimations for join queries and high-quality query plans~\cite{han2021cardinality}. However, these models generally have a long training time, large model size, slow estimation speed, and unscalable performance with the number of joins in the query. 
Furthermore, the learned data-driven approaches currently can not handle self-joins or cyclic-joins and are very ineffective in processing complicated filter predicates such as string pattern matching queries. 

\smallskip
\noindent \underline{\textbf{Learned query-driven methods:}} 
These methods analyze the executed query workload, map the featurized join query directly to its cardinality using supervised models such as neural networks~\cite{kipf2018learned, negi2021flow}, xgboost~\cite{dutt2019selectivity}, tree-LSTM~\cite{sun2019end}, and deep ensembles~\cite{liu2021fauce}.
%The data distributions of different join patterns are implicitly and more efficiently captured through analyzing the executed query workload.
The query-driven methods generally have low estimation latency but the estimation effectiveness is highly dependent on the training workload\revise{; we put a semi-check mark  (\halfcheckmark) for these methods in Table~\ref{tab: summary}.} 
%They can generate highly ineffective estimates in case of workload shift; 
To achieve reasonable estimation quality, they generally need an excessive amount of executed training queries~\cite{han2021cardinality}, which are unavailable for new DB instances. Furthermore, the query-driven models need to retrain in the case of data update or query workload shift. An impractical amount of new executed queries are again needed for this re-training process.

%\smallskip
\noindent \underline{\textbf{Bound-based methods:}} 
Traditional \CE methods in modern DBMS tend to severely under-estimate the cardinality, thus sometimes choosing significantly more expensive query plans~\cite{ngo2018worst, atserias2008size, cai2019pessimistic}.
To avoid under-estimation, the bound-based methods~\cite{cai2019pessimistic, atserias2008size, abo2017shannon, hertzschuch2021simplicity} use information theory to provide an upper bound on the cardinality. 
These bounds are very effective to help generate high-quality query plans as they can avoid expensive join orders and physical operators~\cite{ngo2018worst, han2021cardinality} 
%Their estimation effectiveness is evaluated conceptually~\cite{ngo2018worst} and empirically~\cite{han2021cardinality} that even with significantly worse estimation accuracy (in terms of the relative difference between estimated and true cardinality), these methods can still produce better query plans than most of the other \CE methods.
However, these methods do not build single-table estimators to understand the data distributions. Instead, in presence of base-table filters, they need to materialize the filtered tables and populate the bound during query run-time, which can produce very high latency and overhead. 
Therefore, despite the bound-based methods providing meaningful insights into understanding joins, they can not be practically deployed in a DBMS.

%\srm{Following feels pretty repetitive with the introduction - I think you could cut it down quite a bit.}

%\smallskip
\noindent \underline{\textbf{Summary:}}
None of the existing approaches can simultaneously satisfy the desired properties of \CE, namely effective, efficient and appropriate for system deployment.
However, each of them contains some advantageous techniques, inspiring us to build \Ours that can bridge the gaps amongst all these \CE categories and unify their merits.

\if{0}
\ziniu{I think the inspiration can be removed entirely, given that we have a very detailed intro.}
\smallskip
\noindent \underline{\textbf{Inspiration:}} To summarize, none of the existing approaches can simultaneously satisfy the desired properties of \CE, namely effective,  efficient, and appropriate for system deployment.
However, each of them contains some advantageous techniques that inspire us to build \Ours. 
First, decomposing the join query estimates into single table estimates as  \emph{traditional methods} can minimize estimation latency and make the method friendly for system deployment.
Second, the ML models used in \emph{learned data-driven methods} can accurately understand data distributions.
Third, in addition to data, the executed query workload (if exists) can provide additional useful information in solving the \CE problem.
At last, bounding the cardinalities can produce better query plans by avoiding severely under-estimating potentially large sub-plans.

These observations inspire us to bridge the gaps among all these \CE categories and unify their merits to design a new \CE framework with the desired properties.
%First 
%Specifically, we first accurately formulate the problem of join query \CE into a probabilistic graphical model (PGM) inference problem involving only single-table data distributions (details in Section~\ref{sec: framework}).
To compute this PGM inference, we design a new probabilistic bound-based algorithm by adopting the idea of binning from traditional methods and upper bound techniques from the bound-based methods (details in Section~\ref{sec: bound}).
The estimates of our framework are very effective as they can provide a tight cardinality upper bound most of the time. 
Unlike most of the existing bound-based methods ~\cite{cai2019pessimistic, atserias2008size, abo2017shannon}, it also maintains a very fast estimation speed because it does not need to materialize the filtered tables during query planning.
Furthermore, based on this framework, we designed two techniques to further improve the estimation efficiency, namely causality pattern exploration inspired by data-driven methods and progressive estimation of sub-plan queries inspired by traditional methods (details in Section~\ref{sec: efficiency}).
It is worth noticing that our framework does not rely on query workload information, but it can use workloads to further improve its performance (details in Section~\ref{sec: bound}).
\fi
	
	%\clearpage
	
% !TeX spellcheck = en_US

\section{New framework for join queries}
\label{sec: framework}
Given this background on previous \CE approaches, we now introduce the high-level operation of \Ours. In Section~\ref{subsec: PGM}, we elaborate on \Ours's core technique: accurately estimating join queries with only single-table statistics using factor graph inference techniques.
Then, in Section~\ref{subsec: complexity}, we analyze the complexity of this inference procedure and motivate several optimizations in \Ours.
Finally, we provide a workflow overview in Section~\ref{subsec: overview}.

\subsection{Problem formulation}
\begin{figure}
	\centering
	\includegraphics[width=8.5cm]{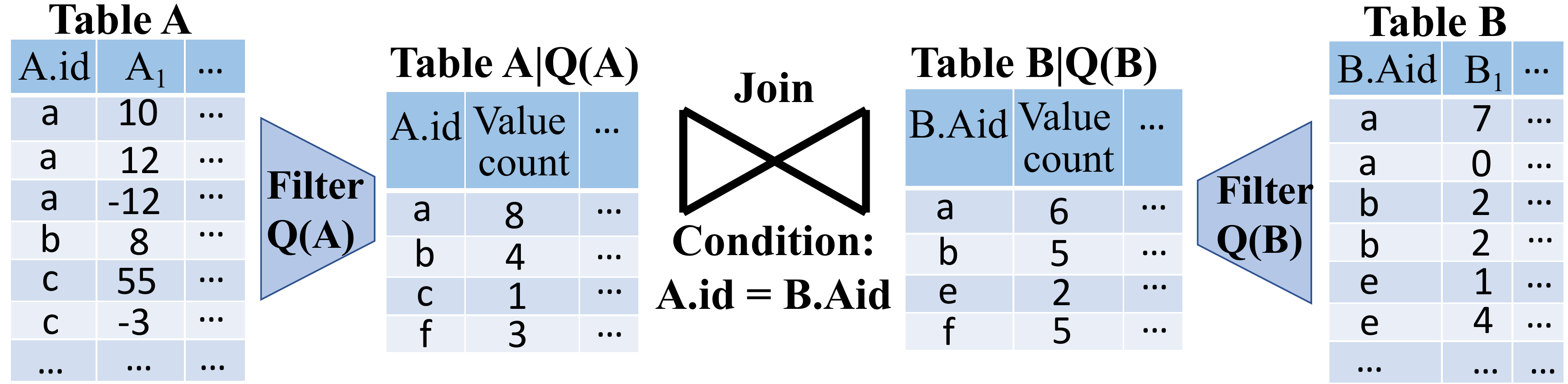}
	\vspace{-1em}
	\caption{Example of query $Q$ joining two tables with join condition $A.id = B.Aid$ and filter predicates $Q(A)$ and $Q(B)$.} %The cardinality of $Q$ is $8\times6 + 4\times5 + 3 \times5 = 83$.}
	\label{fig: join_example}
	%\vspace{-1em}
\end{figure}
\label{subsec: PGM}
We first illustrate the main idea with a simple example of a two-table join query. Then, we formalize the join query estimation problem as a PGM inference problem over single table distributions. 

%\smallskip
\noindent \underline{\textbf{Two-table join query example:}} Figure~\ref{fig: join_example} illustrates a query $Q$ joining tables $A$ and $B$ on the inner join\footnote{In this paper, we only discuss inner and equality joins. \Ours naturally supports left, right, and outer joins. We leave support for \revise{non-equal} joins as future work.} condition $A.id = B.Aid$ with base table filter predicates $Q(A)$ and $Q(B)$.
Table $A$ first goes through the filter $Q(A)$, resulting in an intermediate table $A|Q(A)$ (records in $A$ that satisfy the filter $Q(A)$). The same procedure is applied to table $B$.
Then, the query $Q$ will match the value of join keys $A.id$ and $B.Aid$ from these two intermediate tables. Specifically, the value $a$ appears $8$ times in table $A|Q(A)$ and $6$ times in table $B|Q(B)$, resulting in value $a$ appearing $48$ times in the join result. Therefore, we can calculate the cardinality of this query $Q$ as $8\times 6 + 4\times 5 + 3 \times 5 = 83$.

We can formulate the above procedure for calculating $Q$ as a statistical equation in Equation~\ref{equ: join_example}, where $D(A.id)$ denotes the domain of all unique values of $A.id$. We observe that only single-table distributions $P_A$ and $P_B$ are required to accurately compute the cardinality of this join query. Also in this equation, $P_{A}(A.id \! =  \!  v | Q(A)) * |Q(A)|$ equals to $P_{A}(A.id  \!  =   \! v \land Q(A)) * |A|$, which is exactly what single-table \CE methods estimate. Thus, we can \emph{accurately} calculate the cardinalities of two-table join queries using single-table estimators. Note that the summation over the domain of join key $\mathcal{D}(A.id)$ has the same complexity as computing the join. Therefore, we need to approximate this calculation for \Ours to be practical. The details of our approximation schemes are given in Section~\ref{subsec: complexity} and Section~\ref{sec: bound}.

\vspace{-1em}
{\small
\begin{align}
	|Q| = \sum_{v \in \mathcal{D}(A.id)} & P_{A}(A.id = v | Q(A)) * |Q(A)| * \nonumber \\   
	& P_{B}(B.Aid = v | Q(B)) * |Q(B)|
	\label{equ: join_example}
\end{align}
}

\begin{figure}
	\centering
	\includegraphics[width=8.5cm]{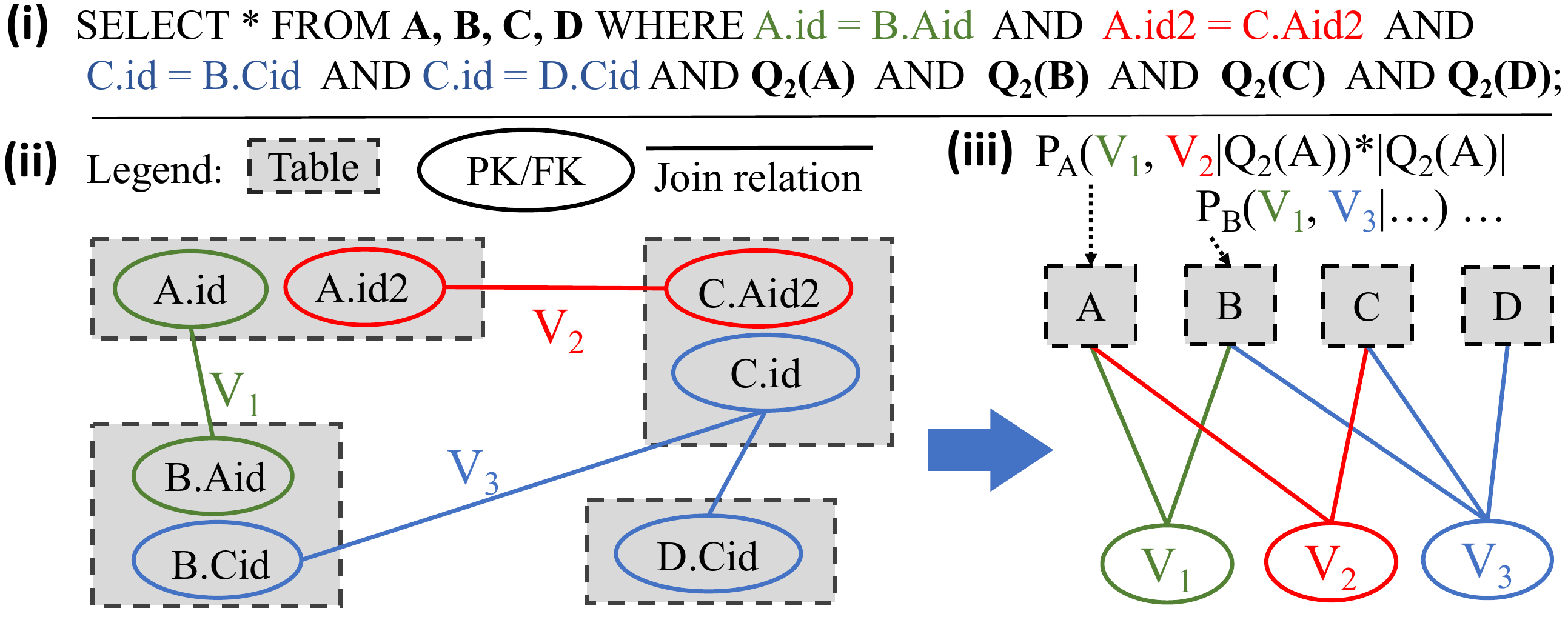}
	\vspace{-1em}
	\caption{Formulating a join query \CE into a factor graph inference problem. Figure (i) shows a SQL query $Q_2$. Figure (ii) visualizes the join template of $Q_2$. Figure (iii) presents the factor graph model to compute $Q_2$'s cardinality. }
	\label{fig: formulation}
	%\vspace{-1.5em}
\end{figure}

\smallskip
\noindent \underline{\textbf{Formal problem formulation:}}  A general join query can involve a combination of different forms of joins (e.g., chain, star, self, or cyclic joins) so its counterpart equation (as Equation~\ref{equ: join_example}) can be very difficult to derive and compute; \revise{we have put these equations/derivations in the supplementary material}.
%\footnote{We put the supplementary material as a pdf document in the home directory of our anonymously submitted code.}. %\srm{give reference - how does reader find this?}. \ziniu{Actually, SIGMOD does not allow appendix so I plan to add this files in the zip for code availability. I'm gonna add a footnote so the reviewers know where to look for.} 
We provide a generalizable formulation that automatically decomposes join queries into single-table estimations using the factor graph model.

Figure~\ref{fig: formulation}-(i) shows a SQL query $Q_2$ joining four tables $A, B, C, D$.
We visualize its join template as a graph in Figure~\ref{fig: formulation}-(ii), where each dashed rectangle represents a table, each ellipse (node) represents a join-key in $Q_2$, and each solid line (edge) represents an equi-join relation between two join keys connected by it. 
Note that both sides of an equi-join relation represent the semantically equivalent join keys.
We call them {\it equivalent key group} variables.
In Figure~\ref{fig: formulation}-(ii), there are three connected components and thus three equivalent key group variables $V_1, \ldots, V_3$. $V_1$ represents $A.id$ and $B.Aid$ in this group since $Q_2$ contains the join condition $A.id = B.Aid$.

% new paragraph below
The exact computation required to compute the cardinality of the SQL query in Figure \ref{fig: formulation} extends Equation \ref{equ: join_example} to sum over the domain $\mathcal{D}$ of each join key.
%--- still utilizing only statistics over single tables ($P_A, P_B, P_C, P_D$).
%C (A.id, v, Q) C (A.id, v, Q)

{\small
\vspace{-1em}
\revise{
\begin{align}
	|Q_2| = & \sum_{a1 \in \mathcal{D}(A.id)} \ \ \ \ \  \sum_{a2 \in \mathcal{D}(A.id2)} \ \ \ \ \  \sum_{c \in \mathcal{D}(C.id)} \nonumber \\  
	& P_{A}(A.id = a1, \ A.id2 = a2 \ | \ Q_2(A)) \ast |Q_2(A)| \nonumber \\   
	&  \ast P_{B}(B.Aid = a1, \ B.Cid = c \ | \ Q_2(B)) \ast |Q_2(B)|  \nonumber \\
	& \ast P_{C}(C.Aid2 = a2, \ C.id = c \ | \ Q_2(C)) \ast |Q_2(C)|  \nonumber \\
	&  \ast P_{D}(D.Cid = c \ | \ Q_2(D)) \ast |Q_2(D)|
	\label{equ: join_example2}
\end{align}
}
}
%\vspace{-1em}

This computation is extremely expensive --- assuming the domains of each id being $n$, a naive implementation would have the complexity $O(n^3)$. We can improve this complexity to $O(n^2)$ by representing this computation as an equivalent factor graph model~\cite{loeliger2004introduction}, which is a particular class of probabilistic graphical models (PGMs)~\cite{koller2009probabilistic}. We have $2$ in the exponent because the maximum number of join keys in $Q_2$ is two in the table $A$. This is still very expensive, but in Section \ref{subsec: complexity}, we describe how we can use principled approximations made possible by the factor graph formulation to make the inference complexity practical.

A factor graph is a bipartite graph with two types of nodes: variable nodes, and factor nodes representing an unnormalized probability distribution w.r.t. the variables connected to it. Figure~\ref{fig: formulation}-(iii) shows the constructed factor graph $\mathcal{F}$ for computing the cardinality of $Q_2$. Specifically, $\mathcal{F}$ contains a variable node for each equivalent key group variable $V_i$, and a factor node for each table touched by $Q_2$. 
A factor node is connected to a variable node if the variable represents a key in the table. In this case, the factor node representing table $A$ is connected to $V_1$ (equivalent to $A.id$) and $V_2$ (equivalent to $A.id2$). 
Each factor node maintains an unnormalized probability distribution for the variable nodes connected to it, for e.g., node $A$ maintains  $P_A(V_1, V_2|(Q_2(A)))*|Q_2(A)|$, which is the same as the distribution $P_A(A.id, A.id2|(Q_2(A)))*|Q_2(A)|$ used in Equation \ref{equ: join_example2}. 
The factor graph model utilizes the graph structure to compute the sum using the well-studied inference algorithms.

We can more formally state this relationship between computing the join cardinalities and factor graphs as the following lemma, whose proof is provided in the supplementary material.

\begin{lemma}
	\label{lemma: PGM}
		Given a join graph $\mathcal{G}$ representing a query $Q$, there exists a factor graph $\mathcal{F}$ such that the variable nodes in $\mathcal{F}$ are the equivalent key group variables of $\mathcal{G}$ and each factor node represents a table $I$ touched by $Q$. A factor node is connected to a variable node if and only if this variable represents a join-key in table $I$. The potential function of a factor node is defined as table $I$'s probability distribution of the connected variables (join keys) conditioned on the filter predicates $Q(I)$. Then, calculating the cardinality of $Q$ is equivalent to computing the partition function of $\mathcal{F}$.
\end{lemma}

%This lemma is proven in the appendix.

%After constructing the factor graph model $\mathcal{F}$, calculating the exact cardinality of $Q$ is equivalent to computing the partition function~\cite{mackay2003information} of $\mathcal{F}$, as stated in Lemma~\ref{lemma: PGM};  the proof of the lemma is given in the appendix.
%At a high level, computing the partition function requires multiplying all the distributions defined by the factor nodes and summing this product over all variables. This procedure is the same as calculating the cardinality of $Q$, which involves summing over all equivalent key group variables.  \srm{I think it would be helpful to show what this computation looks like for the example - after reading I'm not sure I totally understand what exactly is happening.}

\subsection{Inference complexity analysis}
\label{subsec: complexity}
%Calculating the partition function of a 
Inference on factor graphs is a well-studied problem in the domain of PGMs~\cite{mackay2003information, koller2009probabilistic, loeliger2004introduction}. Popular approaches to solving this problem are variable elimination (\emph{VE})~\cite{koller2009probabilistic} and belief propagation algorithms~\cite{kschischang2001factor}. 
In \Ours's implementation, we use the \emph{VE}, which first determines an optimal order of variables and then sums over the distributions defined by the factor graph model in this order. The complexity of \emph{VE} is $O(N*|D|^{max(|JK|)})$, where $N$ is the number of equivalent key groups ($3$ in Figure \ref{fig: formulation}), $|D|$ is the largest domain size of all join keys, and $max(|JK|)$ is the maximum number of join keys in a single table ($2$ in Figure~\ref{fig: formulation}).
Intuitively, this complexity is because the factor nodes need to understand the joint distribution of all join keys in one table. Thus, the largest factor node in the factor graph model maintains a probability distribution of size $|D|^{max(|JK|)}$.
This complexity is not practical for real-world queries,  as $|D|$ can be millions, and $max(|JK|)$  can be larger than $4$ in real-world DB instances, such as IMDB~\cite{leis2015good}. Therefore, instead of calculating the exact cardinality, we propose a new PGM inference algorithm for \Ours that can estimate an upper bound on cardinality.
%\srm{Give intuition - why is an upper bound what we are after/needed?}
Past works~\cite{cai2019pessimistic, atserias2008size, abo2017shannon} have shown that cardinality upper bounds can help avoid very expensive query plans since under-estimating a large result can sometimes be catastrophic. For instance, an optimizer might choose to do a nested loop join if a \CE method under-estimates that the result would easily fit in memory. This would lead to a disastrous plan if the actual result size is much larger than estimated.

\revise{Our approximate inference algorithm performs two approximations: 1) reducing the domain size $|D|$ using binning and 2) decreasing the exponent, $max(|JK|)$, by approximating the distribution of attributes in a single table.}
\revise{Specifically, for binning, we first partition the domain of all join keys into $k$ bins.} Instead of summing over the entire domain, we only need to sum over $k$ summarized values (probabilistic bounds) of each bin. The details are provided in Section~\ref{sec: bound}. 
\revise{Second, we approximate the causal relation among all attributes within a table as a tree structure.}
%by taking into account the pairwise mutual information between attributes~\cite{chow1968approximating}. 
This allows us to factorize the $max(|JK|)$-dimensional joint distribution of all join keys in a table as a product of two-dimensional conditional distributions. 
%Thus, this approximation will reduce the $max(|JK|)$-dimensional distribution of join-keys to multiple two-dimensional conditional distributions.
%\srm{What are the dependencies of an attribute? Is this like following key-foreign key relationships?  Maybe an example would help?} 
This procedure is known as structure learning in Bayesian networks in the PGM domain~\cite{chow1968approximating}. Past work on single table cardinality estimation has shown that such tree approximations do not decrease the modeling accuracy significantly~\cite{wu2020bayescard}. Further details are provided in Section~\ref{subsec: causality}. 

Thereafter, we can directly run the \emph{VE} inference algorithm on the binned domains and factorized two-dimensional distributions to estimate the cardinality upper bound of $Q$.
With these two approximate inference techniques, the complexity of estimating the cardinality bound is reduced to $O(N*k^2)$, which is very efficient as both $N$ and $k$ are very small in practice.
%, or the number of equivalent key groups, is small, and $k$, or the number of bins, can be chosen to be small.

\subsection{Workflow overview}
\label{subsec: overview}

\begin{figure}
	\centering
	\includegraphics[width=8.5cm]{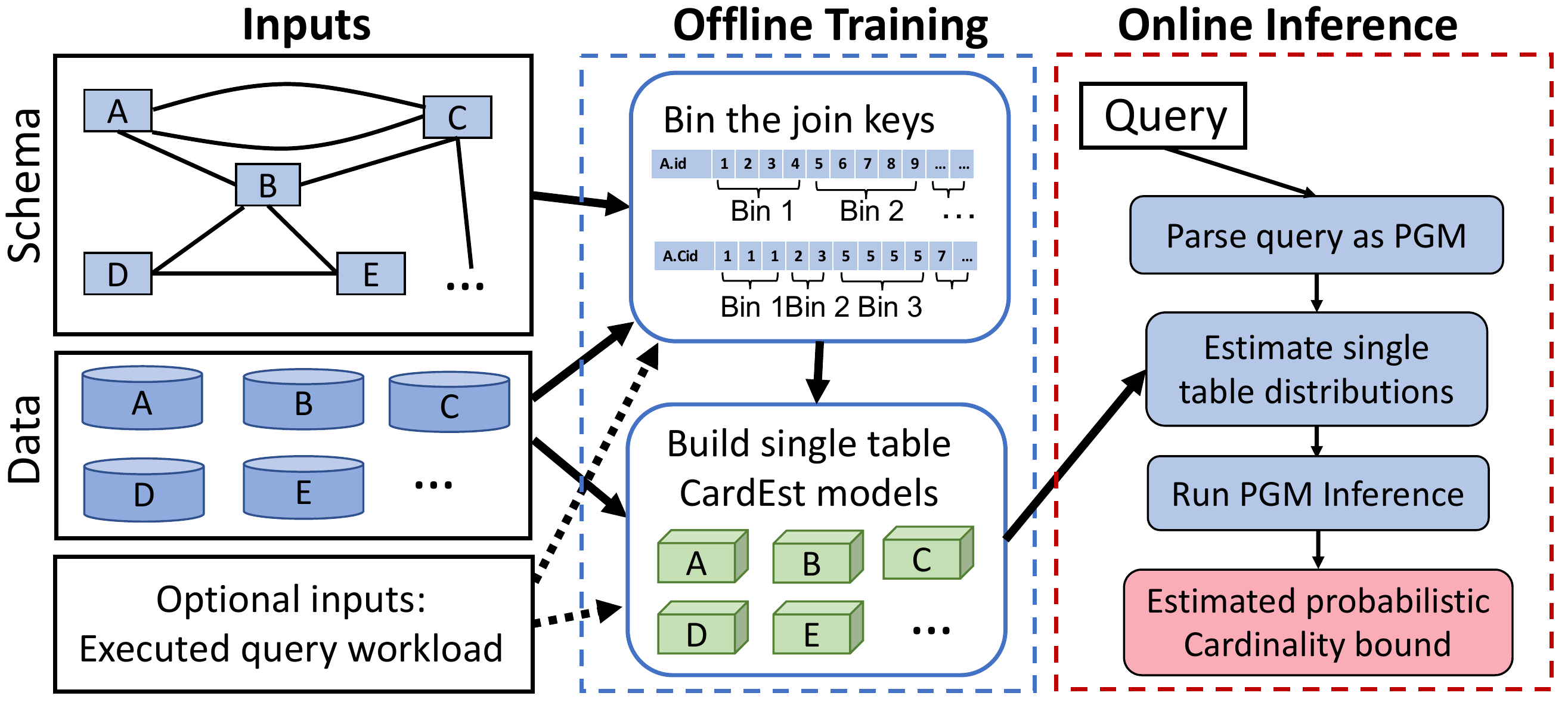}
	\vspace{-1.5em}
	\caption{Workflow of \Ours framework.}
	\label{fig: workflow}
	%\vspace{-2em}
\end{figure}

The workflow of \Ours contains two phases: offline training and online inference (Figure~\ref{fig: workflow}). During the offline phase, our framework analyzes the DB instance, bins the domain of join keys, explores the causal relation, and builds \CE models for every single table. When a join query $Q$ comes in during the online phase, \Ours formulates the estimation of this query as a factor graph involving only single-table distributions, estimates them using (pre-trained) single-table estimators, runs PGM inference, and generates a probabilistic bound for $Q$. We provide the details as follows.

\smallskip
\noindent\underline{\textbf{Offline training phase:}} Given a new DB instance, \Ours first analyzes its DB schema and data tables to get all possible join relations between different join-keys. We consider two join-keys to be semantically equivalent if there exists a join relation between them. After identifying all groups of equivalent join-keys, we bin the domains of all join-keys in each group. \Ours can also optionally use query workload information to optimize this binning procedure (details provided in Section~\ref{subsec: bin selection}).
Based on the data tables and the binned domains, we explore the \revise{causal relation} of these attributes and model them as a tree structure (details are provided in Section~\ref{subsec: causality}). Then, we build the \CE models to understand the data distribution of every single table. 
In principle, any single-table \CE method that is able to provide conditional distributions can be adapted into \Ours. 
In practice, we implement two methods:  traditional random sampling and the learned data-driven BayesCard method~\cite{wu2020bayescard}.
The sampling method is extremely flexible to use and can easily support any complex filter predicates with disjunctions, string pattern matching, or any user-defined functions.
Alternatively, similar to most data-driven methods, BayesCard can only work with queries filtering numeric and categorical attributes, but BayesCard has been shown to consistently produce accurate, fast, and robust estimations for various data distributions~\cite{wu2020bayescard}. 
Users of \Ours can switch between different single-table estimators based on their objectives and knowledge about the DB instances.

\smallskip
\noindent\underline{\textbf{Online inference phase:}} When a join query $Q$ comes in during run-time, \Ours will first parse its join graph and construct the corresponding factor graph $\mathcal{F}$ (as in Figure~\ref{fig: join_example}). 
Then, it uses the trained single-table estimators to estimate the join-key distributions conditioned on the filter predicates of $Q$; \Ours will use these distributions in the corresponding factor nodes of $\mathcal{F}$. 
Finally, we run the approximate version of VE inference algorithms on $\mathcal{F}$ to derive the estimated/probabilistic cardinality bound. 
%The variable elimination algorithm is the most general and well-studied inference algorithm in the field of PGMs, which first generates an optimal elimination order of all variables and then sums over the domain of these variables sequentially in this order to calculate the partition function of $\mathcal{F}$.
We note that the \CE methods working inside a query optimizer will estimate all sub-plan queries of a target query to decide the best query plan. 
Since these sub-plan queries contain a large amount of overlapping information, we provide a progressive estimation algorithm to avoid computation redundancy in Section~\ref{subsec: progressive}.
	
	%\clearpage
	\section{Probabilistic bound algorithm}
\label{sec: bound}
In this section, we design a new probabilistic bound algorithm based on binning to calculate the factor graph inference problem. Specifically, we first explain the main ideas with a two-table join example and formulate the algorithm details within \Ours in Section~\ref{subsec: algorithm}. Then, we discuss how to optimize the bin selection using data and query information in Section~\ref{subsec: bin selection}.

\subsection{Algorithm details}
\label{subsec: algorithm}
As described in Section~\ref{sec: framework}, the main objective of our probabilistic bound-based algorithm is to reduce the domain size of the join keys, i.e., reducing the complexity of summation in Equation~\ref{equ: join_example}. Thus, instead of running the inference algorithm over the entire domain, \Ours only needs to sum over the binned domain.

\smallskip
\noindent \underline{\textbf{Two-table join query example:}} We use the previous example query $Q$ from Section~\ref{subsec: PGM} to illustrate the core idea of our probabilistic upper bound algorithm based on binning. 
The objective is to estimate an upper bound on $Q$, whose exact computation requires summing over the entire domain of $A.id$ according to Equation~\ref{equ: join_example}. 
Assume that we have a set of $k$ bins $\{bin_1, \ldots, bin_k \}$ partitioning the domain of $A.id$ and we apply this same set of bins to $B.Aid$. 
\revise{We need to make sure that a given value in the domain of $A.id$ and $B.Aid$ will always belong to the bin with the same indexes.}
Then, Equation~\ref{equ: join_example} can be equivalently written as Equation~3.
%\srm{I think this should be 3 - we should use a label for this.}. 
We can then replace the sum over all values in a bin $bin_i$ with a probabilistic bound on $bin_i$ to estimate an upper bound for $Q$ as Equation~\ref{equ: bound}.

%\vspace{-1em}
%\begin{alignat}{2}
%|Q| = & $\sum_{i = 1}^k \sum_{v \in bin_i}$ && P_{A}(A.Id = v | Q(A)) * |Q(A)| * \nonumber \\   
% & && P_{B}(B.Aid = v | Q(B)) * |Q(B)|	
%\label{equ: bound}
%\end{alignat}
{\small
\vspace{-1em}
\begin{alignat}{2}
	|Q| &= \sum_{i = 1}^{k}  \sum_{v \in bin_i} && P_{A}(A.Id = v | Q(A)) * |Q(A)| * \nonumber \\   
	& && P_{B}(B.Aid = v | Q(B)) * |Q(B)| \\
	&\lesssim \sum_{i = 1}^{k} && Probabilistic\_bound(A, B, bin_i)
	\label{equ: bound}
\end{alignat}
}

\begin{figure}
	\centering
	\includegraphics[width=8.5cm]{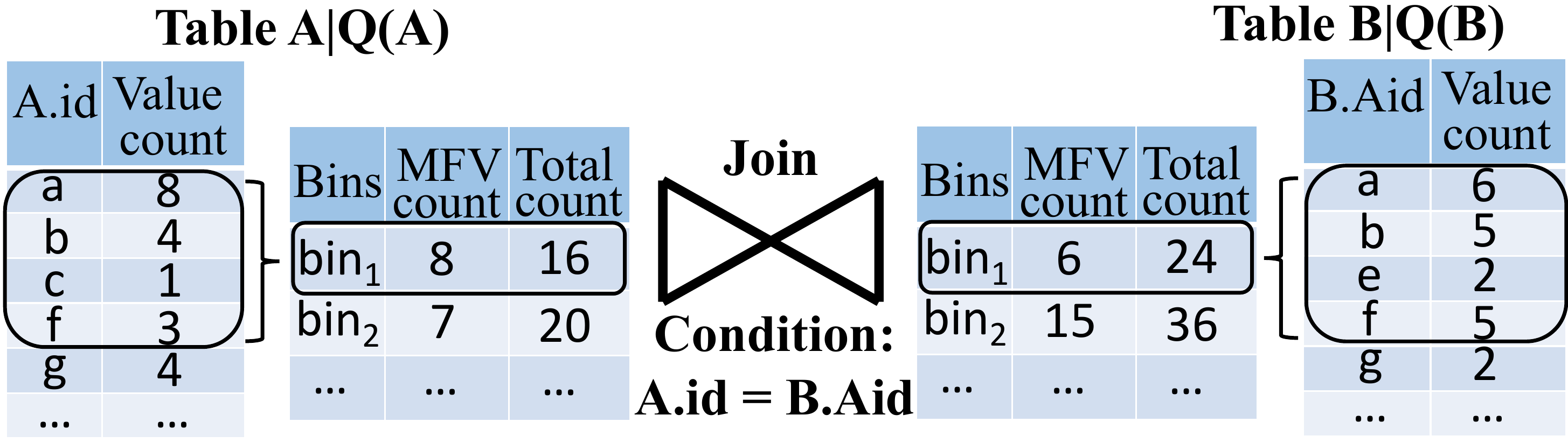}
	\vspace{-1.5em}
	\caption{The bound-based algorithm; each bin is summarized by the most frequent value count and total count.}
	\label{fig: bound}
	%\vspace{-1.5em}
\end{figure}

Motivated by a bound based on the most frequent value (MFV)~\cite{hertzschuch2021simplicity}, we use a simple probabilistic bound method to derive $Probabilistic\break \_bound(A, B, bin_i)$ for a particular bin.
Specifically, assuming that value $\{a, b, c , e, f\}$ of $A.id$ and $B.Aid$ are binned into $bin_1$ as shown in Figure~\ref{fig: bound}. We know the summation of all values in $bin_1$ equals to \revise{$8\times 6 + 4 \times 5 + 3 \times 5 = 83$}. This summation has a dominating term of $8 \times 6$, because the count of MFV of $bin_1$ is $8$ for $A.id$ (denoted as $V^{*}_1(A.id)$) and $6$ for $B.Aid$ (denoted as $V^{*}_1(B.Aid)$) so each value can appear at most $8 \times 6$ times in the denormalized table after the join. Since we know the total count of values in $bin_1$ for $A.id$ is $16$, there can be at most $16/8 = 2$ MFVs. Similarly, there can be at most \revise{$4$} MFVs in $bin_1$ for $B.Aid$. Therefore, we have the summation of all values in $bin_1$ is upper bounded by \revise{$min(2, 4) \times 8 \times 6 = 96$}. 

We formally represent the aforementioned procedure in Equation~\ref{equ: bound2}, where $V^{*}_i(A.id)$ and $P_A(A.id \in bin_i|Q(A)) * |Q(A)|$ are the MFV count and estimated total count of $bin_i$ for $A.id$. Our bound is probabilistic because $P_A(A.id \in bin_i|Q(A)) * |Q(A)|$ is estimated with a single table \CE method, which may have some errors. 

{\small
\vspace{-1em}
\begin{align}
	|Q| & \lesssim \sum_{i = 1}^{k}  min(\frac{P_A(A.id \in bin_i|Q(A)) * |Q(A)|}{V^{*}_i(A.id)}, \nonumber \\ 
	& \frac{P_B(B.Aid \in bin_i|Q(B)) * |Q(B)|}{V^{*}_i(B.Aid)}) * V^{*}_i(A.id) * V^{*}_i(B.Aid)
	\label{equ: bound2}
\end{align}
}

%We provide similar equations for estimating the bound of other forms of joins (e.g. chain, star, self, or cyclic joins) in appendix.

\smallskip
\noindent \underline{\textbf{Working with PGM inference:}} 
Next, we will explain how to generalize this probabilistic bound to make it work inside the factor graph VE inference algorithm mentioned in Section~\ref{subsec: PGM}. 
During the offline training, \Ours computes the MFV counts $V^*$ for all bins of each join-key. During the online phase, \Ours estimates the probability distribution for the binned domain of all join-keys in each table $I$. Then, \Ours puts $I$'s distribution $P_{I}$ and MFV counts $V^*(I)$ into the corresponding factor node. 

Recall that the VE algorithm generates an optimal elimination order of all variables and sums over the domain of these variables sequentially in this order.
Since at each step, the variable elimination algorithm sums out only one variable, its probabilistic bound can be derived using Equation~\ref{equ: bound2} over multiple equivalent join keys.
Thus, this algorithm does not need to sum over the entire value domain of a variable but can only sum over the probabilistic bounds in the binned domain instead, which greatly reduces the complexity. 

\smallskip
\noindent \underline{\textbf{Analysis:}} This probabilistic bound algorithm significantly reduces the domain size of join keys and enables practical approximate PGM inference. 
Further, this algorithm upper-bounds the cardinality most of the time and the bound is very tight with even a small number of bins, such as $k=100$. Therefore, as we empirically verify in our evaluation, the estimated cardinality bound is very effective.

\subsection{Bin selection optimizations}
\label{subsec: bin selection}

In our probabilistic bound algorithm, different bins can result in drastically different bound tightness. 
%The key challenge is that the same bins will be built for all join keys within an equivalent key group, i.e. join keys that share the same semantics. 
\revise{There are two decisions: how many bins to use for each join key group, and what values to put in the same bin.}

%Thus, the problem of how to generate optimal bins is crucial to \Ours. This problem contains two folds: 1) how to decide the number of bins $k$; 2) how to partition the values in the domain into bins $bin_1, \ldots, bin_k$.
%Before going into the details of these two folds recall from Section~\ref{sec: framework} that the same bins will be built for all join keys within an equivalent key group, i.e. join keys that share the same semantics. 

\smallskip
\noindent\textbf{\underline{Deciding $k$ based on query workloads:}} 
\revise{The number of bins, $k$, has a significant effect on the performance of \Ours: fewer bins aggregate more distinct values in the join key domain to each bin and are thus less accurate but more efficient.}
An approach to tune $k$ is to set different values of $k_i$ for different equivalent key groups $Gr_i$.
Suppose that users provide a budget $K$ with $K = \sum k_i$. If \Ours has access to the query workload, it can analyze the join patterns and count the number of times $n_i$ each $Gr_i$ appears in the workload.
Our framework can adaptively set a larger $k_i$ for the frequently visited $Gr_i$ and vice versa. 
In our implementation, we use a simple heuristic --- set $k_i = K * n_i/\sum_j n_j$.
In this way, we can optimize the modeling capacity of \Ours to make sure that it is spent on the important joins.

\smallskip
\noindent\textbf{\underline{Binning strategy:}} For the probabilistic bound algorithm described in Section \ref{sec: bound}, we observe that the upper bound on a particular bin $bin_i$ can be very loose if the MFV count $V^*_i$ is a large outlier in $bin_i$. Taking the two table join query as an example, if $bin_i$ contains only one value that appears $100$ times in $A.id$ but $10,000$ values that only appear once in $B.Aid$, then the bound could be $100$ times larger than the actual cardinality. \revise{Common binning strategies used in DBMS histograms such as equal-depth or equal-width bins can be catastrophic in such cases.} Instead, we want the variance of the join key counts in each bin to be low. This strategy has also been proposed in different contexts~\cite{ioannidis1991propagation, ioannidis1995balancing}, but the key challenge is that the same bins will be built for all join keys within an equivalent key group, i.e. join keys that share the same semantics. For example, for a primary key (e.g. ``title.id'' in IMDB JOB~\cite{leis2015good}), any binning strategy will have zero variance because of uniqueness. However, the same bins will be applied for its equivalent foreign keys (e.g. ``movie\_id'') on other tables, which may lead to a large variance depending on how often each value repeats in the other tables.
\revise{We design a greedy binning selection algorithm (GBSA) to optimize for bins with low variance value counts across all tables, as illustrated in Algorithm~\ref{algo: gbsa}.}
Jointly minimizing the variance of one bin for all join keys has exponential complexity.
Therefore, GBSA uses a greedy algorithm to iteratively minimize the bin variance for all join keys.
\revise{At a high level, GBSA first optimizes the minimal variance bins with half the binning budget of $k/2$ on the domain of one join key (lines 2-4).} 
\revise{Then, it recursively updates these bins by minimizing the variance of other join keys using the other half of the budget (lines 5-13).}
\revise{We put the details of GBSA in the supplementary material.}

\smallskip
\noindent\textbf{\revise{\underline{Discussion:}}}
\revise{As shown in Section~\ref{subsec: ablation}, GBSA has significantly better performance than naive binning strategies.
In the extreme case, if the value counts have zero variance for all equivalent join keys, then our bound can output the exact cardinality.
Admittedly, this binning strategy does have the drawback that after applying the filter predicates, the join key frequencies will change and their variance may become higher within a bin. However, we empirically evaluate that for real-world datasets, this phenomenon does not have a severe impact on \Ours. 
In future work, we will explore an enhanced binning strategy that can efficiently address this issue.}

\begin{algorithm}[t]
 	\small
 	\caption{\revise{Greedy Bin Selection Algorithm (GBSA)}}
 	\label{algo: gbsa}
 	
 	\begin{flushleft}
 		\textbf{Input}: 	
 		Equivalent key groups $Gr_1, \ldots, Gr_m$, where $Gr_i = \{Id_i^1, \ldots, Id_i^{|Gr_i|} \}$;
		
	Column data $\mathcal{D}(Id_i^j)$ of all join keys in the DB instance $\mathcal{D}$;
		
 		Number of bins $k_i$ for each group $Gr_i$.
 	\end{flushleft}
 	\begin{algorithmic}[1]
 		\For{$Gr_i \in \{Gr_1, \ldots, Gr_m\}$}
 		\State $Bin(Gr_i) \gets$ []
 		\State $Gr_i' \gets sort\_key\_based\_on\_domain\_size(\mathcal{D}, Gr_i)$
 		\State $Bin(Gr_i) \gets get\_min\_variance\_bins(\mathcal{D}(Gr_i'[1]), k_i/2)$
 		\State remain\_bins $\gets k_i/2$
 		\For{$j \in \{ 2, \ldots, |Gr_i'| \} $}
 		\State binned\_data $\gets apply\_bin\_to\_data(\mathcal{D}(Gr_i'[j]), Bin(Gr_i))$
 		\State bin\_variance $\gets calculate\_variance$(binned\_data)
 		\State arg\_sort\_idx $\gets arg\_sort\_decreasing(bin\_variance)$
 		\For{$p \in arg\_sort\_idx[1:remain\_bins/2]$}
 		\State $Bin(Gr_i) \gets min\_variance\_dichotomy(Bin(Gr_i)[p]$, binned\_data[p])
 		\EndFor
 		\State remain\_bins $\gets$ remain\_bins/2
 		\EndFor
 		\EndFor
 		\State \textbf{return}  $\{Bin(Gr_i) | i, \ldots, m\}$
 	\end{algorithmic}
 	%\vspace{-1em}
 \end{algorithm}

\revise{\subsection{Incremental model updates}}

\revise{
\Ours is friendly for DBs with frequent data changes because we only need to incrementally update the single table statistics. In the following, we will discuss the algorithm details in the scenario of data insertion and data deletion can be handled similarly.}

Consider we have trained a stale \Ours model $\mathcal{F}$ on data $\mathcal{D}$ and would like to incrementally update $\mathcal{F}$ with the inserted data $\mathcal{D}'$.
\Ours needs to identify the inserted tuple values from $\mathcal{D}'$ for every join key and put them into the original bins optimized on $\mathcal{D}$. 
Then, it will update the total and most-frequent value count in each bin.
At last, \Ours can incrementally update the base table \CE models with off-the-shelf tools (e.g. \emph{BayesCard} provides an efficient and effective approach to incrementally update the single-table models using $\mathcal{D}'$~\cite{wu2020bayescard}; materializing a new sample on $\mathcal{D} \bigcup \mathcal{D}'$ incrementally updates the sampling-based estimator). 

Therefore, the incremental update of \Ours does not need to normalize the joined tables nor need to use executed queries. It is also empirically verified to be very efficient and effective. However, during incremental updates, the \Ours models keep the same set of bins, which is optimized on the previous data and thus might not be optimal after data insertion. This might cause slight degrade in performance and the user can choose to retrain the model in case of massive data updates.

\vspace{2em}

\section{Improving \Ours efficiency}
\label{sec: efficiency}

We describe two techniques to improve the modeling and estimation efficiency of \Ours. 
%The first technique explores the join for attributes on a single table as a tree structure and uses this structure to simplify the PGM inference procedure.
First, we model the joint distribution of attributes on a single table as a tree structure and use it to simplify the PGM inference calculations. 
%The second technique identifies all sub-plan queries needed to optimize a target query. Then it progressively estimates these sub-plan queries by reusing the estimates for intermediate sub-plan queries, which avoids redundant calculation.
Next, we define a systematic way of reusing estimates of intermediate sub-plan queries to estimate a larger query. This is only possible because we decompose cardinality estimates into single table sub-components.

\subsection{\revise{Exploring attributes causal relation}}
\label{subsec: causality}

%Recall in previous sections that using binning and bound to approximate the partition function of the factor graph representing the join query has a complexity of $O(N*k^{max(|JK|)})$ in both storage and inference.
Recall from \revise{Section \ref{subsec: complexity}}, the factor graph computation to calculate the cardinality bound is $O(N*k^{max(|JK|)})$, where $N$ is the number of tables,  $k$ is the number of bins, and $max(|JK|)$ is the maximum number of join keys in a table. 
The exponent term, $max(|JK|)$, represents the factor nodes requiring the joint distribution between all join keys on a single table.
%This is because some factor nodes need to build a distribution of size $O(k^{max(|JK|)})$ to understand the correlation between join keys on a single table.
As $max(|JK|)$ can sometimes be large (e.g., in IMDB it can be up to $4$), we need to reduce the dimensionality of the distributions defined in factor nodes to ensure an efficient \CE procedure.
Fortunately, real-world data has a lot of correlations that can be used to simplify this joint distribution.

%\smallskip
%\noindent\underline{\textbf{Example:}} 
%Consider a table $A$ with $8$ attributes $\{id_1, id_2, id_3, id_4, attr_1, \break attr_2, attr_3, attr_4\}$, where $id_i$ is a join key and $attr_i$ is an attribute to be filtered in a query.
\revise{Consider a table $A$ with $6$ attributes, $\{id_1, id_2, id_3, id_4 attr_1, attr_2\}$, four of which are join keys.} The factor graph computation \revise{may} need to estimate quantities such as:
\begin{align}
    P_A(id_1, id_2, id_3, id_4 | Q(A))
\label{equ: joint_distribution}
\end{align}

\revise{This requires storing a $4$ dimensional probability distribution, and even with binning the domains of all ids into $k$ bins, it still requires $k^4$ space and inference time complexity.}
This joint distribution can be represented as a graph where each attribute is a node, and there are edges between every pair of attributes.
We will use the dependencies and relationships between the attributes and model them as a Bayesian network (BN) ~\cite{koller2009probabilistic}.
%A BN specifies a directed acyclic graph (DAG) to represent the attributes' causality pattern and uses this graph to decompose the high dimensional distributions into a product of conditional distributions. 
Specifically, we will assign each edge in the joint distribution graph a weight proportional to the mutual information between the two attributes connected by the edge. Then, we will remove edges with the least mutual information until only a tree structure graph remains to represent the joint probability over these attributes. \revise{This process is known as the Chow-Liu algorithm for decomposing joint distributions~\cite{chow1968approximating}}. The factorized tree representation of the joint distribution can let us approximate Equation \ref{equ: joint_distribution} with an equation of the form:

\vspace{-.1in}
{\small
\begin{align}
    P_A(id_1 | Q(A)) * P_A(id_2 | Q(A)) * P_A(id_3 | id_1) * P_A(id_4 | id_3)
\end{align}
}
, which is much more efficient to compute because it only involves a product over at most a two-dimensional distribution. This dimension reduction can cause some errors. But it has been shown empirically that the factorized distributions are accurate approximations of the original distribution for most real-world datasets~\cite{wu2020bayescard}.

Using such an optimization, instead of storing the $max(|JK|)$-dimensional distributions in the factor nodes, the factor graph only needs to store a series of one or two-dimensional data distributions without a significant decrease in estimation accuracy. Therefore, the overall complexity for running approximate inference on this factor graph becomes $O(N*k^2)$ in both storage and inference speed. Since $k$ is relatively small (around one hundred), estimating one query using \Ours is very efficient. 

\revise{In principle, \Ours can also use other methods for the single table distribution inference.} This may involve traditional methods, such as sampling, or more powerful graph-structured factorizations of the distributions --- which can increase modeling accuracy at the cost of slower estimation speed.
%\Ours can also use a general DAG-structured causality pattern to factorize distributions. The complexity will be exponential with respect to the tree-width of this DAG~\cite{koller2009probabilistic}. 
%Larger tree-width will have better modeling accuracy but larger model size and slower estimation speed.
Thus, users of \Ours can make trade-offs between estimation accuracy and efficiency. We leave exploring these trade-offs as future work.

% \begin{figure}
% 	\centering
% 	\includegraphics[width=6cm]{fig/BN_tree.pdf}
% 	\vspace{-1.5em}
% 	\caption{Example causality pattern: tree-structured BN.}
% 	\label{fig: BN-tree}
% 	\vspace{-1em}
% \end{figure}

\subsection{Progressive estimates of sub-plan queries}
\label{subsec: progressive}

%The fundamental purpose of a \CE method is to help generate better query plans. 
In order to optimize the plan for one query, the \CE method needs to estimate the cardinalities of hundreds or thousands of sub-plan queries.
Estimating all these queries independently is very inefficient because they contain a lot of redundant computation.
Similar to the traditional \CE methods deployed in real DBMSes, \Ours supports estimating all sub-plan queries of one query progressively, reusing the computation as much as possible.
At a high level, the progressive estimation algorithm will estimate all sub-plan queries in a bottom-up fashion. \Ours first estimates all two-table join sub-plan queries, then it estimates the three-table join sub-plan queries, which will contain the two-table joins as their sub-structures, and so on. 
%In this case, \Ours can estimate these sub-plan queries using the cached results without redundant computation.

\noindent\underline{\textbf{Joining Factor Graphs:}} 
The key insight is that when we estimate the cardinality of a two-table join, we can also combine their factor graphs into a new factor graph for the joined table.
Specifically, as described in Section~\ref{sec: framework}, each base table is represented as a factor node in the factor graph. For instance, the factor node representing $A$ will store the unnormalized probability distribution and the MFV counts for the binned domain of join keys in $A$.
When estimating the cardinality of joining $A$ and $C$, our approximate inference algorithm in Equation~\ref{equ: bound2} computes the bound over all bins and sums them. These bounds essentially define an unnormalized probability distribution over all join keys on the denormalized table $A \Join C$. We can derive a bound on the new MFV counts for a join-key $A \Join C$ as the multiplication of the corresponding MFV counts in $A$ and $C$.
Therefore, we can cache this new probability distribution and MFV counts in a new factor node representing $A \Join C$.
Thus, when estimating the cardinality of $A \Join B \Join C$ we can consider it as a join of two single tables $(A \Join C$) and $B$. 

%\smallskip
%\noindent\underline{\textbf{Analysis:}} 
Since all sub-plans can essentially be considered a join of two sub-plans that we have already estimated, our inference algorithm only needs to compute the join of \revise{two-factor} nodes for each sub-plan query.
Therefore, the progressive estimation algorithm has no redundant computation and maximizes efficiency.
In experiments, with this algorithm, \Ours can estimate the cardinality of $10,000$ sub-plan queries in one second, which is more than ten times faster than estimating all these queries independently. Note that the progressive estimation algorithm is only possible to implement for \Ours and the traditional methods because it requires decomposing join queries into single-table estimates.
None of the existing learned methods can apply this method since their models are built to model the distributions of specific join patterns. 

% \begin{figure}
% 	\centering
% 	\includegraphics[width=7cm]{fig/sub_plans.pdf}
% 	\vspace{-1.3em}
% 	\caption{Sub-plan space of a query.}
% 	\label{fig: sub-plan}
% 	\vspace{-1em}
% \end{figure}

	\section{Experimental Evaluation}
\label{sec: experiments}

In this section, we empirically demonstrate the advantages of our framework over existing \CE methods. We first introduce the datasets, baselines, and experimental environment in Section~\ref{subsec: setups}. Then, we explore the following questions:

\noindent $\bullet$  \textbf{Overall performance} (Section~\ref{subsec: e2e}): how much improvement can \Ours achieve in terms of end-to-end query time? What are the model size and training time compared to SOTA  methods?

\noindent $\bullet$ \textbf{Detailed analysis} (Section~\ref{subsec: practicality}):
Why does \Ours attain good overall performance? \revise{When does it perform better or worse than existing baselines? How will it perform in terms of data updates?}

\noindent $\bullet$ \textbf{Ablation study} (Section~\ref{subsec: ablation}):
How much does each optimization technique of \Ours contribute to the overall performance?

\subsection{Experimental Setup}
\label{subsec: setups}

\noindent\textbf{\underline{Datasets:}} We evaluate the performance of our \CE framework on two well-established benchmarks: \emph{STATS-CEB}~\cite{han2021cardinality} and \emph{IMDB-JOB}~\cite{leis2015good}, whose statistics are summarized in Table~\ref{tab: dataset}. 

The \emph{STATS-CEB} benchmark uses the real-world dataset STATS, which is an anonymized dump of user-contributed content on
 Stack Exchange (Stack Overflow and related websites). 
This dataset consists of $8$ tables with  $34$ columns and more than one million rows. 
It contains a large number of attribute correlations and skewed data distributions.
The \emph{STATS-CEB} query workload contains $146$ queries involving $70$ different join templates with real-world semantics.
The true cardinalities of these queries range from $200$ to $2 \cdot 10^{10}$, with some queries taking more than one hour to execute on Postgres.
The \emph{STATS-CEB} benchmark only has star or chain join templates and numerical or categorical filtered attributes --- therefore it is possible to evaluate all existing \CE baselines on it.

The \emph{IMDB-JOB} benchmark uses the real-world Internet movie database (IMDB). This dataset contains $21$ tables with more than fifty million rows.
The \emph{IMDB-JOB} query workload contains $113$ queries with $33$ different join templates.
Some queries have more than $10,000$ sub-plan queries, posing a significant challenge to the efficiency of \CE methods.
Because \emph{IMDB-JOB} contains cyclic joins and string pattern matching filters, the \emph{JoinHist} and existing learned data-driven methods do not support this benchmark.

The queries in these two benchmarks cover a wide range of joins, filter predicates, cardinalities, and runtimes. To the best of our knowledge, they represent the most comprehensive and challenging real-world benchmarks for \CE evaluation.

%\vspace{.1in}
\noindent\textbf{\underline{Baselines:}} We compare our framework with the most competitive baselines in each \CE method category.

1) \emph{PostgreSQL}~\cite{psql2020} refers to the histogram-based \CE method
used in PostgreSQL. 

2) \emph{JoinHist}~\cite{dell2007join} is the classical join-histogram method whose details are explained in Section~\ref{subsect: related}. 

3) \emph{WJSample}~\cite{li2016wander} uses a random walk based-method called wander join to generate samples from the join of multiple tables. Its superior performance over traditional sampling-based methods has been demonstrated in recent studies~\cite{li2016wander, zhao2018random, park2020g}. \revise{We limit the amount of time spent for each random walk of \emph{WJSample} so the overall time spent for estimating cardinalities is comparable to other methods. If we allow \emph{WJSample} longer inference time, it can generate more effective query plans --- but it is impractical since its total estimation time will exceed the query execution time.}

4) \emph{MSCN}~\cite{kipf2018learned} is a learned query-driven method based on a multi-set convolutional network model. 
On both datasets, we train MSCN on roughly $100K$ sub-plan training queries that have similar distributions to the testing query workloads.

5) \emph{BayesCard}~\cite{wu2020bayescard}, 6) \emph{DeepDB}~\cite{hilprecht2019deepdb}, and 7) \emph{FLAT}~\cite{zhu2020flat} are the learned data-driven \CE methods. They use fanout methods to understand the data distributions of all join templates. Then, they apply different distribution models (Bayesian networks, sum-product networks~\cite{poon2011sum}, factorized sum-product networks~\cite{wu2020fspn}, respectively) to capture the data distribution on single tables or denormalized join tables.
We test these methods on \emph{STATS-CEB} using the optimally-tuned model parameters, as used in~\cite{han2021cardinality}. %As noted above, they do not support \emph{IMDB-JOB}.

8) \emph{PessEst}~\cite{cai2019pessimistic} is the SOTA bound-based method, which leverages randomized hashing and data sketching to tighten the bound for join queries. Its estimation has been verified to be very effective in real-world DBMSes.
We use the best-tuned hyperparameters to evaluate PessEst on both benchmarks.

9) \emph{U-Block}~\cite{hertzschuch2021simplicity} uses top-k statistics to estimate cardinality bounds. The original paper designed a new plan enumeration strategy to work with this bound and showed promising results on \emph{IMDB-JOB}. However, for a fair comparison, we only evaluate the standalone U-Block \CE method.

10) \emph{TrueCard} outputs true cardinality for given queries with no estimation latency. This baseline does not refer to any specific method but represents the optimal \CE performance.

We omit comparisons with many other \CE methods~\cite{bruno2001stholes, srivastava2006isomer, khachatryan2015improving, fuchs2007compressed, stillger2001leo, wu2018towards, heimel2015self,kiefer2017estimating, leis2017cardinality, hasan2020, yang2020neurocard, wu2021uae, dutt2019selectivity} \revise{because the baselines above have demonstrated clear advantages over these methods in various aspects~\cite{han2021cardinality}.} In addition, some newly proposed methods~\cite{sun2019end, wang2021face, liu2021fauce} do not have open-source  implementations yet.

We implemented \Ours in Python mainly using the NumPy package. Regarding the hyperparameters of \Ours, we assume no knowledge of the query workload and set the bin size $k$ to $100$ for both datasets. 
We use the BN-based \CE method~\cite{wu2020bayescard} as the single-table estimators for \emph{STATS-CEB} benchmark because this method has been shown to be very accurate, fast, and robust for single-table estimations.
For \emph{IMDB-JOB}, we use random sampling with $1\%$ sampling rate as the single-table estimator because this benchmark contains very complicated filter predicates such as disjunctions and string pattern matching, which are not supported by the learned data-driven methods. 
We discuss and compare other hyperparameters of our framework in Section~\ref{subsec: ablation}.

\begin{table}[t]
	\scalebox{0.78}{
		\begin{tabular}{c|c|cc}
			\thickhline
			Category &Statistics &  STATS-CEB & IMDB-JOB \\ \thickhline
			\multirow{5}{*}{Data} & \# of tables & 8 & 21  \\ \cline{2-4}
			& \# of rows in each table & $10^3$ --- $4 \cdot 10^5$ & 6 --- $3 \cdot 10^7$ \\ \cline{2-4}
			& \# of columns in each table & 3 --- 10 & 2-12 \\ \cline{2-4}
			& \# of join keys & 13 & 36 \\ \cline{2-4}
			& \# of equivalent key groups & 2 & 11 \\ \thickhline
			\multirow{7}{*}{Query} & \# of queries & 146 & 113 \\ \cline{2-4}
			&\# of join templates & 70 & 33  \\ \cline{2-4}
			&join template type & star \& chain & +cyclic\\ \cline{2-4}
			&\# of filter predicates & 1 --- 16 & 1 --- 13  \\ \cline{2-4}
			&filter attributes & numerical \&  categorical & +string LIKE \\ \cline{2-4}
			&\# of sub-plan queries & 1 --- 75 & 8 --- $1 \cdot 10^4$ \\ \cline{2-4}
			&true cardinality range & 200 --- $2\cdot10^{10}$  & 1 --- $4 \cdot 10^6$  \\ \cline{2-4}
			&Postgres runtime (s) & 0.05 --- $10^{4}$  & $5 \cdot 10^{-4}$ --- 450 \\
			\thickhline
		\end{tabular}
	}
	\caption{Summary of STATS-CEB and IMDB-JOB benchmark.}
	%\vspace{-2em}
	\label{tab: dataset}
\end{table}

\begin{figure*}
	\centering
	\includegraphics[width=17cm]{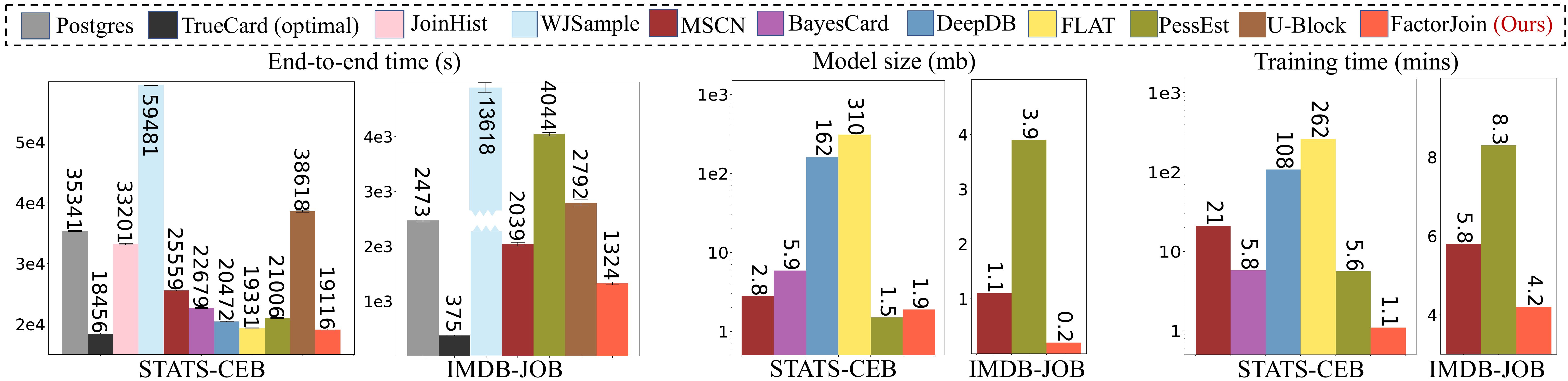}
	\vspace{-1.2em}
	\caption{Overall performance on STATS-CEB and IMDB-JOB.}
	\label{fig: overall}
	%\vspace{-1em}
\end{figure*}

\smallskip
\noindent\textbf{\underline{Environment:}}
\revise{We use the open-source implementation of all the baselines and tune their parameters to achieve the best possible performance.}
All experiments are performed on a Ubuntu server with Intel Core Processor CPU with 20 cores, 57GB DDR4 main memory, and 300GB SSD.
For end-to-end evaluation, we integrate our framework along with all the baselines into the query optimizer of PostgreSQL 13.1 following the procedure described in the recent work~\cite{han2021cardinality, han2021code}. 
We build the primary key and foreign key indices and disable parallel execution in PostgreSQL.
Specifically, we inject into PostgreSQL all sub-plan query cardinalities estimated by each method, 
so the PostgreSQL optimizer uses the injected cardinalities to optimize the query plan, and record the end-to-end query runtime (planning plus execution time). 
To eliminate the effects of cache, we first execute each workload multiple times. Then, we execute the queries three times for each method and report the average \revise{and the standard errors}.

\subsection{Overall performance}
\label{subsec: e2e}

The most straightforward criterion to evaluate the performance of a \CE method is to test how much each method improves the end-to-end query performance inside the query optimizer. Therefore, we compare the end-to-end performance along with the model size and training time of existing baselines on \emph{STATS-CEB} and \emph{IMDB-JOB}.

We summarize the overall performance of all baselines in Figure~\ref{fig: overall}.
Recall that \emph{JoinHist}, \emph{BayesCard}, \emph{DeepDB}, and \emph{FLAT} can not support the \emph{IMDB-JOB} benchmark.
In addition, we do not provide the model size and training time for \emph{Postgres}, \emph{TrueCard}, \emph{JoinHist}, \emph{WJSample}, and \emph{U-Block} as they are negligible.

\begin{table}[t]
	\scalebox{0.78}{
		\begin{tabular}{c|ccc}
			\thickhline
			Method &  End-to-end time & Exec. + plan time & Improvement \\ \thickhline
			Postgres & 35,341s  &  35,316s + 25s & -- \\ \hline
			TrueCard (Optimal)  & 18,456s & 18,432s + 24s & 47.8\% \\ \hline
			JoinHist  & 33,201s & 33,173s + 28s & 6.1\% \\ \hline
			WJSample & 59,481s & 59,436s + 45s & -68.4\% \\ \hline
			MSCN & 25,559s & 25,524s + 35s & 27.7\% \\ \hline
			BayesCard & 22,679s & 22,644s + 35s & 35.9\% \\ \hline
			DeepDB & 20,472s & 20,304s + 168s &  42.0\% \\ \hline
			FLAT & 19,331s & 18,934s + 437s & 45.3\% \\ \hline
			PessEst & 21,006s & 19,872s + 1,135s & 40.5\% \\ \hline
			U-Block & 38,618s & 38,592s + 26s & -9.3\% \\ \hline
			\textbf{FactorJoin (Ours)} & \textbf{19,116s} & 19,080s + 36s & \textbf{45.9\%} \\ \thickhline
		\end{tabular}
	}
	\caption{End-to-end performance on STATS-CEB.}
	%\vspace{-3em}
	\label{tab: STATS-CEB}
\end{table}

\smallskip
\noindent\textbf{\underline{Performance on STATS-CEB:}} 
In addition to Figure~\ref{fig: overall}, we provide a detailed comparison of query execution and planning time (sum over all queries in the \emph{STATS-CEB}) in Table~\ref{tab: STATS-CEB}, with the relative end-to-end improvement over \emph{Postgres} shown in the last column, i.e. (\emph{Postgres} time - \emph{method} time) / \emph{Postgres} time.

\Ours achieves the best end-to-end query runtime ($19,116s$), which is close to optimal performance ($18,456s$ for \emph{TrueCard}). 
When compared with the traditional methods (\emph{Postgres}, \emph{JoinHist}, and \emph{WJSample}), \Ours has significantly better query execution time, indicating that our estimates are much more effective at generating high-quality query plans. 
Meanwhile, the planning time of our framework is close to \emph{Postgres}. 
%\Ours efficient and accurate estimates.

Judging from the pure query execution time in Table~\ref{tab: STATS-CEB}, \Ours has comparable estimation effectiveness as the previous SOTA method \emph{FLAT} on this benchmark. 
This learned data-driven method uses the fanout method to understand the distribution of all join patterns, which is accurate but very inefficient. Specifically, \Ours can achieve better end-to-end performance than \emph{FLAT}, with $10 \times$ smaller planning time, $160 \times$ smaller model size, and $240 \times$ faster training time.
When compared to the other learned methods: \emph{BayesCard},  \emph{DeepDB}, and \emph{MSCN}, \Ours can simultaneously achieve better query execution and planning time, smaller model size, and faster training time.
Therefore, we empirically verify that instead of modeling the distributions of all join patterns, accurately decomposing the join query into single-table estimates can generate equally effective but more efficient estimation with much smaller model size and training time.

\Ours and the SOTA bound-based method \emph{PessEst} can both generate very effective estimates. This observation verifies that a bound-based method can produce effective estimates because it can help avoid very expensive query plans~\cite{cai2019pessimistic, atserias2008size, abo2017shannon, han2021cardinality}.
However, the planning time of \emph{PessEst} (1135s) is undesirable because it needs to materialize tables after applying the filters and populate the upper bound during run-time.
\Ours has $35 \times$ less planning latency since instead of using exact bounds, we use probabilistic bounds derived from single-table estimates, which is much more efficient.

\begin{table}[t]
	\scalebox{0.78}{
		\begin{tabular}{c|ccc}
			\thickhline
			Method &  End-to-end time & Exec. + plan time & Improvement \\ \thickhline
			Postgres & 2,472s  &  2,451s + 21s & -- \\ \hline
			TrueCard (Optimal)  & 375s & 358s + 17s & 84.8\% \\ \hline
			WJSample &  13,618s & 12,445s + 1,173s & -450.9\% \\ \hline
			MSCN & 2,039s  & 1,937s + 102s  & 18.1\% \\ \hline
			PessEst & 4,044s & 1,230s + 2,814s & -63.6\% \\ \hline
			U-Block & 2,792s & 2,762s + 30s & -12.9\% \\ \hline
			\textbf{FactorJoin (Ours)} & \textbf{1,324s} & 1,272s + 42s & \textbf{46.4\%} \\ \thickhline
		\end{tabular}
	}
	\caption{End-to-end performance on IMDB-JOB.}
	%\vspace{-2.5em}
	\label{tab: IMDB-JOB}
\end{table}

\smallskip
\noindent\textbf{\underline{Performance on IMDB-JOB:}} 
Table~\ref{tab: IMDB-JOB} shows a detailed end-to-end time comparison for the \emph{IMDB-JOB} workload.

\Ours achieves the best end-to-end query runtime ($1324s$) amongst all baselines.
Similar observations hold as in the \emph{STATS-CEB} workload when compared to the traditional methods. 
Like STATS-CEB, \emph{PessEst} achieves the best performance in terms of pure execution time but has a much larger overhead and planning latency.
\Ours has comparable performance as \emph{PessEst} in terms of pure execution time. However, our planning time is $50 \times$ shorter than \emph{PessEst}, making our framework much faster in overall end-to-end query time. Specifically, our framework can estimate $10,000$ sub-plan queries within one second, which is close to \emph{Postgres}.

However, unlike in \emph{STATS-CEB} where we have near-optimal performance, there still exists a large gap between \Ours (1324s) and the optimal \emph{TrueCard} (375s). There are two possible explanations. First, our probabilistic bound error can accumulate with the number of joins. Compared to \emph{STATS-CEB}, the \emph{IMDB-JOB} contains more than twice the number of joins in the queries. Therefore, the relative performance of our framework drops severely when shifting from \emph{STATS-CEB} to \emph{IMDB-JOB}. 
Second, the base-table filters in \emph{IMDB-JOB} contain a large amount of highly-selective predicates, making our sampling-based single-table \CE less accurate. 
These observations suggest room for improving \Ours, which we discuss further in the future work section.

\smallskip
\noindent\textbf{\underline{Summary:}}
\Ours achieves the best performance among all the baselines on both benchmarks. 
Specifically, \Ours is as effective as the previous SOTA methods, and simultaneously as efficient and practical as the traditional \CE methods. 

%We leave the exploration of our model update as future work.\pari{again -- can move to future work section, or maybe even a stronger: It is much easier to update these models, while past work in CE models such as NeuroCard, MSCN etc. (...), typically suggest to retrain the model as data is updated.}

\begin{figure}
	\centering
	\includegraphics[width=8.5cm]{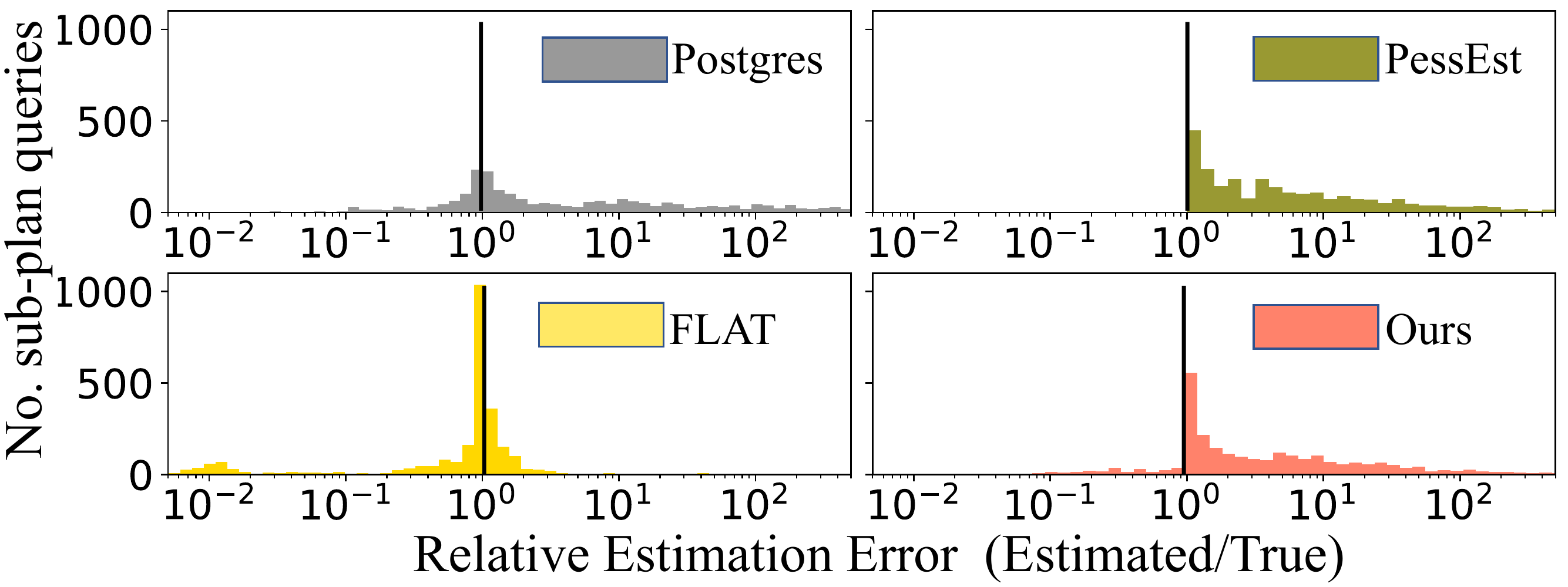}
	\vspace{-2.5em}
	\caption{Relative estimation errors on \emph{STATS-CEB}.}
	\label{fig: errors}
	%\vspace{-1em}
\end{figure}

\subsection{Detailed Analysis}
\label{subsec: practicality}
In this section, we first study why \Ours is able to achieve  SOTA performance by analyzing the tightness of our probabilistic bound. 
Then, we exhaustively compare the \CE method's end-to-end performance on queries with different runtimes to investigate when can \Ours perform better or worse than the competitive baselines. For compactness, here, we only report the results of the most representative methods (\emph{Postgres}, \emph{FLAT}, \emph{PessEst}) on \emph{STATS-CEB};  the remaining results are in the appendix. 

\smallskip
\noindent\underline{\textbf{Bound tightness:}} 
We show the relative errors between the baselines' estimates and the true cardinalities (estimate / true) for all sub-plan queries on \emph{STATS-CEB} in Figure~\ref{fig: errors}.
Overall, all three SOTA methods significantly outperform \emph{Postgres} in terms of estimation accuracy. \emph{PessEst} generates exact upper bounds and never underestimates. \emph{FLAT} uses a much larger model to understand the distributions of join patterns and produces the most accurate estimates.

\Ours can output an upper bound on cardinality for more than $90\%$ of the sub-plan queries. Most of the marginal underestimates are very close to the true cardinality, which can still generate relatively effective plans.
Our generated bounds are slightly tighter than the bounds from \emph{PessEst}, which explains why our query plans achieve slightly better results.

\begin{figure}
	\centering
	\includegraphics[width=8.5cm]{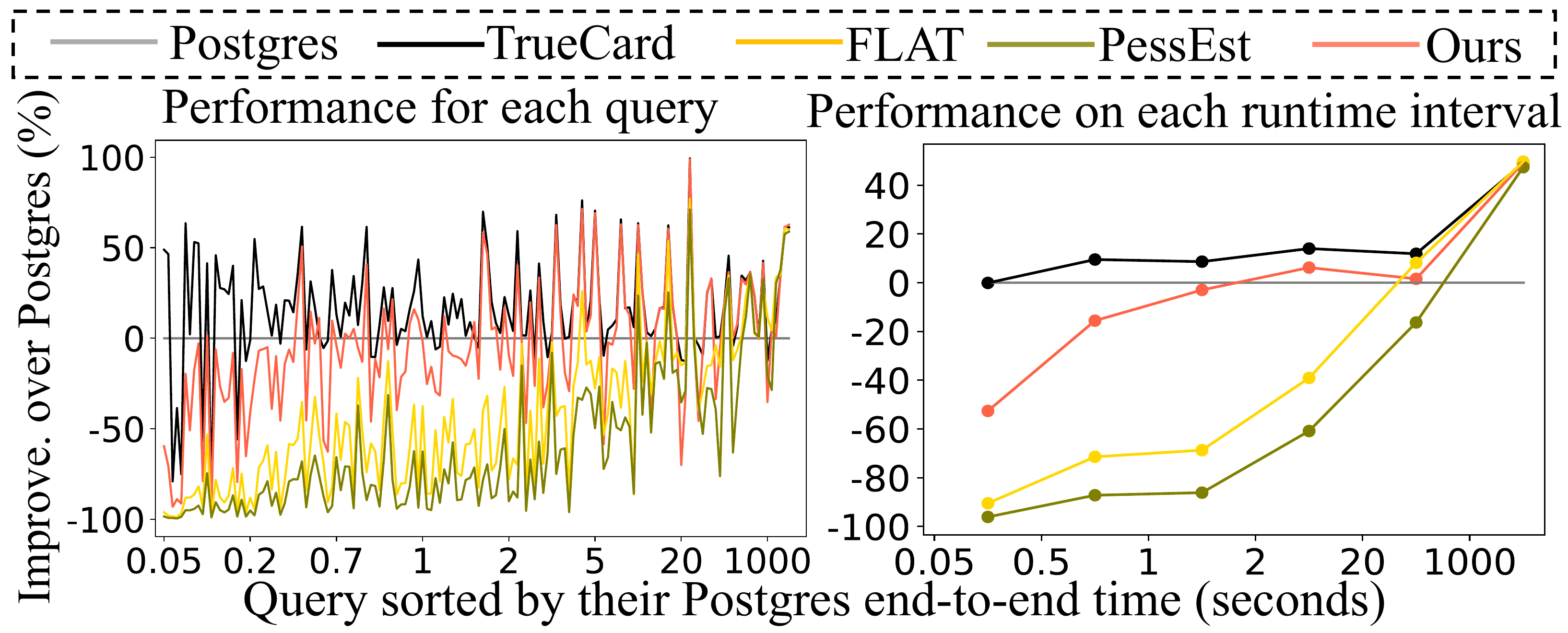}
	\vspace{-2.5em}
	\caption{Per query performance of \emph{STATS-CEB}.}
	\label{fig: intervals}
	%\vspace{-1em}
\end{figure}

\smallskip
\noindent\underline{\textbf{Detailed comparision:}} We sort the $146$ queries of \emph{STATS-CEB} based on their Postgres end-to-end runtime and cluster them into $6$ different runtime intervals.
Figure~\ref{fig: intervals} reports relative improvements of the most competitive baselines over \emph{Postgres} for each query on the left and each cluster of queries on the right.   

For the very short-running queries (which represent an OLTP-like workload), \emph{Postgres} is the best amongst all baselines. These baselines perform worse because the estimation latency plays a significant role in these queries.
We observe that improving the estimation accuracy of these queries has a very limited effect on the query plan quality. 
This also explains why the optimal \emph{TrueCard} only marginally outperforms \emph{Postgres} on queries with less than $1s$ of runtime.  
However, these queries only contribute a negligible proportion of total runtime in \emph{STATS-CEB}. The query optimizer should fall back to default traditional \CE methods for OLTP-like workload.
%\srm{I'm concerned a review might ding us because we actually make many of the queries slower, so they might say our system isn't useful except in the case of the longest running queries.  I think we should note that for these short running time queries the average change in actually latency is very slow - i.e., we only hurt queries < 1s and the average difference in runtimes is x ms for these queries.  Alternatively, we can say we can fallback on the traditional optimizer for queries where we estimate the runtime to be < 1s?}
%The estimation latency plays a significant role in these queries.
\emph{FLAT} and \emph{PessEst} are significantly worse than \Ours because their planning latency surpasses the execution time, which overshadows the minor improvement in query plans.

For the extremely long-running queries, the advantages of the SOTA methods over \emph{Postgres} gradually appear. The reason is that estimation latency becomes increasingly insignificant for queries with a long execution time. Thus, \emph{FLAT} and \emph{PessEst} which spend a long time planning can generate significantly better query plans. \Ours has comparable performance on these queries. 
%\srm{I don't love this graph because it makes it look like the number of queries where we help is very small.  Maybe a scatter plot would be better?}

\smallskip
\noindent\revise{\underline{\textbf{Incremental updates:}}}
%\Ours is very friendly for dynamic DBs with frequent data changes because we only need to update the single table statistics.
Following the same evaluation setup as previous work~\cite{han2021cardinality}, we train a stale \Ours model on STATS data created before 2014 (roughly $50\%$) and
insert the rest of the data to incrementally update this model.

We provide the end-to-end query time of the updated \Ours and the update time on \emph{STATS-CEB} queries in Table~\ref{tab: update}. Since \Ours only needs to update single-table statistics, it is extremely efficient, i.e. taking only $2.5s$ to update millions of tuples. 
This update speed up to $168 \times$ faster than the learned data-driven methods with better end-to-end query performance after model update.
We cannot fairly compare with the learned query-driven methods because there does not exist a training query workload after the data insertion.
It worth noticing that the update times reported for other methods in Table~\ref{tab: update} include the time to re-compute the denormalized join tables for the inserted data, so numbers are larger than the ones reported in the original paper~\cite{han2021cardinality}.

However, we do see a slight drop in the end-to-end improvement when compared to the \Ours trained on the entire dataset ($43.4\%$ in Table~\ref{tab: update} versus $45.9\%$ in Table~\ref{tab: STATS-CEB}). This small performance difference is due to the fact that the bins are decided on the data before 2014 and remain fixed during incremental updates. Therefore, after inserting the new data, the min-variance property of the bins could be violated, resulting in less accurate predictions.

\begin{table}[t]
\revise{
	\scalebox{0.9}{
		\begin{tabular}{c|c|c|c}
			\thickhline
			Method & Update time & End-to-end time & Improvement \\ \hline
			BayesCard & 84s & 22,679s & 35.9\% \\ \hline
			DeepDB & 310s & 21,352s & 39.6\%
			\\ \hline
			FLAT & 422s & 22,120s & 37.4\%
			\\ \hline
			\textbf{FactorJoin (Ours)} & \textbf{2.5s} & \textbf{20,015s} & \textbf{43.4\%}
			  \\ \thickhline
		\end{tabular}
	}
	}
	\caption{\revise{Incremental update performance on STATS-CEB.}}
	\vspace{-2em}
	\label{tab: update}
\end{table}

%This update procedure is clearly more efficient than existing learned methods, which either need to re-denormalize the join tables and re-calculate the fanout attributes~\cite{wu2020bayescard, hilp2019deepdb, zhu2020flat, yang2020neurocard} or need to gather new queries to retrain the model~\cite{kipf2018learned, sun2019end}. We leave the detailed comparison on update as future work.}

\subsection{Ablation study}
\label{subsec: ablation}

We conduct a series of ablation studies to analyze different optimization techniques inside \Ours. Specifically, we first investigate how the number of bins (k) affects the overall performance. Next, we analyze the effectiveness of our new bin selection algorithm (GBSA in Section~\ref{subsec: bin selection}) over the traditional equal-width and equal-depth algorithms. Then, we study the performance of plugging-in different single-table \CE methods into \Ours. Due to space limitations, we only provide the results on \emph{STATS-CEB}. At last, we investigate how removing each of the simplifying assumptions improves the performance of the original joining histogram.

\begin{figure}
	\centering
	\includegraphics[width=8.3cm]{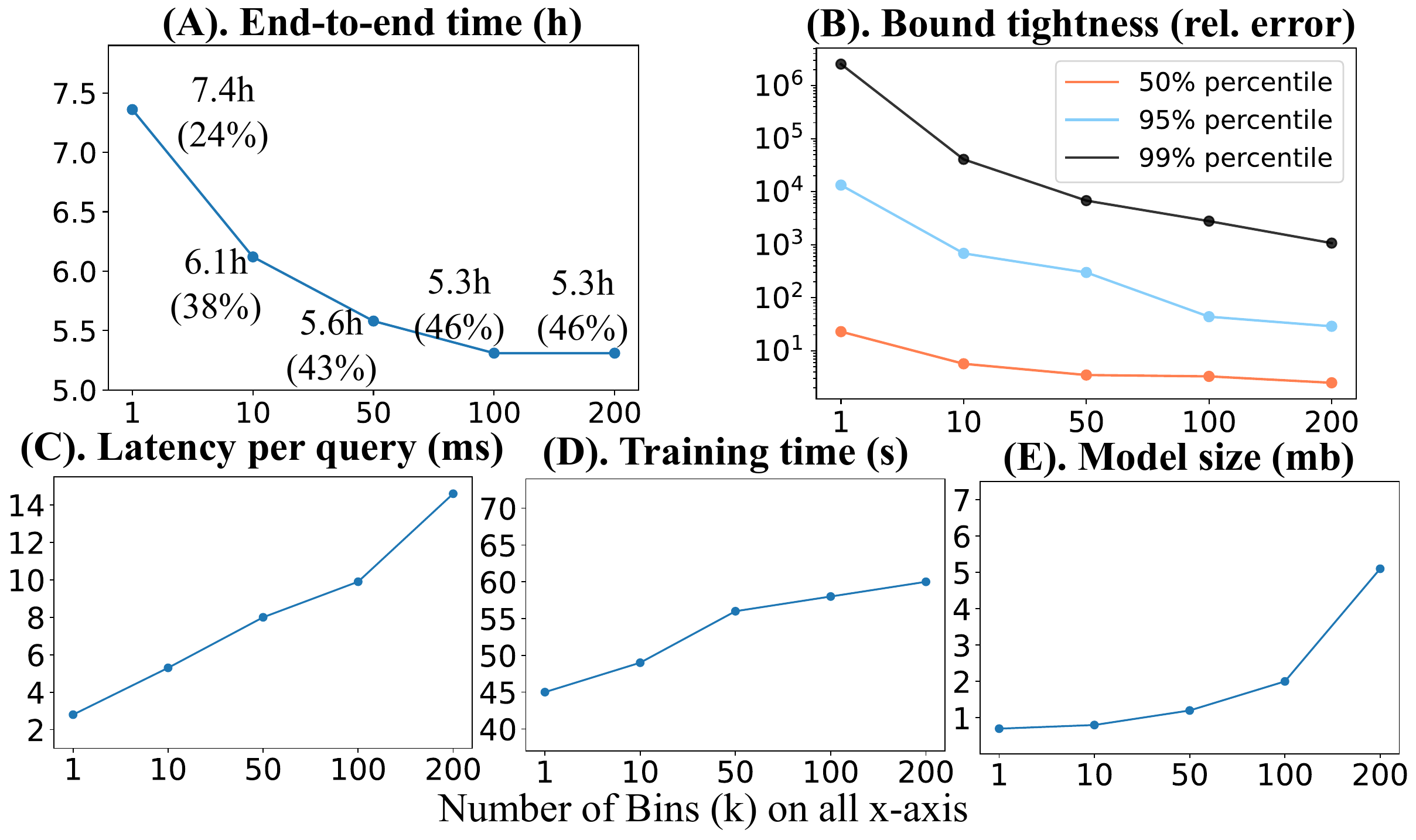}
	\vspace{-1.5em}
	\caption{Performance for different number of bins.}
	\label{fig: n_bins}
	%\vspace{-1em}
\end{figure}

\smallskip
\noindent\underline{\textbf{Number of bins (k):}} 
We train and evaluate \Ours with five different number of bins ($1$, $10$, $50$, $100$, $200$).
Figure~\ref{fig: n_bins} shows the end-to-end query time (A), bound tightness (B), estimation latency per query (C), training time (D), and model size (E) of \Ours.
For bound tightness, we report the $50\%, 95\%, 99\%$ percentiles of all queries' relative error: estimated/true, as the one used in Figure~\ref{fig: errors}.

Figure~\ref{fig: n_bins}-(A-B) shows that larger numbers of bins ($k$) will generate tighter cardinality bounds and thus more effective query plans.
Figure~\ref{fig: n_bins}-(C-E) shows that more bins also increase estimation latency, training time, and model size, as expected.
There are several interesting observations from these results.

$\bullet$ According to Figure~\ref{fig: n_bins}-(B), with $k=1$ (i.e. bin the entire domain of each join key into one bin) \Ours does not generate tight bounds but it still outperforms \emph{Postgres} baseline by $24.9\%$. 
This highlights the importance of upper bounds over underestimation and verifies the effectiveness of bound-based algorithms.

$\bullet$ Increasing $k$ can substantially increase the bound tightness. However, at some point (from $k=100$ to $k=200$) the framework stops achieving better end-to-end performance. One explanation is that this fine-grained improvement in estimation accuracy can generate a slightly better query plan but the slower estimation latency cancels out the improvements in query plans. 

$\bullet$ The model size increases quadratically with $k$ because \Ours has $O(N*k^2)$ complexity in both storage and inference. However, estimation latency increases linearly with $k$ because the overall inference time is dominated by the single-table \CE methods' inference time, whose complexity is linear with $k$.

\begin{table}[t]
	\scalebox{0.78}{
		\begin{tabular}{c|c|c|ccc}
			\thickhline
			\multirow{2}{*}{Algorithm} &  \multirow{2}{*}{End-to-end time} &  \multirow{2}{*}{Improvement} & \multicolumn{3}{c}{Bound Tightness (rel. error)} \\ \cline{4-6}
			& & & 50\% & 95\% & 99\% \\ \hline
			Equal-width & 23,868s & 32.4\%& 8.7 & 3,135 & $2\times 10^5$ \\ \hline
			Equal-depth & 23,436s & 33.6\% & 8.4 & 2,050 & $7 \times 10^4$ \\ \hline
			GBSA & \textbf{19,116s} & \textbf{45.9\%} &\textbf{3.3} & \textbf{44} & \textbf{2,782}  \\ \thickhline
		\end{tabular}
	}
	\caption{Performance of different binning techniques.}
	\vspace{-1em}
	\label{tab: ablation-binning}
\end{table}

%\smallskip
\noindent\underline{\textbf{Different bin selection algorithms:}} Recall that in Section~\ref{subsec: bin selection}, we design a new algorithm (GBSA) to optimize the binning process for \Ours. Here, we compare it with the traditional equal-width and equal-depth binning algorithms. We set $k = 100$ for all three algorithms.
They have approximately the same estimation latency, training time, and model size. 
Thus, we only report the end-to-end performance and bound tightness in Figure~\ref{tab: ablation-binning}. 
We observe that GBSA generates much tighter upper bounds, leading to significantly better end-to-end performance. This demonstrates the effectiveness of our GBSA algorithm.

%\smallskip
\noindent\underline{\textbf{Varying single-table \CE methods:}} 
To compare the importance of different single-table \CE methods in \Ours, we tried three different estimators: 1) Bayesian network (\revise{\emph{BayesCard}}) for single-table, which is the SOTA learned single-table method~\cite{wu2020bayescard}, 2) \emph{Sampling}, which draws a random sample (5\%) from single tables on-the-fly and estimates the cardinalities of filter predicates, and 3) \emph{TrueScan} which scans and filters the tables during query time and calculate the true cardinalities.
The end-to-end performance of different single table estimators with bin size $k=100$ is shown in Table~\ref{tab: single}. The \revise{\emph{BayesCard}} method performs significantly better than \emph{Sampling} because of more accurate estimates and slightly faster estimation speeds. \revise{\emph{TrueScan}} produces an exact upper bound on the cardinalities and thus generates more effective query plans. However, the estimation latency is too high, making its overall end-to-end performance worse than \revise{\emph{BayesCard}}.

\begin{table}[t]
	\scalebox{0.8}{
		\begin{tabular}{c|ccc}
			\thickhline
			Method &  End-to-end time & Exec. + plan time & Improvement \\ \hline
			\revise{BayesCard} & \textbf{19,116s}  &  19,080s + 36s & \textbf{45.9\%} \\ \hline
			Sampling  & 20,633s & 20,592s+ 41s & 41.6\% \\ \hline
			TrueScan & 19,334s & 18,756s + 578s & 45.3\% \\ \thickhline
		\end{tabular}
	}
	\caption{\Ours, varying single-table \CE methods.}
	\vspace{-2em}
	\label{tab: single}
\end{table}

\smallskip
\noindent\underline{\textbf{Improvement over joining histograms:}}
Since \Ours follows the convention of \emph{JoinHist}, we investigate how much each component of \Ours help improve this method. Specifically, apart from the original \emph{JoinHist} method, we evaluate and compare the following variants. We incorporate our new probabilistic bound  into \emph{JoinHist} to remove its \emph{join uniformity assumption} (denote as \emph{with Bound}).
We incorporate into \emph{JoinHist} the histograms learned from \emph{BayesNet} that represent conditional distributions of join keys given the filter predicates (denote as \emph{with Conditional}). This variant avoids the \emph{attribute independent assumption}.
\Ours reduces to the \emph{JoinHist} with both bound and conditional techniques on non-cyclic join templates.
The results in Table~\ref{tab: joinhist} demonstrate that we can achieve significant improvement over the \emph{JoinHist} method by removing either or all of its simplifying assumptions.

\smallskip
\noindent\underline{\textbf{Summary:}} 
With a reasonable bin size $k$, \Ours generates tight bounds and effective query plans. The GBSA bin selection algorithm is very effective at generating high-quality plans. Single-table estimators do have an impact on \Ours's overall performance and we should choose one that can generate effective and efficient estimates.
Overall, all different settings/hyperparameters of \Ours can significantly outperform \emph{Postgres} as well as all existing SOTA methods, demonstrating the robustness and stability of \Ours.

\begin{table}[t]
	\scalebox{0.8}{
		\begin{tabular}{c|ccc}
			\thickhline
			Method &  End-to-end time & Exec. + plan time & Improvement \\ \hline
			JoinHist & 33,201s &  33,173s + 28s & 6.1\% \\ \hline
			with Bound  & 29,175s & 29147s+ 28s & 17.46\% \\ \hline
			with Conditional & 23,450s & 23,414s + 36s & 31.7\% \\ \hline 
			with Both (Ours) & \textbf{19,116s} & 19,080s + 36s & \textbf{45.9\%} \\
			\thickhline
			
		\end{tabular}
	}
	\caption{Performance comparison with JoinHist.}
	%\vspace{-2em}
	\label{tab: joinhist}
\end{table}
	
	\section{Related work}
\label{sec: related}

We briefly review literatures on single-table \CE methods and the learned query optimizers.

\smallskip
\noindent\underline{\textbf{Single-table \CE:}} 
%Traditional approaches including histograms~\cite{selinger1979access} and sampling~\cite{lipton1990practical}, are very efficient and perfect for system deployment but can have very ineffective estimates. 
%Multi-dimensional histograms~\cite{poosala1997selectivity, deshpande2001independence, gunopulos2000approximating, gunopulos2005selectivity, muralikrishna1988equi, wang2003multi, liu2021lhist}, self-tuning histograms~\cite{bruno2001stholes, srivastava2006isomer, khachatryan2015improving, fuchs2007compressed}, and kernel density estimations~\cite{heimel2015self,kiefer2017estimating} are popular improvements of the traditional methods. 
Machine learning approaches have been use to solve single-table \CE problem, which can achieve accurate estimation with very low latency and overhead~\cite{yang2019deep, wu2020bayescard, zhu2020flat}.
Specifically, the data-driven learned methods build statistical models to understand data distributions, such as deep autoregressive models~\cite{yang2019deep, hasan2019multi}, Bayesian networks~\cite{wu2020bayescard, dasfaa2019, getoor2001selectivity}, sum-product-networks~\cite{hilprecht2019deepdb}, factorized sum-product-networks~\cite{zhu2020flat}, and normalizing flow models~\cite{wang2021face}. 
As for the query-driven methods, the first approach using neural networks for \CE has been proposed for UDF predicates~\cite{lakshmi1998selectivity}. Later on, a semi-automatic alternative~\cite{malik2007black} and a
regression-based model~\cite{akdere2012learning}  were used for general predicates.
Recently, more sophisticated supervised models such as multi-set convolutional networks~\cite{kipf2018learned}, XG-boost~\cite{dutt2019selectivity}, tree-LSTM~\cite{sun2019end}, and deep ensembles~\cite{liu2021fauce}, are used to provide accurate estimation.

These works are orthogonal to our framework and in principle, we can support plugging in any one of them for single table \CE.

\smallskip
\noindent\underline{\textbf{Learned query optimizer:}} Apart from \CE, there exist a large number of works that use ML to solve other tasks of query optimizer, such as cost estimation and join order selection. 
Learned cost estimation methods use tree convolution~\cite{marcus2019plan} and tree-LSTM~\cite{sun2019end} to encode a query plan as a tree and map the encoding to its estimated costs.
Active learning~\cite{ma2020active} and zero-shot learning~\cite{hilprecht2021one} are proposed to tackle this problem from a new perspective.
Some deep reinforcement learning approaches~\cite{yu2020reinforcement, marcus2018deep} have been proposed to determine the optimal join order.
Recently, many methods~\cite{marcus2019neo, marcus2020bao, negi2021steering, wu2021unified} propose to learn the entire query optimizer end-to-end without a clear separation of these components.

\section{Conclusions}
	\label{sec: conclusion}
	In this paper, we propose \Ours, a framework for cardinality estimation of join queries. It combines classical join-histogram methods with learned single table cardinality estimates into a factor graph model.  This framework converts \CE problem into an inference problem over the factor graph involving only single-table distributions. We further propose several optimizations to make this problem tractable for large join graphs.
	Our experiments show that \Ours generates effective and efficient estimates and is suitable for system deployment on large real-world join benchmarks.
	Specifically, \Ours produces more effective estimates than the previous SOTA learned methods, with 40x less estimation latency, 100x smaller model size, 100x faster updating speed, and 100x faster training speed, while matching or exceeding them in terms of query execution time.
	%\srm{Give a performance takeaway.}s
	We believe this work points out a new direction for estimating join queries, which would enable truly practical learned \CE methods as a system component. 
	
	%As future work, we plan to integrate \Ours into commercial DBMSes and further explore the merits and limits of our framework on various real-world DB instances. %Moreover, we would like to further optimize \Ours by improving its modeling accuracy, efficiency, and updatability.

%\clearpage

\bibliographystyle{ACM-Reference-Format}
\bibliography{main}


\begin{thebibliography}{77}

%%% the Web address.

\ifx \showCODEN    \undefined \def \showCODEN     #1{\unskip}     \fi
\ifx \showDOI      \undefined \def \showDOI       #1{#1}\fi
\ifx \showISBNx    \undefined \def \showISBNx     #1{\unskip}     \fi
\ifx \showISBNxiii \undefined \def \showISBNxiii  #1{\unskip}     \fi
\ifx \showISSN     \undefined \def \showISSN      #1{\unskip}     \fi
\ifx \showLCCN     \undefined \def \showLCCN      #1{\unskip}     \fi
\ifx \shownote     \undefined \def \shownote      #1{#1}          \fi
\ifx \showarticletitle \undefined \def \showarticletitle #1{#1}   \fi
\ifx \showURL      \undefined \def \showURL       {\relax}        \fi
% invisible to TeX
\providecommand\bibfield[2]{#2}
\providecommand\bibinfo[2]{#2}
\providecommand\natexlab[1]{#1}
\providecommand\showeprint[2][]{arXiv:#2}

\bibitem[\protect\citeauthoryear{Abo~Khamis, Ngo, and Suciu}{Abo~Khamis
  et~al\mbox{.}}{2017}]%
        {abo2017shannon}
\bibfield{author}{\bibinfo{person}{Mahmoud Abo~Khamis}, \bibinfo{person}{Hung~Q
  Ngo}, {and} \bibinfo{person}{Dan Suciu}.} \bibinfo{year}{2017}\natexlab{}.
\newblock \showarticletitle{What do Shannon-type Inequalities, Submodular
  Width, and Disjunctive Datalog have to do with one another?}. In
  \bibinfo{booktitle}{\emph{Proceedings of the 36th ACM SIGMOD-SIGACT-SIGAI
  Symposium on Principles of Database Systems}}. \bibinfo{pages}{429--444}.
\newblock


\bibitem[\protect\citeauthoryear{Akdere, Cetintemel, Riondato, Upfal, and
  Zdonik}{Akdere et~al\mbox{.}}{2012}]%
        {akdere2012learning}
\bibfield{author}{\bibinfo{person}{Mert Akdere}, \bibinfo{person}{Ugur
  Cetintemel}, \bibinfo{person}{Matteo Riondato}, \bibinfo{person}{Eli Upfal},
  {and} \bibinfo{person}{Stanley~B Zdonik}.} \bibinfo{year}{2012}\natexlab{}.
\newblock \showarticletitle{Learning-based query performance modeling and
  prediction}. In \bibinfo{booktitle}{\emph{2012 IEEE 28th International
  Conference on Data Engineering}}. IEEE, \bibinfo{pages}{390--401}.
\newblock


\bibitem[\protect\citeauthoryear{Atserias, Grohe, and Marx}{Atserias
  et~al\mbox{.}}{2008}]%
        {atserias2008size}
\bibfield{author}{\bibinfo{person}{Albert Atserias}, \bibinfo{person}{Martin
  Grohe}, {and} \bibinfo{person}{D{\'a}niel Marx}.}
  \bibinfo{year}{2008}\natexlab{}.
\newblock \showarticletitle{Size bounds and query plans for relational joins}.
  In \bibinfo{booktitle}{\emph{2008 49th Annual IEEE Symposium on Foundations
  of Computer Science}}. IEEE, \bibinfo{pages}{739--748}.
\newblock


\bibitem[\protect\citeauthoryear{Bruno, Chaudhuri, and Gravano}{Bruno
  et~al\mbox{.}}{2001}]%
        {bruno2001stholes}
\bibfield{author}{\bibinfo{person}{Nicolas Bruno}, \bibinfo{person}{Surajit
  Chaudhuri}, {and} \bibinfo{person}{Luis Gravano}.}
  \bibinfo{year}{2001}\natexlab{}.
\newblock \showarticletitle{STHoles: a multidimensional workload-aware
  histogram}. In \bibinfo{booktitle}{\emph{SIGMOD}}. \bibinfo{pages}{211--222}.
\newblock


\bibitem[\protect\citeauthoryear{Cai, Balazinska, and Suciu}{Cai
  et~al\mbox{.}}{2019}]%
        {cai2019pessimistic}
\bibfield{author}{\bibinfo{person}{Walter Cai}, \bibinfo{person}{Magdalena
  Balazinska}, {and} \bibinfo{person}{Dan Suciu}.}
  \bibinfo{year}{2019}\natexlab{}.
\newblock \showarticletitle{Pessimistic cardinality estimation: Tighter upper
  bounds for intermediate join cardinalities}. In
  \bibinfo{booktitle}{\emph{SIGMOD}}. \bibinfo{pages}{18--35}.
\newblock


\bibitem[\protect\citeauthoryear{Chow and Liu}{Chow and Liu}{1968}]%
        {chow1968approximating}
\bibfield{author}{\bibinfo{person}{C. Chow} {and} \bibinfo{person}{Cong Liu}.}
  \bibinfo{year}{1968}\natexlab{}.
\newblock \showarticletitle{Approximating discrete probability distributions
  with dependence trees}.
\newblock \bibinfo{journal}{\emph{IEEE transactions on Information Theory}}
  \bibinfo{volume}{14}, \bibinfo{number}{3} (\bibinfo{year}{1968}),
  \bibinfo{pages}{462--467}.
\newblock


\bibitem[\protect\citeauthoryear{Dell'Era}{Dell'Era}{2007}]%
        {dell2007join}
\bibfield{author}{\bibinfo{person}{Alberto Dell'Era}.}
  \bibinfo{year}{2007}\natexlab{}.
\newblock \showarticletitle{Join Over Histograms}.
\newblock \bibinfo{journal}{\emph{Available on www. adellera.
  it/investigations/join\_over\_histograms}} (\bibinfo{year}{2007}).
\newblock


\bibitem[\protect\citeauthoryear{Dell'Era}{Dell'Era}{2017}]%
        {oracle}
\bibfield{author}{\bibinfo{person}{Alberto Dell'Era}.}
  \bibinfo{year}{2017}\natexlab{}.
\newblock \showarticletitle{Oracle Database Online Documentation 12c Release 1
  (12.1)}.
\newblock \bibinfo{journal}{\emph{Available at:
  <http://docs.oracle.com/database/121/index.html>}} (\bibinfo{year}{2017}).
\newblock


\bibitem[\protect\citeauthoryear{Deshpande, Garofalakis, and Rastogi}{Deshpande
  et~al\mbox{.}}{2001}]%
        {deshpande2001independence}
\bibfield{author}{\bibinfo{person}{Amol Deshpande}, \bibinfo{person}{Minos
  Garofalakis}, {and} \bibinfo{person}{Rajeev Rastogi}.}
  \bibinfo{year}{2001}\natexlab{}.
\newblock \showarticletitle{{Independence is good: Dependency-based histogram
  synopses for high-dimensional data}}.
\newblock \bibinfo{journal}{\emph{ACM SIGMOD Record}} \bibinfo{volume}{30},
  \bibinfo{number}{2} (\bibinfo{year}{2001}), \bibinfo{pages}{199--210}.
\newblock


\bibitem[\protect\citeauthoryear{Documentation~12}{Documentation~12}{2020}]%
        {psql2020}
\bibfield{author}{\bibinfo{person}{Postgresql Documentation~12}.}
  \bibinfo{year}{2020}\natexlab{}.
\newblock \showarticletitle{Chapter 70.1. Row Estimation Examples}.
\newblock
  \bibinfo{journal}{\emph{https://www.postgresql.org/docs/current/row-estimation-examples.html}}
  (\bibinfo{year}{2020}).
\newblock


\bibitem[\protect\citeauthoryear{Dutt, Wang, Nazi, Kandula, Narasayya, and
  Chaudhuri}{Dutt et~al\mbox{.}}{2019}]%
        {dutt2019selectivity}
\bibfield{author}{\bibinfo{person}{Anshuman Dutt}, \bibinfo{person}{Chi Wang},
  \bibinfo{person}{Azade Nazi}, \bibinfo{person}{Srikanth Kandula},
  \bibinfo{person}{Vivek Narasayya}, {and} \bibinfo{person}{Surajit
  Chaudhuri}.} \bibinfo{year}{2019}\natexlab{}.
\newblock \showarticletitle{Selectivity estimation for range predicates using
  lightweight models}.
\newblock \bibinfo{journal}{\emph{PVLDB}} \bibinfo{volume}{12},
  \bibinfo{number}{9} (\bibinfo{year}{2019}), \bibinfo{pages}{1044--1057}.
\newblock


\bibitem[\protect\citeauthoryear{Fuchs, He, and Lee}{Fuchs
  et~al\mbox{.}}{2007}]%
        {fuchs2007compressed}
\bibfield{author}{\bibinfo{person}{Dennis Fuchs}, \bibinfo{person}{Zhen He},
  {and} \bibinfo{person}{Byung~Suk Lee}.} \bibinfo{year}{2007}\natexlab{}.
\newblock \showarticletitle{{Compressed histograms with arbitrary bucket
  layouts for selectivity estimation}}.
\newblock \bibinfo{journal}{\emph{Information Sciences}} \bibinfo{volume}{177},
  \bibinfo{number}{3} (\bibinfo{year}{2007}), \bibinfo{pages}{680--702}.
\newblock


\bibitem[\protect\citeauthoryear{Getoor, Taskar, and Koller}{Getoor
  et~al\mbox{.}}{2001}]%
        {getoor2001selectivity}
\bibfield{author}{\bibinfo{person}{Lise Getoor}, \bibinfo{person}{Benjamin
  Taskar}, {and} \bibinfo{person}{Daphne Koller}.}
  \bibinfo{year}{2001}\natexlab{}.
\newblock \showarticletitle{Selectivity estimation using probabilistic models}.
  In \bibinfo{booktitle}{\emph{SIGMOD}}. \bibinfo{pages}{461--472}.
\newblock


\bibitem[\protect\citeauthoryear{Gunopulos, Kollios, Tsotras, and
  Domeniconi}{Gunopulos et~al\mbox{.}}{2000}]%
        {gunopulos2000approximating}
\bibfield{author}{\bibinfo{person}{Dimitrios Gunopulos},
  \bibinfo{person}{George Kollios}, \bibinfo{person}{Vassilis~J Tsotras}, {and}
  \bibinfo{person}{Carlotta Domeniconi}.} \bibinfo{year}{2000}\natexlab{}.
\newblock \showarticletitle{{Approximating multi-dimensional aggregate range
  queries over real attributes}}. In \bibinfo{booktitle}{\emph{SIGMOD}}.
  \bibinfo{pages}{463--474}.
\newblock


\bibitem[\protect\citeauthoryear{Gunopulos, Kollios, Tsotras, and
  Domeniconi}{Gunopulos et~al\mbox{.}}{2005}]%
        {gunopulos2005selectivity}
\bibfield{author}{\bibinfo{person}{Dimitrios Gunopulos},
  \bibinfo{person}{George Kollios}, \bibinfo{person}{Vassilis~J Tsotras}, {and}
  \bibinfo{person}{Carlotta Domeniconi}.} \bibinfo{year}{2005}\natexlab{}.
\newblock \showarticletitle{Selectivity estimators for multidimensional range
  queries over real attributes}.
\newblock \bibinfo{journal}{\emph{The VLDB Journal}} \bibinfo{volume}{14},
  \bibinfo{number}{2} (\bibinfo{year}{2005}), \bibinfo{pages}{137--154}.
\newblock


\bibitem[\protect\citeauthoryear{Halford, Saint-Pierre, and Morvan}{Halford
  et~al\mbox{.}}{2019}]%
        {dasfaa2019}
\bibfield{author}{\bibinfo{person}{Max Halford}, \bibinfo{person}{Philippe
  Saint-Pierre}, {and} \bibinfo{person}{Franck Morvan}.}
  \bibinfo{year}{2019}\natexlab{}.
\newblock \showarticletitle{An approach based on bayesian networks for query
  selectivity estimation.}
\newblock \bibinfo{journal}{\emph{DASFAA}}  \bibinfo{volume}{2}
  (\bibinfo{year}{2019}).
\newblock


\bibitem[\protect\citeauthoryear{Han}{Han}{2021}]%
        {han2021code}
\bibfield{author}{\bibinfo{person}{Yuxing Han}.}
  \bibinfo{year}{2021}\natexlab{}.
\newblock \showarticletitle{Github repository: E2E benchmark}.
\newblock
  \bibinfo{journal}{\emph{https://github.com/Nathaniel-Han/End-to-End-CardEst-Benchmark}}
  (\bibinfo{year}{2021}).
\newblock


\bibitem[\protect\citeauthoryear{Han, Wu, Wu, Zhu, Yang, Tan, Zeng, Cong, Qin,
  Pfadler, et~al\mbox{.}}{Han et~al\mbox{.}}{2021}]%
        {han2021cardinality}
\bibfield{author}{\bibinfo{person}{Yuxing Han}, \bibinfo{person}{Ziniu Wu},
  \bibinfo{person}{Peizhi Wu}, \bibinfo{person}{Rong Zhu},
  \bibinfo{person}{Jingyi Yang}, \bibinfo{person}{Liang~Wei Tan},
  \bibinfo{person}{Kai Zeng}, \bibinfo{person}{Gao Cong},
  \bibinfo{person}{Yanzhao Qin}, \bibinfo{person}{Andreas Pfadler},
  {et~al\mbox{.}}} \bibinfo{year}{2021}\natexlab{}.
\newblock \showarticletitle{Cardinality Estimation in DBMS: A Comprehensive
  Benchmark Evaluation}.
\newblock \bibinfo{journal}{\emph{VLDB Endowment}} (\bibinfo{year}{2021}).
\newblock


\bibitem[\protect\citeauthoryear{Hasan, Thirumuruganathan, Augustine, Koudas,
  and Das}{Hasan et~al\mbox{.}}{2019}]%
        {hasan2019multi}
\bibfield{author}{\bibinfo{person}{Shohedul Hasan}, \bibinfo{person}{Saravanan
  Thirumuruganathan}, \bibinfo{person}{Jees Augustine}, \bibinfo{person}{Nick
  Koudas}, {and} \bibinfo{person}{Gautam Das}.}
  \bibinfo{year}{2019}\natexlab{}.
\newblock \showarticletitle{Multi-attribute selectivity estimation using deep
  learning}. In \bibinfo{booktitle}{\emph{SIGMOD}}.
\newblock


\bibitem[\protect\citeauthoryear{Hasan, Thirumuruganathan, Augustine, Koudas,
  and Das}{Hasan et~al\mbox{.}}{2020}]%
        {hasan2020}
\bibfield{author}{\bibinfo{person}{Shohedul Hasan}, \bibinfo{person}{Saravanan
  Thirumuruganathan}, \bibinfo{person}{Jees Augustine}, \bibinfo{person}{Nick
  Koudas}, {and} \bibinfo{person}{Gautam Das}.}
  \bibinfo{year}{2020}\natexlab{}.
\newblock \showarticletitle{Deep Learning Models for Selectivity Estimation of
  Multi-Attribute Queries}. In \bibinfo{booktitle}{\emph{Proceedings of the
  2020 ACM SIGMOD International Conference on Management of Data}}.
  \bibinfo{pages}{1035–1050}.
\newblock


\bibitem[\protect\citeauthoryear{Heimel, Kiefer, and Markl}{Heimel
  et~al\mbox{.}}{2015}]%
        {heimel2015self}
\bibfield{author}{\bibinfo{person}{Max Heimel}, \bibinfo{person}{Martin
  Kiefer}, {and} \bibinfo{person}{Volker Markl}.}
  \bibinfo{year}{2015}\natexlab{}.
\newblock \showarticletitle{Self-tuning, gpu-accelerated kernel density models
  for multidimensional selectivity estimation}. In
  \bibinfo{booktitle}{\emph{SIGMOD}}. \bibinfo{pages}{1477--1492}.
\newblock


\bibitem[\protect\citeauthoryear{Hertzschuch, Hartmann, Habich, and
  Lehner}{Hertzschuch et~al\mbox{.}}{2021}]%
        {hertzschuch2021simplicity}
\bibfield{author}{\bibinfo{person}{Axel Hertzschuch}, \bibinfo{person}{Claudio
  Hartmann}, \bibinfo{person}{Dirk Habich}, {and} \bibinfo{person}{Wolfgang
  Lehner}.} \bibinfo{year}{2021}\natexlab{}.
\newblock \showarticletitle{Simplicity Done Right for Join Ordering.}
\newblock \bibinfo{journal}{\emph{CIDR}} (\bibinfo{year}{2021}).
\newblock


\bibitem[\protect\citeauthoryear{Hilprecht}{Hilprecht}{2019}]%
        {hilp2019deepdb}
\bibfield{author}{\bibinfo{person}{Benjamin Hilprecht}.}
  \bibinfo{year}{2019}\natexlab{}.
\newblock \showarticletitle{Github repository: deepdb public}.
\newblock
  \bibinfo{journal}{\emph{https://github.com/DataManagementLab/deepdb-public}}
  (\bibinfo{year}{2019}).
\newblock


\bibitem[\protect\citeauthoryear{Hilprecht and Binnig}{Hilprecht and
  Binnig}{2022}]%
        {hilprecht2021one}
\bibfield{author}{\bibinfo{person}{Benjamin Hilprecht} {and}
  \bibinfo{person}{Carsten Binnig}.} \bibinfo{year}{2022}\natexlab{}.
\newblock \showarticletitle{One Model to Rule them All: Towards Zero-Shot
  Learning for Databases}.
\newblock \bibinfo{journal}{\emph{CIDR}} (\bibinfo{year}{2022}).
\newblock


\bibitem[\protect\citeauthoryear{Hilprecht, Schmidt, Kulessa, Molina, Kersting,
  and Binnig}{Hilprecht et~al\mbox{.}}{2019}]%
        {hilprecht2019deepdb}
\bibfield{author}{\bibinfo{person}{Benjamin Hilprecht},
  \bibinfo{person}{Andreas Schmidt}, \bibinfo{person}{Moritz Kulessa},
  \bibinfo{person}{Alejandro Molina}, \bibinfo{person}{Kristian Kersting},
  {and} \bibinfo{person}{Carsten Binnig}.} \bibinfo{year}{2019}\natexlab{}.
\newblock \showarticletitle{DeepDB: learn from data, not from queries!}. In
  \bibinfo{booktitle}{\emph{PVLDB}}.
\newblock


\bibitem[\protect\citeauthoryear{Ioannidis}{Ioannidis}{2003}]%
        {ioannidis2003history}
\bibfield{author}{\bibinfo{person}{Yannis Ioannidis}.}
  \bibinfo{year}{2003}\natexlab{}.
\newblock \showarticletitle{The history of histograms (abridged)}. In
  \bibinfo{booktitle}{\emph{Proceedings 2003 VLDB Conference}}. Elsevier,
  \bibinfo{pages}{19--30}.
\newblock


\bibitem[\protect\citeauthoryear{Ioannidis}{Ioannidis}{1993}]%
        {ioannidis1993universality}
\bibfield{author}{\bibinfo{person}{Yannis~E Ioannidis}.}
  \bibinfo{year}{1993}\natexlab{}.
\newblock \showarticletitle{Universality of serial histograms}. In
  \bibinfo{booktitle}{\emph{VLDB}}, Vol.~\bibinfo{volume}{93}. Citeseer,
  \bibinfo{pages}{256--267}.
\newblock


\bibitem[\protect\citeauthoryear{Ioannidis and Christodoulakis}{Ioannidis and
  Christodoulakis}{1991}]%
        {ioannidis1991propagation}
\bibfield{author}{\bibinfo{person}{Yannis~E Ioannidis} {and}
  \bibinfo{person}{Stavros Christodoulakis}.} \bibinfo{year}{1991}\natexlab{}.
\newblock \showarticletitle{On the propagation of errors in the size of join
  results}. In \bibinfo{booktitle}{\emph{Proceedings of the 1991 ACM SIGMOD
  International Conference on Management of data}}. \bibinfo{pages}{268--277}.
\newblock


\bibitem[\protect\citeauthoryear{Ioannidis and Christodoulakis}{Ioannidis and
  Christodoulakis}{1993}]%
        {ioannidis1993optimal}
\bibfield{author}{\bibinfo{person}{Yannis~E Ioannidis} {and}
  \bibinfo{person}{Stavros Christodoulakis}.} \bibinfo{year}{1993}\natexlab{}.
\newblock \showarticletitle{Optimal histograms for limiting worst-case error
  propagation in the size of join results}.
\newblock \bibinfo{journal}{\emph{ACM Transactions on Database Systems (TODS)}}
  \bibinfo{volume}{18}, \bibinfo{number}{4} (\bibinfo{year}{1993}),
  \bibinfo{pages}{709--748}.
\newblock


\bibitem[\protect\citeauthoryear{Ioannidis and Poosala}{Ioannidis and
  Poosala}{1995}]%
        {ioannidis1995balancing}
\bibfield{author}{\bibinfo{person}{Yannis~E Ioannidis} {and}
  \bibinfo{person}{Viswanath Poosala}.} \bibinfo{year}{1995}\natexlab{}.
\newblock \showarticletitle{Balancing histogram optimality and practicality for
  query result size estimation}.
\newblock \bibinfo{journal}{\emph{Acm Sigmod Record}} \bibinfo{volume}{24},
  \bibinfo{number}{2} (\bibinfo{year}{1995}), \bibinfo{pages}{233--244}.
\newblock


\bibitem[\protect\citeauthoryear{Khachatryan, M{\"u}ller, Stier, and
  B{\"o}hm}{Khachatryan et~al\mbox{.}}{2015}]%
        {khachatryan2015improving}
\bibfield{author}{\bibinfo{person}{Andranik Khachatryan},
  \bibinfo{person}{Emmanuel M{\"u}ller}, \bibinfo{person}{Christian Stier},
  {and} \bibinfo{person}{Klemens B{\"o}hm}.} \bibinfo{year}{2015}\natexlab{}.
\newblock \showarticletitle{{Improving accuracy and robustness of self-tuning
  histograms by subspace clustering}}.
\newblock \bibinfo{journal}{\emph{IEEE TKDE}} \bibinfo{volume}{27},
  \bibinfo{number}{9} (\bibinfo{year}{2015}), \bibinfo{pages}{2377--2389}.
\newblock


\bibitem[\protect\citeauthoryear{Kiefer, Heimel, Bre{\ss}, and Markl}{Kiefer
  et~al\mbox{.}}{2017}]%
        {kiefer2017estimating}
\bibfield{author}{\bibinfo{person}{Martin Kiefer}, \bibinfo{person}{Max
  Heimel}, \bibinfo{person}{Sebastian Bre{\ss}}, {and} \bibinfo{person}{Volker
  Markl}.} \bibinfo{year}{2017}\natexlab{}.
\newblock \showarticletitle{Estimating join selectivities using
  bandwidth-optimized kernel density models}.
\newblock \bibinfo{journal}{\emph{PVLDB}} \bibinfo{volume}{10},
  \bibinfo{number}{13} (\bibinfo{year}{2017}), \bibinfo{pages}{2085--2096}.
\newblock


\bibitem[\protect\citeauthoryear{Kipf, Kipf, Radke, Leis, Boncz, and
  Kemper}{Kipf et~al\mbox{.}}{2019}]%
        {kipf2018learned}
\bibfield{author}{\bibinfo{person}{Andreas Kipf}, \bibinfo{person}{Thomas
  Kipf}, \bibinfo{person}{Bernhard Radke}, \bibinfo{person}{Viktor Leis},
  \bibinfo{person}{Peter Boncz}, {and} \bibinfo{person}{Alfons Kemper}.}
  \bibinfo{year}{2019}\natexlab{}.
\newblock \showarticletitle{Learned cardinalities: Estimating correlated joins
  with deep learning}. In \bibinfo{booktitle}{\emph{CIDR}}.
\newblock


\bibitem[\protect\citeauthoryear{Koller and Friedman}{Koller and
  Friedman}{2009}]%
        {koller2009probabilistic}
\bibfield{author}{\bibinfo{person}{Daphne Koller} {and} \bibinfo{person}{Nir
  Friedman}.} \bibinfo{year}{2009}\natexlab{}.
\newblock \bibinfo{booktitle}{\emph{Probabilistic graphical models: principles
  and techniques}}.
\newblock \bibinfo{publisher}{MIT press}.
\newblock


\bibitem[\protect\citeauthoryear{Kschischang, Frey, and Loeliger}{Kschischang
  et~al\mbox{.}}{2001}]%
        {kschischang2001factor}
\bibfield{author}{\bibinfo{person}{Frank~R Kschischang},
  \bibinfo{person}{Brendan~J Frey}, {and} \bibinfo{person}{H-A Loeliger}.}
  \bibinfo{year}{2001}\natexlab{}.
\newblock \showarticletitle{Factor graphs and the sum-product algorithm}.
\newblock \bibinfo{journal}{\emph{IEEE Transactions on information theory}}
  \bibinfo{volume}{47}, \bibinfo{number}{2} (\bibinfo{year}{2001}),
  \bibinfo{pages}{498--519}.
\newblock


\bibitem[\protect\citeauthoryear{Lakshmi and Zhou}{Lakshmi and Zhou}{1998}]%
        {lakshmi1998selectivity}
\bibfield{author}{\bibinfo{person}{Seetha Lakshmi} {and}
  \bibinfo{person}{Shaoyu Zhou}.} \bibinfo{year}{1998}\natexlab{}.
\newblock \showarticletitle{Selectivity estimation in extensible databases-A
  neural network approach}. In \bibinfo{booktitle}{\emph{VLDB}},
  Vol.~\bibinfo{volume}{98}. \bibinfo{pages}{24--27}.
\newblock


\bibitem[\protect\citeauthoryear{Lauritzen}{Lauritzen}{1996}]%
        {lauritzen1996graphical}
\bibfield{author}{\bibinfo{person}{Steffen~L Lauritzen}.}
  \bibinfo{year}{1996}\natexlab{}.
\newblock \bibinfo{booktitle}{\emph{Graphical models}}.
  Vol.~\bibinfo{volume}{17}.
\newblock \bibinfo{publisher}{Clarendon Press}.
\newblock


\bibitem[\protect\citeauthoryear{Leis, Gubichev, Mirchev, Boncz, Kemper, and
  Neumann}{Leis et~al\mbox{.}}{2015}]%
        {leis2015good}
\bibfield{author}{\bibinfo{person}{Viktor Leis}, \bibinfo{person}{Andrey
  Gubichev}, \bibinfo{person}{Atanas Mirchev}, \bibinfo{person}{Peter Boncz},
  \bibinfo{person}{Alfons Kemper}, {and} \bibinfo{person}{Thomas Neumann}.}
  \bibinfo{year}{2015}\natexlab{}.
\newblock \showarticletitle{How good are query optimizers, really?}
\newblock \bibinfo{journal}{\emph{PVLDB}} \bibinfo{volume}{9},
  \bibinfo{number}{3} (\bibinfo{year}{2015}), \bibinfo{pages}{204--215}.
\newblock


\bibitem[\protect\citeauthoryear{Leis, Radke, Gubichev, Kemper, and
  Neumann}{Leis et~al\mbox{.}}{2017}]%
        {leis2017cardinality}
\bibfield{author}{\bibinfo{person}{Viktor Leis}, \bibinfo{person}{Bernhard
  Radke}, \bibinfo{person}{Andrey Gubichev}, \bibinfo{person}{Alfons Kemper},
  {and} \bibinfo{person}{Thomas Neumann}.} \bibinfo{year}{2017}\natexlab{}.
\newblock \showarticletitle{Cardinality Estimation Done Right: Index-Based Join
  Sampling}. In \bibinfo{booktitle}{\emph{CIDR}}.
\newblock


\bibitem[\protect\citeauthoryear{Li, Wu, Yi, and Zhao}{Li
  et~al\mbox{.}}{2016}]%
        {li2016wander}
\bibfield{author}{\bibinfo{person}{Feifei Li}, \bibinfo{person}{Bin Wu},
  \bibinfo{person}{Ke Yi}, {and} \bibinfo{person}{Zhuoyue Zhao}.}
  \bibinfo{year}{2016}\natexlab{}.
\newblock \showarticletitle{Wander join: Online aggregation via random walks}.
  In \bibinfo{booktitle}{\emph{SIGMOD}}. \bibinfo{pages}{615--629}.
\newblock


\bibitem[\protect\citeauthoryear{Lipton, Naughton, and Schneider}{Lipton
  et~al\mbox{.}}{1990}]%
        {lipton1990practical}
\bibfield{author}{\bibinfo{person}{Richard~J Lipton},
  \bibinfo{person}{Jeffrey~F Naughton}, {and} \bibinfo{person}{Donovan~A
  Schneider}.} \bibinfo{year}{1990}\natexlab{}.
\newblock \showarticletitle{Practical selectivity estimation through adaptive
  sampling}. In \bibinfo{booktitle}{\emph{Proceedings of the 1990 ACM SIGMOD
  international conference on Management of data}}. \bibinfo{pages}{1--11}.
\newblock


\bibitem[\protect\citeauthoryear{Liu, Dong, Zhou, and Li}{Liu
  et~al\mbox{.}}{2021a}]%
        {liu2021fauce}
\bibfield{author}{\bibinfo{person}{Jie Liu}, \bibinfo{person}{Wenqian Dong},
  \bibinfo{person}{Qingqing Zhou}, {and} \bibinfo{person}{Dong Li}.}
  \bibinfo{year}{2021}\natexlab{a}.
\newblock \showarticletitle{Fauce: fast and accurate deep ensembles with
  uncertainty for cardinality estimation}.
\newblock \bibinfo{journal}{\emph{Proceedings of the VLDB Endowment}}
  \bibinfo{volume}{14}, \bibinfo{number}{11} (\bibinfo{year}{2021}),
  \bibinfo{pages}{1950--1963}.
\newblock


\bibitem[\protect\citeauthoryear{Liu, Shen, and Chen}{Liu
  et~al\mbox{.}}{2021b}]%
        {liu2021lhist}
\bibfield{author}{\bibinfo{person}{Qiyu Liu}, \bibinfo{person}{Yanyan Shen},
  {and} \bibinfo{person}{Lei Chen}.} \bibinfo{year}{2021}\natexlab{b}.
\newblock \showarticletitle{LHist: Towards Learning Multi-dimensional Histogram
  for Massive Spatial Data}. In \bibinfo{booktitle}{\emph{2021 IEEE 37th
  International Conference on Data Engineering (ICDE)}}. IEEE,
  \bibinfo{pages}{1188--1199}.
\newblock


\bibitem[\protect\citeauthoryear{Loeliger}{Loeliger}{2004}]%
        {loeliger2004introduction}
\bibfield{author}{\bibinfo{person}{H-A Loeliger}.}
  \bibinfo{year}{2004}\natexlab{}.
\newblock \showarticletitle{An introduction to factor graphs}.
\newblock \bibinfo{journal}{\emph{IEEE Signal Processing Magazine}}
  \bibinfo{volume}{21}, \bibinfo{number}{1} (\bibinfo{year}{2004}),
  \bibinfo{pages}{28--41}.
\newblock


\bibitem[\protect\citeauthoryear{Ma, Ding, Das, and Swaminathan}{Ma
  et~al\mbox{.}}{2020}]%
        {ma2020active}
\bibfield{author}{\bibinfo{person}{Lin Ma}, \bibinfo{person}{Bailu Ding},
  \bibinfo{person}{Sudipto Das}, {and} \bibinfo{person}{Adith Swaminathan}.}
  \bibinfo{year}{2020}\natexlab{}.
\newblock \showarticletitle{Active learning for ML enhanced database systems}.
  In \bibinfo{booktitle}{\emph{Proceedings of the 2020 ACM SIGMOD International
  Conference on Management of Data}}. \bibinfo{pages}{175--191}.
\newblock


\bibitem[\protect\citeauthoryear{MacKay, Mac~Kay, et~al\mbox{.}}{MacKay
  et~al\mbox{.}}{2003}]%
        {mackay2003information}
\bibfield{author}{\bibinfo{person}{David~JC MacKay}, \bibinfo{person}{David~JC
  Mac~Kay}, {et~al\mbox{.}}} \bibinfo{year}{2003}\natexlab{}.
\newblock \bibinfo{booktitle}{\emph{Information theory, inference and learning
  algorithms}}.
\newblock \bibinfo{publisher}{Cambridge university press}.
\newblock


\bibitem[\protect\citeauthoryear{Malik, Burns, and Chawla}{Malik
  et~al\mbox{.}}{2007}]%
        {malik2007black}
\bibfield{author}{\bibinfo{person}{Tanu Malik}, \bibinfo{person}{Randal~C
  Burns}, {and} \bibinfo{person}{Nitesh~V Chawla}.}
  \bibinfo{year}{2007}\natexlab{}.
\newblock \showarticletitle{A Black-Box Approach to Query Cardinality
  Estimation.}. In \bibinfo{booktitle}{\emph{CIDR}}. Citeseer,
  \bibinfo{pages}{56--67}.
\newblock


\bibitem[\protect\citeauthoryear{Marcus, Negi, Mao, Tatbul, Alizadeh, and
  Kraska}{Marcus et~al\mbox{.}}{2021}]%
        {marcus2020bao}
\bibfield{author}{\bibinfo{person}{Ryan Marcus}, \bibinfo{person}{Parimarjan
  Negi}, \bibinfo{person}{Hongzi Mao}, \bibinfo{person}{Nesime Tatbul},
  \bibinfo{person}{Mohammad Alizadeh}, {and} \bibinfo{person}{Tim Kraska}.}
  \bibinfo{year}{2021}\natexlab{}.
\newblock \showarticletitle{Bao: Learning to steer query optimizers}.
\newblock \bibinfo{journal}{\emph{SIGMOD}} (\bibinfo{year}{2021}).
\newblock


\bibitem[\protect\citeauthoryear{Marcus, Negi, Mao, Zhang, Alizadeh, Kraska,
  Papaemmanouil, and Tatbul}{Marcus et~al\mbox{.}}{2019}]%
        {marcus2019neo}
\bibfield{author}{\bibinfo{person}{Ryan Marcus}, \bibinfo{person}{Parimarjan
  Negi}, \bibinfo{person}{Hongzi Mao}, \bibinfo{person}{Chi Zhang},
  \bibinfo{person}{Mohammad Alizadeh}, \bibinfo{person}{Tim Kraska},
  \bibinfo{person}{Olga Papaemmanouil}, {and} \bibinfo{person}{Nesime Tatbul}.}
  \bibinfo{year}{2019}\natexlab{}.
\newblock \showarticletitle{Neo: A learned query optimizer}.
\newblock \bibinfo{journal}{\emph{arXiv preprint arXiv:1904.03711}}
  (\bibinfo{year}{2019}).
\newblock


\bibitem[\protect\citeauthoryear{Marcus and Papaemmanouil}{Marcus and
  Papaemmanouil}{2018}]%
        {marcus2018deep}
\bibfield{author}{\bibinfo{person}{Ryan Marcus} {and} \bibinfo{person}{Olga
  Papaemmanouil}.} \bibinfo{year}{2018}\natexlab{}.
\newblock \showarticletitle{Deep reinforcement learning for join order
  enumeration}. In \bibinfo{booktitle}{\emph{Proceedings of the First
  International Workshop on Exploiting Artificial Intelligence Techniques for
  Data Management}}. \bibinfo{pages}{1--4}.
\newblock


\bibitem[\protect\citeauthoryear{Marcus and Papaemmanouil}{Marcus and
  Papaemmanouil}{2019}]%
        {marcus2019plan}
\bibfield{author}{\bibinfo{person}{Ryan Marcus} {and} \bibinfo{person}{Olga
  Papaemmanouil}.} \bibinfo{year}{2019}\natexlab{}.
\newblock \showarticletitle{Plan-structured deep neural network models for
  query performance prediction}.
\newblock \bibinfo{journal}{\emph{arXiv preprint arXiv:1902.00132}}
  (\bibinfo{year}{2019}).
\newblock


\bibitem[\protect\citeauthoryear{Muralikrishna and DeWitt}{Muralikrishna and
  DeWitt}{1988}]%
        {muralikrishna1988equi}
\bibfield{author}{\bibinfo{person}{M Muralikrishna} {and}
  \bibinfo{person}{David~J DeWitt}.} \bibinfo{year}{1988}\natexlab{}.
\newblock \showarticletitle{Equi-depth multidimensional histograms}. In
  \bibinfo{booktitle}{\emph{Proceedings of the 1988 ACM SIGMOD international
  conference on Management of data}}. \bibinfo{pages}{28--36}.
\newblock


\bibitem[\protect\citeauthoryear{Negi, Interlandi, Marcus, Alizadeh, Kraska,
  Friedman, and Jindal}{Negi et~al\mbox{.}}{2021a}]%
        {negi2021steering}
\bibfield{author}{\bibinfo{person}{Parimarjan Negi}, \bibinfo{person}{Matteo
  Interlandi}, \bibinfo{person}{Ryan Marcus}, \bibinfo{person}{Mohammad
  Alizadeh}, \bibinfo{person}{Tim Kraska}, \bibinfo{person}{Marc Friedman},
  {and} \bibinfo{person}{Alekh Jindal}.} \bibinfo{year}{2021}\natexlab{a}.
\newblock \showarticletitle{Steering query optimizers: A practical take on big
  data workloads}. In \bibinfo{booktitle}{\emph{Proceedings of the 2021
  International Conference on Management of Data}}.
  \bibinfo{pages}{2557--2569}.
\newblock


\bibitem[\protect\citeauthoryear{Negi, Marcus, Kipf, Mao, Tatbul, Kraska, and
  Alizadeh}{Negi et~al\mbox{.}}{2021b}]%
        {negi2021flow}
\bibfield{author}{\bibinfo{person}{Parimarjan Negi}, \bibinfo{person}{Ryan
  Marcus}, \bibinfo{person}{Andreas Kipf}, \bibinfo{person}{Hongzi Mao},
  \bibinfo{person}{Nesime Tatbul}, \bibinfo{person}{Tim Kraska}, {and}
  \bibinfo{person}{Mohammad Alizadeh}.} \bibinfo{year}{2021}\natexlab{b}.
\newblock \showarticletitle{Flow-Loss: Learning Cardinality Estimates That
  Matter}.
\newblock \bibinfo{journal}{\emph{arXiv preprint arXiv:2101.04964}}
  (\bibinfo{year}{2021}).
\newblock


\bibitem[\protect\citeauthoryear{Ngo, Porat, R{\'e}, and Rudra}{Ngo
  et~al\mbox{.}}{2018}]%
        {ngo2018worst}
\bibfield{author}{\bibinfo{person}{Hung~Q Ngo}, \bibinfo{person}{Ely Porat},
  \bibinfo{person}{Christopher R{\'e}}, {and} \bibinfo{person}{Atri Rudra}.}
  \bibinfo{year}{2018}\natexlab{}.
\newblock \showarticletitle{Worst-case optimal join algorithms}.
\newblock \bibinfo{journal}{\emph{Journal of the ACM (JACM)}}
  \bibinfo{volume}{65}, \bibinfo{number}{3} (\bibinfo{year}{2018}),
  \bibinfo{pages}{1--40}.
\newblock


\bibitem[\protect\citeauthoryear{Park, Ko, Bhowmick, Kim, Hong, and Han}{Park
  et~al\mbox{.}}{2020}]%
        {park2020g}
\bibfield{author}{\bibinfo{person}{Yeonsu Park}, \bibinfo{person}{Seongyun Ko},
  \bibinfo{person}{Sourav~S Bhowmick}, \bibinfo{person}{Kyoungmin Kim},
  \bibinfo{person}{Kijae Hong}, {and} \bibinfo{person}{Wook-Shin Han}.}
  \bibinfo{year}{2020}\natexlab{}.
\newblock \showarticletitle{G-CARE: A Framework for Performance Benchmarking of
  Cardinality Estimation Techniques for Subgraph Matching}. In
  \bibinfo{booktitle}{\emph{Proceedings of the 2020 ACM SIGMOD International
  Conference on Management of Data}}. \bibinfo{pages}{1099--1114}.
\newblock


\bibitem[\protect\citeauthoryear{Poon and Domingos}{Poon and Domingos}{2011}]%
        {poon2011sum}
\bibfield{author}{\bibinfo{person}{Hoifung Poon} {and} \bibinfo{person}{Pedro
  Domingos}.} \bibinfo{year}{2011}\natexlab{}.
\newblock \showarticletitle{Sum-product networks: A new deep architecture}. In
  \bibinfo{booktitle}{\emph{ICCV Workshops}}. \bibinfo{pages}{689--690}.
\newblock


\bibitem[\protect\citeauthoryear{Poosala and Ioannidis}{Poosala and
  Ioannidis}{1997}]%
        {poosala1997selectivity}
\bibfield{author}{\bibinfo{person}{Viswanath Poosala} {and}
  \bibinfo{person}{Yannis~E Ioannidis}.} \bibinfo{year}{1997}\natexlab{}.
\newblock \showarticletitle{Selectivity estimation without the attribute value
  independence assumption}. In \bibinfo{booktitle}{\emph{VLDB}},
  Vol.~\bibinfo{volume}{97}. \bibinfo{pages}{486--495}.
\newblock


\bibitem[\protect\citeauthoryear{Reference~Manual}{Reference~Manual}{2020}]%
        {mysql2020}
\bibfield{author}{\bibinfo{person}{MySQL~8.0 Reference~Manual}.}
  \bibinfo{year}{2020}\natexlab{}.
\newblock \showarticletitle{Chapter 15.8.10.2 Configuring Non-Persistent
  Optimizer Statistics Parameters}.
\newblock
  \bibinfo{journal}{\emph{https://dev.mysql.com/doc/refman/8.0/en/innodb-statistics-estimation.html}}
  (\bibinfo{year}{2020}).
\newblock


\bibitem[\protect\citeauthoryear{Selinger, Astrahan, Chamberlin, Lorie, and
  Price}{Selinger et~al\mbox{.}}{1979}]%
        {selinger1979access}
\bibfield{author}{\bibinfo{person}{P~Griffiths Selinger},
  \bibinfo{person}{Morton~M Astrahan}, \bibinfo{person}{Donald~D Chamberlin},
  \bibinfo{person}{Raymond~A Lorie}, {and} \bibinfo{person}{Thomas~G Price}.}
  \bibinfo{year}{1979}\natexlab{}.
\newblock \showarticletitle{Access path selection in a relational database
  management system}. In \bibinfo{booktitle}{\emph{SIGMOD}}.
  \bibinfo{pages}{23--34}.
\newblock


\bibitem[\protect\citeauthoryear{Srivastava, Haas, Markl, Kutsch, and
  Tran}{Srivastava et~al\mbox{.}}{2006}]%
        {srivastava2006isomer}
\bibfield{author}{\bibinfo{person}{Utkarsh Srivastava},
  \bibinfo{person}{Peter~J Haas}, \bibinfo{person}{Volker Markl},
  \bibinfo{person}{Marcel Kutsch}, {and} \bibinfo{person}{Tam~Minh Tran}.}
  \bibinfo{year}{2006}\natexlab{}.
\newblock \showarticletitle{Isomer: Consistent histogram construction using
  query feedback}. In \bibinfo{booktitle}{\emph{ICDE}}.
  \bibinfo{pages}{39--39}.
\newblock


\bibitem[\protect\citeauthoryear{Stillger, Lohman, Markl, and Kandil}{Stillger
  et~al\mbox{.}}{2001}]%
        {stillger2001leo}
\bibfield{author}{\bibinfo{person}{Michael Stillger}, \bibinfo{person}{Guy~M
  Lohman}, \bibinfo{person}{Volker Markl}, {and} \bibinfo{person}{Mokhtar
  Kandil}.} \bibinfo{year}{2001}\natexlab{}.
\newblock \showarticletitle{LEO-DB2's learning optimizer}. In
  \bibinfo{booktitle}{\emph{PVLDB}}, Vol.~\bibinfo{volume}{1}.
  \bibinfo{pages}{19--28}.
\newblock


\bibitem[\protect\citeauthoryear{Sun and Li}{Sun and Li}{2019}]%
        {sun2019end}
\bibfield{author}{\bibinfo{person}{Ji Sun} {and} \bibinfo{person}{Guoliang
  Li}.} \bibinfo{year}{2019}\natexlab{}.
\newblock \showarticletitle{An end-to-end learning-based cost estimator}.
\newblock \bibinfo{journal}{\emph{VLDB}} (\bibinfo{year}{2019}).
\newblock


\bibitem[\protect\citeauthoryear{Tzoumas, Deshpande, and Jensen}{Tzoumas
  et~al\mbox{.}}{2011}]%
        {tzoumas2011lightweight}
\bibfield{author}{\bibinfo{person}{Kostas Tzoumas}, \bibinfo{person}{Amol
  Deshpande}, {and} \bibinfo{person}{Christian~S Jensen}.}
  \bibinfo{year}{2011}\natexlab{}.
\newblock \showarticletitle{Lightweight graphical models for selectivity
  estimation without independence assumptions}.
\newblock \bibinfo{journal}{\emph{PVLDB}} \bibinfo{volume}{4},
  \bibinfo{number}{11} (\bibinfo{year}{2011}), \bibinfo{pages}{852--863}.
\newblock


\bibitem[\protect\citeauthoryear{Tzoumas, Deshpande, and Jensen}{Tzoumas
  et~al\mbox{.}}{2013}]%
        {tzoumas2013vldb}
\bibfield{author}{\bibinfo{person}{Kostas Tzoumas}, \bibinfo{person}{Amol
  Deshpande}, {and} \bibinfo{person}{Christian~S Jensen}.}
  \bibinfo{year}{2013}\natexlab{}.
\newblock \showarticletitle{Efficiently adapting graphical models for
  selectivity estimation}.
\newblock \bibinfo{journal}{\emph{Proceedings of the VLDB Endowment}}
  \bibinfo{volume}{1}, \bibinfo{number}{22} (\bibinfo{year}{2013}).
\newblock


\bibitem[\protect\citeauthoryear{Wang and Sevcik}{Wang and Sevcik}{2003}]%
        {wang2003multi}
\bibfield{author}{\bibinfo{person}{Hai Wang} {and} \bibinfo{person}{Kenneth~C
  Sevcik}.} \bibinfo{year}{2003}\natexlab{}.
\newblock \showarticletitle{{A multi-dimensional histogram for selectivity
  estimation and fast approximate query answering}}. In
  \bibinfo{booktitle}{\emph{Proceedings of the 2003 conference of the Centre
  for Advanced Studies on Collaborative research}}. \bibinfo{pages}{328--342}.
\newblock


\bibitem[\protect\citeauthoryear{Wang, Chai, Liu, and Li}{Wang
  et~al\mbox{.}}{2021}]%
        {wang2021face}
\bibfield{author}{\bibinfo{person}{Jiayi Wang}, \bibinfo{person}{Chengliang
  Chai}, \bibinfo{person}{Jiabin Liu}, {and} \bibinfo{person}{Guoliang Li}.}
  \bibinfo{year}{2021}\natexlab{}.
\newblock \showarticletitle{FACE: a normalizing flow based cardinality
  estimator}.
\newblock \bibinfo{journal}{\emph{Proceedings of the VLDB Endowment}}
  \bibinfo{volume}{15}, \bibinfo{number}{1} (\bibinfo{year}{2021}),
  \bibinfo{pages}{72--84}.
\newblock


\bibitem[\protect\citeauthoryear{Wu, Jindal, Amizadeh, Patel, Le, Qiao, and
  Rao}{Wu et~al\mbox{.}}{2018}]%
        {wu2018towards}
\bibfield{author}{\bibinfo{person}{Chenggang Wu}, \bibinfo{person}{Alekh
  Jindal}, \bibinfo{person}{Saeed Amizadeh}, \bibinfo{person}{Hiren Patel},
  \bibinfo{person}{Wangchao Le}, \bibinfo{person}{Shi Qiao}, {and}
  \bibinfo{person}{Sriram Rao}.} \bibinfo{year}{2018}\natexlab{}.
\newblock \showarticletitle{Towards a learning optimizer for shared clouds}.
\newblock \bibinfo{journal}{\emph{PVLDB}} \bibinfo{volume}{12},
  \bibinfo{number}{3} (\bibinfo{year}{2018}), \bibinfo{pages}{210--222}.
\newblock


\bibitem[\protect\citeauthoryear{Wu and Cong}{Wu and Cong}{2021}]%
        {wu2021uae}
\bibfield{author}{\bibinfo{person}{Peizhi Wu} {and} \bibinfo{person}{Gao
  Cong}.} \bibinfo{year}{2021}\natexlab{}.
\newblock \showarticletitle{A Unified Deep Model of Learning from both Data and
  Queries for Cardinality Estimation}. In \bibinfo{booktitle}{\emph{Proceedings
  of the 2021 ACM SIGMOD International Conference on Management of Data}}.
\newblock


\bibitem[\protect\citeauthoryear{Wu, Shaikhha, Zhu, Zeng, Han, and Zhou}{Wu
  et~al\mbox{.}}{2020a}]%
        {wu2020bayescard}
\bibfield{author}{\bibinfo{person}{Ziniu Wu}, \bibinfo{person}{Amir Shaikhha},
  \bibinfo{person}{Rong Zhu}, \bibinfo{person}{Kai Zeng},
  \bibinfo{person}{Yuxing Han}, {and} \bibinfo{person}{Jingren Zhou}.}
  \bibinfo{year}{2020}\natexlab{a}.
\newblock \showarticletitle{BayesCard: Revitilizing Bayesian Frameworks for
  Cardinality Estimation}.
\newblock \bibinfo{journal}{\emph{arXiv e-prints}} (\bibinfo{year}{2020}),
  \bibinfo{pages}{arXiv--2012}.
\newblock


\bibitem[\protect\citeauthoryear{Wu, Yang, Yu, Zhu, Han, Li, Lian, Zeng, and
  Zhou}{Wu et~al\mbox{.}}{2022}]%
        {wu2021unified}
\bibfield{author}{\bibinfo{person}{Ziniu Wu}, \bibinfo{person}{Peilun Yang},
  \bibinfo{person}{Pei Yu}, \bibinfo{person}{Rong Zhu}, \bibinfo{person}{Yuxing
  Han}, \bibinfo{person}{Yaliang Li}, \bibinfo{person}{Defu Lian},
  \bibinfo{person}{Kai Zeng}, {and} \bibinfo{person}{Jingren Zhou}.}
  \bibinfo{year}{2022}\natexlab{}.
\newblock \showarticletitle{A Unified Transferable Model for ML-Enhanced DBMS}.
\newblock \bibinfo{journal}{\emph{CIDR}} (\bibinfo{year}{2022}).
\newblock


\bibitem[\protect\citeauthoryear{Wu, Zhu, Pfadler, Han, Li, Qian, Zeng, and
  Zhou}{Wu et~al\mbox{.}}{2020b}]%
        {wu2020fspn}
\bibfield{author}{\bibinfo{person}{Ziniu Wu}, \bibinfo{person}{Rong Zhu},
  \bibinfo{person}{Andreas Pfadler}, \bibinfo{person}{Yuxing Han},
  \bibinfo{person}{Jiangneng Li}, \bibinfo{person}{Zhengping Qian},
  \bibinfo{person}{Kai Zeng}, {and} \bibinfo{person}{Jingren Zhou}.}
  \bibinfo{year}{2020}\natexlab{b}.
\newblock \showarticletitle{FSPN: A New Class of Probabilistic Graphical
  Model}.
\newblock \bibinfo{journal}{\emph{arXiv preprint arXiv:2011.09020}}
  (\bibinfo{year}{2020}).
\newblock


\bibitem[\protect\citeauthoryear{Yang, Kamsetty, Luan, Liang, Duan, Chen, and
  Stoica}{Yang et~al\mbox{.}}{2021}]%
        {yang2020neurocard}
\bibfield{author}{\bibinfo{person}{Zongheng Yang}, \bibinfo{person}{Amog
  Kamsetty}, \bibinfo{person}{Sifei Luan}, \bibinfo{person}{Eric Liang},
  \bibinfo{person}{Yan Duan}, \bibinfo{person}{Xi Chen}, {and}
  \bibinfo{person}{Ion Stoica}.} \bibinfo{year}{2021}\natexlab{}.
\newblock \showarticletitle{NeuroCard: One Cardinality Estimator for All
  Tables}.
\newblock \bibinfo{journal}{\emph{PVLDB}} \bibinfo{volume}{14},
  \bibinfo{number}{1} (\bibinfo{year}{2021}), \bibinfo{pages}{61--73}.
\newblock


\bibitem[\protect\citeauthoryear{Yang, Liang, Kamsetty, Wu, Duan, Chen, Abbeel,
  Hellerstein, Krishnan, and Stoica}{Yang et~al\mbox{.}}{2019}]%
        {yang2019deep}
\bibfield{author}{\bibinfo{person}{Zongheng Yang}, \bibinfo{person}{Eric
  Liang}, \bibinfo{person}{Amog Kamsetty}, \bibinfo{person}{Chenggang Wu},
  \bibinfo{person}{Yan Duan}, \bibinfo{person}{Xi Chen},
  \bibinfo{person}{Pieter Abbeel}, \bibinfo{person}{Joseph~M Hellerstein},
  \bibinfo{person}{Sanjay Krishnan}, {and} \bibinfo{person}{Ion Stoica}.}
  \bibinfo{year}{2019}\natexlab{}.
\newblock \showarticletitle{Deep unsupervised cardinality estimation}.
\newblock \bibinfo{journal}{\emph{PVLDB}} (\bibinfo{year}{2019}).
\newblock


\bibitem[\protect\citeauthoryear{Yu, Li, Chai, and Tang}{Yu
  et~al\mbox{.}}{2020}]%
        {yu2020reinforcement}
\bibfield{author}{\bibinfo{person}{Xiang Yu}, \bibinfo{person}{Guoliang Li},
  \bibinfo{person}{Chengliang Chai}, {and} \bibinfo{person}{Nan Tang}.}
  \bibinfo{year}{2020}\natexlab{}.
\newblock \showarticletitle{Reinforcement learning with tree-lstm for join
  order selection}. In \bibinfo{booktitle}{\emph{2020 IEEE 36th International
  Conference on Data Engineering (ICDE)}}. IEEE, \bibinfo{pages}{1297--1308}.
\newblock


\bibitem[\protect\citeauthoryear{Zhao, Christensen, Li, Hu, and Yi}{Zhao
  et~al\mbox{.}}{2018}]%
        {zhao2018random}
\bibfield{author}{\bibinfo{person}{Zhuoyue Zhao}, \bibinfo{person}{Robert
  Christensen}, \bibinfo{person}{Feifei Li}, \bibinfo{person}{Xiao Hu}, {and}
  \bibinfo{person}{Ke Yi}.} \bibinfo{year}{2018}\natexlab{}.
\newblock \showarticletitle{Random sampling over joins revisited}. In
  \bibinfo{booktitle}{\emph{SIGMOD}}. \bibinfo{pages}{1525--1539}.
\newblock


\bibitem[\protect\citeauthoryear{Zhu, Wu, Han, Zeng, Pfadler, Qian, Zhou, and
  Cui}{Zhu et~al\mbox{.}}{2021}]%
        {zhu2020flat}
\bibfield{author}{\bibinfo{person}{Rong Zhu}, \bibinfo{person}{Ziniu Wu},
  \bibinfo{person}{Yuxing Han}, \bibinfo{person}{Kai Zeng},
  \bibinfo{person}{Andreas Pfadler}, \bibinfo{person}{Zhengping Qian},
  \bibinfo{person}{Jingren Zhou}, {and} \bibinfo{person}{Bin Cui}.}
  \bibinfo{year}{2021}\natexlab{}.
\newblock \showarticletitle{FLAT: Fast, Lightweight and Accurate Method for
  Cardinality Estimation}.
\newblock \bibinfo{journal}{\emph{VLDB}} \bibinfo{volume}{14},
  \bibinfo{number}{9} (\bibinfo{year}{2021}), \bibinfo{pages}{1489--1502}.
\newblock


\end{thebibliography}
	
\clearpage
\appendix
\section{Equations and proofs}

\subsection{Equations for decomposing joins into single-tables}

In this section, we provide the case-by-case equations on how to accurately formulate join estimation problem into single-table \CE problem without any simplified assumptions. 

\smallskip
\noindent \underline{\textbf{Two-table join case:}}

Assume that we have two tables $A$ and $B$ with join keys $A.id$ and $B.Aid$ and a query $Q = Q(A) \fullouterjoin Q(B)$ where $Q(A)$ is the filter on table $A$ and same for $Q(B)$. The join condition is $A.id = B.Aid$. Then, the join cardinality of $Q$ can be expressed as follows where $D(A.id)$ represents the domain of $A$.

\begin{align}
	|Q| = \sum_{v \in D(A.id)} &P_{A}(A.id = v | Q(A)) * |Q(A)| * \nonumber \\   
	 & P_{B}(B.Aid = v | Q(B)) * |Q(B)|
\end{align}

\smallskip
\noindent \underline{\textbf{Chain join cases:}}

\noindent \emph{\textbf{Case1 A.id = B.Aid = C.Aid:}}
Assume we have three tables $A$, $B$ and $C$ with join keys $A.id$, $B.Aid$, and $C.Aid$. There exists query $Q = Q(A) \fullouterjoin Q(B) \fullouterjoin Q(C)$ on join condition $A.id = B.Aid \ \ \text{AND}  \ \ B.Aid = C.Aid$.
 Then, the join cardinality can be expressed as follows:

\begin{align}
	|Q| &= \sum_{v \in D(A.id)} P_{A}(A.id = v | Q(A)) * |Q(A)| *  \nonumber \\   
	& P_{B}(B.Aid = v | Q(B)) * |Q(B)| * P_{C}(C.Aid = v | Q(C)) * |Q(C)|
\end{align}

\noindent \emph{\textbf{Case2 A.id = B.Aid, B.id = C.Bid:}}
Assume we have three tables $A$, $B$ and $C$ with join keys $A.id$, $B.Aid$, $B.id$, and $C.Bid$. There exists query $Q = Q(A) \fullouterjoin Q(B) \fullouterjoin Q(C)$ on join condition $A.id = B.Aid \ \ \text{AND} \ \  B.id = C.Bid$.
Then, the join cardinality can be expressed as follows:
\begin{align}
	|Q| = &\sum_{v_1 \in D(A.id)} \sum_{v_2 \in D(B.id)} P_{A}(A.id = v_1 | Q(A)) * |Q(A)| * \nonumber \\ 
	&P_{B}(B.Aid = v_1, B.id = v_2 | Q(B)) * |Q(B)| * \nonumber \\ 
	& P_{C}(C.Bid = v_2 | Q(C)) * |Q(C)|
\end{align}

\smallskip
\noindent \underline{\textbf{Self join case:}}

Assume we have one table $A$ with join keys $A.id, A.id_2$. There exists a query $Q = Q(A) \fullouterjoin Q(A')$ on join condition $A.id = A.id_2$. Then, the join cardinality can be expressed as follows:

\begin{align}
	|Q| = \sum_{v \in D(A.id)} &P_{A}(A.id = v | Q(A)) * |Q(A)| * \nonumber \\   
	& P_{A}(A.id_2 = v | Q(A')) * |Q(A')|
\end{align}

\smallskip
\noindent \underline{\textbf{Cyclic join case:}}

Assume we have two tables $A$ and $B$ with join keys $A.id, A.id_2$ and $B.Aid, B.Aid_2$. There exists a query $Q = Q(A) \fullouterjoin Q(B)$ on join condition $A.id = B.Aid \ \ \text{AND} \ \  A.id_2 = B.Aid_2$. Then, the join cardinality can be expressed as follows:

\begin{align}
	|Q| = & \sum_{v_1 \in D(A.id)} \sum_{v_2 \in D(A.id_2)} P_{A}(A.id = v_1| Q(A)) * |Q(A)| * \nonumber \\ 
	&P_{B}(B.Aid = v_1, B.Aid_2 = v_2 | Q(B)) * |Q(B)| *  \nonumber \\ 
	& P_{A}(A.id_2 = v_2| Q(A')) * |Q(A')| \nonumber \\ 
	= & \sum_{v_1 \in D(A.id)} \sum_{v_2 \in D(A.id_2)} P_{B}(B.Aid = v_1| Q(B)) * |Q(B)| \nonumber \\ 
	&P_{A}(A.id = v_1, A.id_2 = v_2| Q(A)) * |Q(A)| * \nonumber \\ 
	&P_{B}(B.Aid_2 = v_2 | Q(B')) * |Q(B')| 
\end{align}

\subsection{Proof of lamma 1}

\begin{lemma}
	\label{lemma: PGM-appendix}
	Given a join graph $\mathcal{G}$ representing a query $Q$, there exists a factor graph $\mathcal{F}$ such that the variable nodes in $\mathcal{F}$ are the equivalent key group variables of $\mathcal{G}$ and each factor node represents a table $I$ touched by $Q$. A factor node is connected to a variable node if and only if this variable represents a join key in table $I$. The potential function of a factor node is defined as table $I$'s probability distribution of the connected variables (join keys) conditioned on the filter predicates $Q(I)$. Then, calculating the cardinality of $Q$ is equivalent to computing the partition function of $\mathcal{F}$.
\end{lemma}

\smallskip
\noindent \underline{\textbf{Proof:}}
Assume that we have a query $Q$ joining $m$ tables $A, B, \ldots, M$.
We represent $Q$ as a join graph $G$, where each node represents a join key and each edge represents a join relation between two keys connected by it. We further define hyper-nodes in $G$ as a set of nodes whose corresponding join key lies in the same table. Thus, each hyper-nodes naturally represents a table in $Q$. A visualization of such $G$ can be found in Figure~\ref{fig: formulation}.
Each connected components of $G$ connects a group of join keys with equal join relation, suggesting that they share the same semantics. Therefore, we consider all join keys in a connected component as a equivalent key group variable, denoted as $V_i$.
Therefore, $G$ defines $n$ equivalent variables $V_1, \ldots, V_n$.

Then, assume that we have a set of unnormalized single table distributions $P_A(V_A|Q(A)) * |Q(A)|, \ldots, P_M(V_M|Q(M)) * |Q(M)|$ and $P_A(V_{A'}|Q(A')) * |Q(A')|, \ldots, P_M(V_{M'}|Q(M')) * |Q(M')|$.
Each $V_I$ presents a set of equivalent variables that represent a join key in table $I$. The $I'$ is the duplicated table introduced by query $Q$ only if there exists a cyclic join situation. 
Therefore, the cardinality of $Q$ can be writen as:

\begin{align}
	\label{equ: lemma1}
	|Q| = & \sum_{v_1 \in D(V_1)} \sum_{v_2 \in D(V_2)} \cdots \sum_{v_n \in D(V_n)}  (P_{A}(V_A| Q(A)) * |Q(A)|) * \nonumber \\ 
	&(P_{B}(V_B | Q(B)) * |Q(B)|) * \cdots (P_{M}(V_M | Q(M)) * |Q(M)|) *  \nonumber \\ 
	&  (P_{A}(V_{A'}| Q(A')) * |Q(A')|) * \cdots (P_{M'}(V_{M'} | Q(M')) * |Q(M')|)
\end{align}

Next, let's construct a factor graph $\mathcal{F}$, such that the variable nodes in $\mathcal{F}$ are the equivalent key group variables $V_i$ of $\mathcal{G}$ and the factor nodes in $\mathcal{F}$ represent the tables $A, \ldots, M$ touched by $Q$ and the duplicated tables if exist $A', \ldots, M'$. A factor node representing table $I$ is connected to a variable node if and only if this variable represents a join key in table $I$. 
The potential function of a factor node is defined as the unnormalized probability distribution $P_I(V_I|Q(I)) * |Q(I)|$.
The partition function of $\mathcal{F}$ is defined exactly the same as Equation~\ref{equ: lemma1}.
Therefore, instead of summing a nested loop of $n$ variables $V_i$ by brute-force, the partition function (cardinality) can be computed using the well-established variable elimination and belief propogation techniques in PGM domain.

\subsection{Probabilistic bound}

\Ours approximate the exact cardinality computation with a probabilistic bound based inference algorithm. An example on two table join queries can found below. In this section we discuss how to derive an upper bound ($Probabilistic\_bound(A, B, bin_i)$ in this equation) for different case of join.

\begin{alignat}{2}
	|Q| &= \sum_{i = 1}^{k}  \sum_{v \in bin_i} && P_{A}(A.Id = v | Q(A)) * |Q(A)| * \nonumber \\   
	& && P_{B}(B.Aid = v | Q(B)) * |Q(B)| \nonumber \\
	&\lesssim \sum_{i = 1}^{k} && Probabilistic\_bound(A, B, bin_i)
\end{alignat}

\smallskip
\noindent \underline{\textbf{Case1: Joining two tables$A.id = B.Aid$:}}

Let $V_i^{*}(A.id) = MAX_{v \in bin_i} |A.id = v|$ be the most frequent value (MFV) count of $A.id$ in a bin $bin_i$, and same for $V_i^{*}(B.Aid)$. We have the following bound:

\begin{alignat}{2}
	|Q| & = \sum_{v = D(A.id)} && P_{A}(A.Id = v | Q(A)) * |Q(A)| * \nonumber \\   
	& && P_{B}(B.Aid = v | Q(B)) * |Q(B)| \nonumber \\
	&  \leq \sum_{i = 1}^{k} min( &&\frac{P_A(A.id \in bin_i|Q(A)) * |Q(A)|}{V^{*}_i(A.id)}, \nonumber \\ 
	& && \frac{P_B(B.Aid \in bin_i|Q(B)) * |Q(B)|}{V^{*}_i(B.Aid)}) * \nonumber \\ 
	 & && V^{*}_i(A.id) * V^{*}_i(B.Aid)
	\label{equ: bound2join}
\end{alignat}

\smallskip
\noindent \underline{\textbf{Case2: Joining three tables $A.id = B.Aid = C.Aid$:}} 

Let $V_i^{*}(A.id) = MAX_{v \in bin_i} |A.id = v|$ be the most frequent value (MFV) count of $A.id$ in a bin $bin_i$, and same for $V_i^{*}(B.Aid)$, $V_i^{*}(C.Aid)$. We have the following bound:

\begin{align}
	|Q| & = \sum_{v \in D(A.id)} P_{A}(A.id = v | Q(A)) * |Q(A)| *  \nonumber \\   
	& \hspace{2em} P_{B}(B.Aid = v | Q(B)) * |Q(B)| * P_{C}(C.Aid = v | Q(C)) * |Q(C)| \nonumber \\
	& \leq \sum_{i = 1}^{k} min\{ \frac{P_{A}(A.id \in bin_i | Q(A)) * |Q(A)|}{V_i^{*}(A.id)}, \nonumber \\ 
	& \hspace{5em} \frac{P_{B}(B.Aid \in bin_i | Q(B)) * |Q(B)|}{V_i^{*}(B.Aid)} \nonumber \\ 
	& \hspace{5em} \frac{P_{C}(C.Aid \in bin_i | Q(C)) * |Q(C)|}{V_i^{*}(C.Aid) }\} * \nonumber \\ 
	& \hspace{5em} {V_i^{*}(A.id)} * {V_i^{*}(B.Aid) } * {V_i^{*}(C.Aid)} 
\end{align}

\smallskip
\noindent \underline{\textbf{Case3: Joining three tables $A.id = B.Aid$, $B.id = C.Bid$:}}

Let $V_i^{*}(A.id) = MAX_{v \in bin_i} |A.id = v|$ be the most frequent value (MFV) count of $A.id$ in a bin $bin_i$, and same for $V_i^{*}(B.Aid)$, $V_i^{*}(B.id)$, and $V_i^{*}(C.Bid)$.
We have the following bound, where $Upper(A \fullouterjoin B)$ is derived from Equation~\ref{equ: bound2join}.

\begin{align}
	|Q| &= \sum_{v_1 \in D(A.id)} \sum_{v_2 \in D(B.id)} P_{A}(A.id = v_1 | Q(A)) * |Q(A)| * \nonumber \\ 
	&\hspace{6em} P_{B}(B.Aid = v_1, B.id = v_2 | Q(B)) * |Q(B)| *  \nonumber \\ 
	& \hspace{6em} P_{C}(C.Bid = v_2 | Q(C)) * |Q(C)| \nonumber \\ 
	&\leq \sum_{i = 1}^{k} min\{ \frac{Upper(Q(A) \fullouterjoin Q(B))}{V_i^{*}(A.id) * V_i^{*}(B.id)}, \nonumber \\ 
	& \hspace{5em}  \frac{P_{C}(C.Aid \in bin_i | Q(C)) * |Q(C)|}{V_i^{*}(C.Bid)}\} *  \nonumber \\ 
	& \hspace{5em} * V_i^{*}(A.id) * V_i^{*}(B.id) * {V_i^{*}(C.Bid)}
\end{align}

\smallskip
\noindent \underline{\textbf{Case4: Self join of one table $A.id = A.id2$:}}

Let $V_i^{*}(A.id) = MAX_{v \in bin_i} |A.id = v|$ be the most frequent value (MFV) count of $A.id$ in a bin $bin_i$, and same for $V_i^{*}(A.id_2)$. We have the following bound:

\begin{alignat}{2}
	|Q| & = \sum_{v = D(A.id)} && P_{A}(A.Id = v | Q(A)) * |Q(A)| * \nonumber \\   
	& && P_{A}(A.id_2 = v | Q(A')) * |Q(A')| \nonumber \\
	&  \leq \sum_{i = 1}^{k} min( &&\frac{P_A(A.id \in bin_i|Q(A)) * |Q(A)|}{V^{*}_i(A.id)}, \nonumber \\ 
	& && \frac{P_A(A.id_2 \in bin_i|Q(A')) * |Q(A')|}{V^{*}_i(A.id_2)}) * \nonumber \\ 
	& && V^{*}_i(A.id) * V^{*}_i(A.id_2)
\end{alignat}

\smallskip
\noindent \underline{\textbf{Case5: Cyclic join of two tables $A.id = B.Aid$, $A.id_2 = B.Aid_2$:}}

Let $V_i^{*}(A.id) = MAX_{v \in bin_i} |A.id = v|$ be the most frequent value (MFV) count of $A.id$ in a bin $bin_i$, and same for $V_i^{*}(B.Aid)$, $V_i^{*}(A.id_2)$, and $V_i^{*}(B.Aid_2)$. We have the following bound:

\begin{align}
	|Q| &= \sum_{ v_1 \in D(A.id)} \sum_{v_2 \in D(A.id_2)} P_{A}(A.id = v_1, A.id_2 = v_2 | Q(A)) * \nonumber \\ 
	& \hspace{3em} |Q(A)| * P_{B}(B.Aid = v_1, B.Aid_2 = v_2 | Q(B)) * |Q(B)| * \nonumber \\ 
	& \hspace{3em} P_{A}(A.id_2 = v_2| Q(A')) * |Q(A')| \nonumber \\ 
	& \leq \sum_{i = 1}^{k} min\{\frac{Upper(Q(A) \fullouterjoin Q(B)}{V_i^{*}(A.id_2) * V_i^{*}(B.Aid)} , \frac{Upper(Q(A') \fullouterjoin Q(B))}{V_i^{*}(A.id) * V_i^{*}(B.Aid_2)}\} * \nonumber \\ 
	& \hspace{5em} V_i^{*}(A.id_2) * V_i^{*}(B.Aid) * V_i^{*}(A.id) * V_i^{*}(B.Aid_2)
\end{align}

\section{Greedy bin selection Algorithm details}

We observe that the bound on a particular bin $bin_i$ can be very loose if the MFV count $V^*_i$ is a large outlier in $bin_i$. Taking the two table join query as an example, if $bin_i$ contains only one value that appears $100$ times in $A.id$ but $10,000$ values that only appear once in $B.Aid$, then the bound could be $100$ times larger than the actual cardinality. 
The existing equal-width or equal-depth binning strategy can generate very large estimation errors, so we design a new binning strategy called the greedy bin selection algorithm (GBSA) to optimize our bound tightness.

The objective of GBSA is to minimize the variance of the value counts within $bin_i$.
In the extreme case, if the value counts have zero variance for all join keys in one equivalent key group, then our bound can output the exact cardinality (if with perfect single table \CE models). 
However, as the same bin partition will be applied for all equivalent join keys, minimizing the value counts variance of $bin_i$ on the domain of one key may result in a bad bin for other keys.
Jointly minimizing the variance of one bin for all join keys has exponential complexity.
There, GBSA uses a greedy algorithm to iteratively minimize the bin variance for all join keys.
At a high level, GBSA first optimizes the minimal variance bins with half number of binning budget $k/2$ on the domain of one join key. Then, it recursively updates these bins by minimizing the variance of other join keys using the rest half of the budget.
The details of GBSA are provided in Algorithm~\ref{algo: gbsa}.

\begin{algorithm}[t]
	\small
	\caption{Greedy Bin Selection Algorithm (GBSA)}
	\label{algo: gbsa}
	\begin{flushleft}
		\textbf{Input}: 	
		Equivalent key groups $Gr_1, \ldots, Gr_m$, where $Gr_i = \{Id_i^1, \ldots, Id_i^{|Gr_i|} \}$;
		
		Column data $\mathcal{D}(Id_i^j)$ of all join keys in the DB instance $\mathcal{D}$;
		
		Number of bins $k_i$ for each group $Gr_i$.
	\end{flushleft}
	\begin{algorithmic}[1]
		\For{$Gr_i \in \{Gr_1, \ldots, Gr_m\}$}
		\State $Bin(Gr_i) \gets$ []
		\State $Gr_i' \gets sort\_key\_based\_on\_domain\_size(\mathcal{D}, Gr_i)$
		\State $Bin(Gr_i) \gets get\_min\_variance\_bins(\mathcal{D}(Gr_i'[1]), k_i/2)$
		\State remain\_bins $\gets k_i/2$
		\For{$j \in \{ 2, \ldots, |Gr_i'| \} $}
		\State binned\_data $\gets apply\_bin\_to\_data(\mathcal{D}(Gr_i'[j]), Bin(Gr_i))$
		\State bin\_variance $\gets calculate\_variance$(binned\_data)
		\State arg\_sort\_idx $\gets arg\_sort\_decreasing(bin\_variance)$
		\For{$p \in arg\_sort\_idx[1:remain\_bins/2]$}
		\State $Bin(Gr_i) \gets min\_variance\_dichotomy(Bin(Gr_i)[p]$, binned\_data[p])
		\EndFor
		\State remain\_bins $\gets$ remain\_bins/2
		\EndFor
		\EndFor
		\State \textbf{return}  $\{Bin(Gr_i) | i, \ldots, m\}$
	\end{algorithmic}
	%\vspace{-1em}
\end{algorithm}

\Ours analyzes the schema of DB instance $\mathcal{D}$ to derive $m$ equivalent key groups $Gr_1, \ldots, Gr_m$, each of which contains $|Gr_i|$ join keys $Id_i^1, \ldots, Id_i^{|Gr_i|}$. Let $\mathcal{D}(Id_i^j)$ represents the column data (domain) of join key $Id_i^j$.
GBSA will derive the sub-optimal binning with $k_i$ bins $Bin(Gr_i) = \{bin_1, \ldots, bin_{k_i} \}$ for each the equivalent key group $Gr_i$ (line 1). We explain the procedure of this algorithm for binning one group $Gr_i$ (line 2-14). 

First, GBSA sorts the join keys $\{d_i^1, \ldots, Id_i^{|Gr_i|}\}$ in decreasing order based on their domain size (line 3) to get $Gr_i'=\{Id_i^{1'}, \ldots, Id_i^{|Gr_i|'}\}$. 
We apply half of the binning budget to generate $k_i/2$ minimal variance bins on the domain of $d_i^{1'}$ (line 4). Note that minimal variance bins on a single attribute can be easily obtained by sorting the value counts of $\mathcal{D}(d_i^{1'})$ and applying equal-depth binning over the sorted values.
The remaining number of bins available $remain_bins$ is $k_i/2$ (line 5).
Then for each rest of the join key $d_i^j$, GBSA applies the current bins $Bin(Gr_i)$ to its data column $\mathcal{D}(d_i^j)$, calculates the variance for each bin, and sorts these bins in decreasing order (line 6-9).
For the top $remain_bins/2$ bins with the largest variance, we dichotomize each bin into two bins to minimize the variance on join key $d_i^j$ (lines 10 -12). 
We iterative the above procedure until all join keys are optimized.

According to our evaluation results, the GBSA has a significant impact on improving our probabilistic bound tightness and estimation effectiveness.

\begin{figure*}
	\centering
	\includegraphics[width=17.8cm]{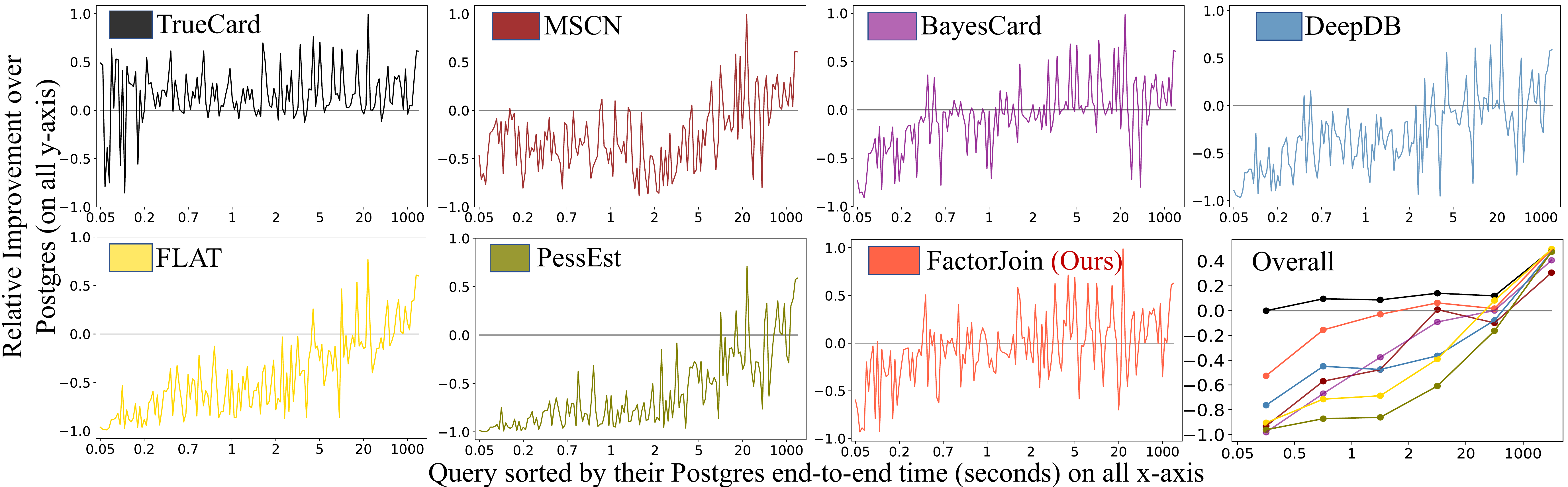}
	\vspace{-2.3em}
	\caption{Per query performance of STATS-CEB.}
	\label{fig: intervals-STATS}
\end{figure*}

\begin{figure*}
	\centering
	\includegraphics[width=17.8cm]{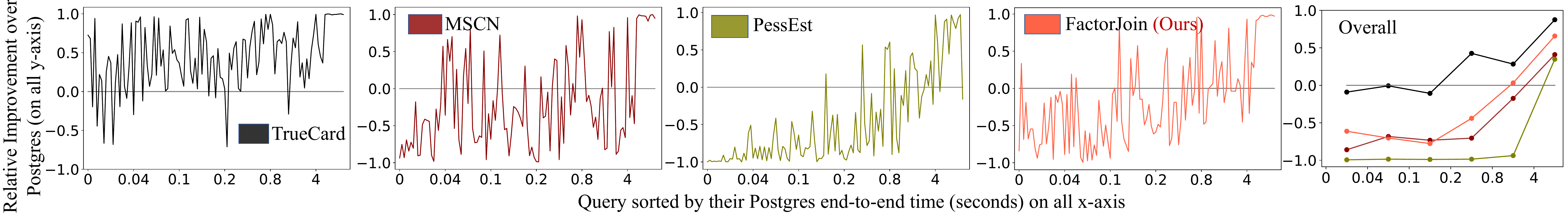}
	\vspace{-2.3em}
	\caption{Per query performance of STATS-CEB.}
	\label{fig: intervals-IMDB}
\end{figure*}

\section{Additional experiments and details}
In this section, we provide the detailed comparision of all baselines on \emph{STATS-CEB} and \emph{IMDB-JOB} benchmarks.

\smallskip
\noindent \underline{\textbf{Performance on STATS-CEB:}}
We sort the $146$ queries of \emph{STATS-CEB} based on their Postgres end-to-end runtime and cluster them into $6$ different runtime intervals.
Figure~\ref{fig: intervals-STATS} reports relative improvements of all competitive baselines over \emph{Postgres} for each query and the overall performance comparision on each cluster of queries on the last figure. 

For the very short-running queries (which represent an OLTP-like workload), \emph{Postgres} is the best among all baselines. These baselines perform worse because the estimation latency plays a significant role in these queries.
We observe that improving estimation accuracy has a very limited effect on the query plan quality of short-running queries. This also explains why the optimal \emph{TrueCard} only marginally outperforms \emph{Postgres} on queries with less than $2s$ of runtime. 
%The estimation latency plays a significant role in these queries.
Overall, \Ours has the best performance among all baselines except for \emph{Postgres}.

For the extremely long-running queries, the advantage of the learning-based methods over \emph{Postgres} gradually appear. The reason is that estimation latency becomes increasingly insignificant for queries with a long execution time. \Ours has comparable performance as the SOTA learning-based methods on these queries.

\smallskip
\noindent \underline{\textbf{Performance on IMDB-JOB:}} 
Since a large proportion of the queries in  \emph{IMDB-JOB} workload are short-running queries, \emph{Postgres} has a better performance than all baselines for more than half of the queries. Similar to STATS-CEB queries, we do not observe any performance gain for using the learned \CE method over \emph{Postgres} on queries with less on $1s$ runtime. In fact, the optimal \emph{TrueCard} also barely improves the \emph{Postgres}.
This suggests that the learned methods shoulf fall back to \emph{Postgres} for OLTP workloads.

Similar to \emph{STATS-CEB}, \Ours also has a significantly better performance over all other baselines except \emph{Postgres}.

\end{document}